\documentclass[floatfix,a4paper,preprint]{revtex4-2}
\usepackage[utf8]{inputenc}
\usepackage{amsfonts}
\usepackage{amssymb}
\usepackage{amsmath}
\usepackage{graphicx}
\usepackage{subcaption}
\usepackage{xcolor}
\usepackage[width=1\linewidth]{caption}

\usepackage{dcolumn}
\newcolumntype{d}[1]{D{.}{.}{#1}}

\begin{document}

\hfill\today

\title{Collision-induced three-body polarizability of helium}
\author{J. Lang}
\email[Corresponding author: ]{jakub.lang@chem.uw.edu.pl}
\author{M. Przybytek}
\author{M. Lesiuk}
\author{B. Jeziorski}
\affiliation{Faculty of Chemistry, University of Warsaw, Pasteura 1, 02-093 Warsaw, Poland}
\begin{abstract}
% the coupled-cluster theory with approximate account of three-electron excitations (the CC3 method).
We  present  first-principles theoretical determination of the three-body polarizability  and the third dielectric virial coefficient of helium.  
%with rigorous uncertainty estimates. 
Coupled-cluster theory and the full configuration interaction procedure 
were used to perform required electronic structure  calculations. 
The mean absolute relative uncertainty of the trace of the three-body polarizability tensor, resulting from the incompleteness of orbital basis set, was determined using extrapolation techniques and found to be 4.7\%. Additional uncertainty due to approximate treatment of triple and the neglect of higher excitations was estimated at 5.7\% using full configuration interaction calculations. 
% for a subset ($\sim$10\%) of the considered trimer configurations.
An analytic function %representing the three-body polarizability
was developed to correctly describe the short-range behavior of the polarizability and its asymptotic decay for trimer configurations corresponding to both the three-atom and the atom-diatom fragmentation channels.
We also developed an %short-range 
analytic function describing the local behavior of the total uncertainty of our calculations.
Using both fits we calculated the third dielectric virial coefficient and its uncertainty using the classical and semiclassical Feynman-Hibbs approaches. 
%using the quadratic effective potential 
The results of our calculations were compared with available experimental data and with recent   Path-Integral Monte Carlo (PIMC) calculations [G. Garberoglio \emph{et al.}, J. Chem. Phys. \textbf{155}, 234103 (2021)] employing the so-called superposition approximation of the three-body polarizability.
For temperatures above 200~K we observed significant discrepancy between the classical results  obtained  using either the superposition approximation or the \emph{ab initio} computed polarizability.
While the superposition approximation may appear to be a crude one, for temperatures from 10~K up to 200~K, the differences between PIMC calculations %of Garberoglio \emph{et al.} 
and our semiclassical %
Feynman-Hibbs calculations 
employing the \emph{ab initio} polarizability 
%with an accurate 3B polarizability and the 
%semiclassical FH approximation% 
are several times smaller than uncertainties of our results. 
This shows that both the superposition approximation and the semiclassical Feynman-Hibbs approach work very well in this temperature range. 
Except at low temperatures our results agree very well with the available experimental data but have much smaller uncertainties. The theoretical data reported in this work  eliminate  the
main accuracy bottleneck of the development of optical pressure standard
[C. Gaiser \emph{et al.},
Ann. Phys. (Berlin) {\bf 534}, 2200336 (2022)]
and are expected to facilitate  further %open new perspectives for  
progress in the field of quantum thermal metrology.
\end{abstract}

\maketitle

%%%%%%%%%%%%%%%%%%%%%%%%%%%%%%%%%%%%%%%%%%%%%%%%%%%%%%%%%%%%%%%%%%%%%%
%%%%%%%%%%%%%%%%%%%%%%%%%%%%%%%%%%%%%%%%%%%%%%%%%%%%%%%%%%%%%%%%%%%%%%
%%%%%%%%%%%%%%%%%%%%%%%%%%%%%%%%%%%%%%%%%%%%%%%%%%%%%%%%%%%%%%%%%%%%%%
\section{Introduction}
%%%%%%%%%%%%%%%%%%%%%%%%%%%%%%%%%%%%%%%%%%%%%%%%%%%%%%%%%%%%%%%%%%%%%%
%%%%%%%%%%%%%%%%%%%%%%%%%%%%%%%%%%%%%%%%%%%%%%%%%%%%%%%%%%%%%%%%%%%%%%
%%%%%%%%%%%%%%%%%%%%%%%%%%%%%%%%%%%%%%%%%%%%%%%%%%%%%%%%%%%%%%%%%%%%%%

Interactions of atoms induce small changes in the polarizability of atomic clusters.
In turn, these small changes affect the measured dielectric and optical properties of gases \cite{1989borysowcollision,1995CPL...247..440M,1968mctaguecollision}.
In particular, the isotropic interaction-induced polarizability determines the polarized Raman scattering spectrum \cite{Moszynski1996raman}, while the polarizability anisotropy determines the depolarized spectrum and contributes to the density dependence of the Kerr constant \cite{Skomorowski2013}.
The mean polarizability is also related to the dielectric virial coefficients of gases through the Clausius-Mossotti equation \cite{1955buckinghamstatistical}.
The dielectric virial expansion has been extensively used in metrology, formerly in the determination of the Boltzmann constant \cite{2017Metro..54..280G} and recently in the development of new temperature \cite{gaiser2015dielectric,2017Metro..54..141G,gaiser2020thermodynamic,2021Metro..58b5008R} and pressure standards \cite{schmidt2007polarizability,2020NatPh..16..177G}.
Nowadays, the dielectric-constant gas thermometry (DCGT) \cite{gaiser2015dielectric,2017Metro..54..141G,gaiser2020thermodynamic} and capacitance measurements enable the determination of thermodynamic temperature with accuracy of a few parts per million.
These experiments, performed mostly on atomic gases, require accurate knowledge of the dielectric virial coefficients which can be obtained from experiments and, more recently, from theory. 
%Recently, a gas pressure standard based on  capacitance  measurement and \emph{ab initio} 
%theory data passed stress  test \cite{Geiser_Annalen_2022}. 

In the case of helium, the polarizability of a single atom can be calculated with an unprecedented accuracy \cite{2015PhRvL.114q3004P,2020PhRvA.101b2505P}.
In effect, a vast majority of the uncertainty in the determination of thermophysical properties comes from the error in the second- and higher-order virial coefficients \cite{2020NatPh..16..177G}.
The second dielectric virial coefficient requires the knowledge of the two-body interaction potential and the two-body interaction-induced polarizability.
The third dielectric virial coefficient requires the same quantities and, additionally, the knowledge of three-body properties such as the non-additive three-body interaction potential and the three-body interaction-induced polarizability.

Two- and three-body interaction potentials and two-body interaction-induced polarizabilities for noble gas atoms with their contributions to the virial coefficients were already reported in the literature \cite{czachorowski2020second,cencek2009threehe,2011JChPh.135a4301C,2010MolPh.108.3335V,2010JChPh.133i4304P,cencek2007three,cencek2013three}. 
On the contrary, the effects of three-body interaction-induced polarizabilities were studied only experimentally for various systems \cite{bafile1987three,bafile1991third,pestelli1994three,barocchi1988interaction,van1988density} and no complete theoretical description of the three-body polarizability surface is available in the literature. 
The first attempt to develop a theoretical model describing the three-body polarizabilities was made by Buckingham and Hands \cite{1991CPL...185..544B}. 
They used the dipole-induced-dipole (DID) model, but were unable to assess its accuracy as no sufficiently accurate three-body polarizability calculations were available at that time.
Their comparison to the self-consistent electron pair calculations (SCEP) of P\'erez \emph{et al.} \cite{perez1984three} showed a substantial difference between the DID model and the \emph{ab initio} results.
A few years later, Champagne \emph{et al.} \cite{champagne2000nonadditive} used the non-local response theory \cite{Li96nonlocal,Li97nonlocal} and derived analytical behavior of the three-body interaction-induced polarizability for inert gases and simple molecules valid when all distances between interacting atoms or molecules are large. 
Accounting for the DID interaction, quadrupolar induction, dispersion, and induction-dispersion effects they obtained all asymptotic terms behaving at large distances as $r_{12}^{-l} r_{23}^{-m} r_{13}^{-n}$, with $l+m+n \le 9$, where $r_{ij}$ are interatomic or intermolecular distances.
Unfortunately, they did not compare their results with finite distance calculations of P\'erez \emph{et al.} \cite{perez1984three} which were, to the best of our knowledge, the only \emph{ab initio} calculations of the three-body polarizability of helium available at that time in the literature. 
Other \emph{ab initio} calculations of the three-body polarizability were performed by Aleman \emph{et al.} \cite{aleman1992ab} at the SCF level of theory and concerned neon atoms.

For the purpose of calculating the third dielectric virial coefficient, Heller and Gelbart \cite{heller1974superpos} suggested the so-called superposition approximation to obtain an estimate of the three-body polarizability using only two-body polarizabilities. 
However, they used crude two-body polarizability and interaction potential for helium in their third dielectric virial estimation. 
Recently, Garberoglio \emph{et al.} \cite{garberoglio20213dielec} used this approximation in Path-Integral Monte Carlo (PIMC) \cite{feynman1965pathintegrals,garberoglio2009firstprincip,2011JChPh.134m4106G,shaul2012path} calculations to obtain the third dielectric virial coefficient for helium employing the more accurate two-body polarizability reported by Cencek \emph{et al.} \cite{2011JChPh.135a4301C}.
The values of the third dielectric virial coefficient for helium were also given in the paper of Alder \emph{et al.} \cite{1980PNAS...77.3098A}.
However, as noted in Ref.~\onlinecite{garberoglio20213dielec}, these authors arbitrarily adjusted the two-body polarizability in order to improve the agreement with the experimental data.

In the present work, we report accurate \emph{ab initio} calculations of the three-body polarizability for helium performed using the coupled-cluster theory with approximate account of three-electron excitations (the CC3 method) \cite{hald2003calculation}.
We also present an accurate analytical fit of the calculated three-dimensional isotropic polarizability surface.
For metrology applications we provide a three-dimensional fit of the corresponding local uncertainties of this surface.
Finally, the reported polarizability surface is employed in classical and semiclassical calculations to obtain the third dielectric virial coefficient of helium.
When applied in a fully quantum path-integral calculations \cite{Garberoglio_at_al_2022}, our three-body polarizability surface was utilized in the development of a new primary pressure standard based on capacitance measurements, which recently passed stress test with 2~parts per million (ppm) accuracy level achieved at 7~MPa \cite{Geiser_Annalen_2022}. 

%%%%%%%%%%%%%%%%%%%%%%%%%%%%%%%%%%%%%%%%%%%%%%%%%%%%%%%%%%%%%%%%%%%%%%
%%%%%%%%%%%%%%%%%%%%%%%%%%%%%%%%%%%%%%%%%%%%%%%%%%%%%%%%%%%%%%%%%%%%%%
%%%%%%%%%%%%%%%%%%%%%%%%%%%%%%%%%%%%%%%%%%%%%%%%%%%%%%%%%%%%%%%%%%%%%%
\section{Three-body polarizability of helium and its asymptotics at large interatomic distances}
\label{theory}
%%%%%%%%%%%%%%%%%%%%%%%%%%%%%%%%%%%%%%%%%%%%%%%%%%%%%%%%%%%%%%%%%%%%%%
%%%%%%%%%%%%%%%%%%%%%%%%%%%%%%%%%%%%%%%%%%%%%%%%%%%%%%%%%%%%%%%%%%%%%%
%%%%%%%%%%%%%%%%%%%%%%%%%%%%%%%%%%%%%%%%%%%%%%%%%%%%%%%%%%%%%%%%%%%%%%

 The components $\alpha_{\mu\nu}$ of a polarizability tensor $\boldsymbol{\alpha}$ 
 %of an arbitrary system in a specific quantum state is
 are defined as   
\begin{equation}\label{polar_definition}
\alpha_{\mu\nu} = -\left(
\frac{\partial^2 E}{\partial F_\mu\,\partial F_\nu}
\right)_{F_\mu=F_\nu=0},
\end{equation}
where $\mu,\nu \in \{x,y,z\}$ and $E$ is the energy of a considered quantum state in the presence of a static and uniform electric field $\mathbf{F}$ with components $F_\mu$. 
In this work, we consider a system of three identical closed-shell atoms at fixed positions in space specified by vectors $\mathbf{r}_i$, $i=1,2,3$. 
The energy and, consequently, the components of the polarizability tensor depend on these atomic positions.
For weakly interacting systems, like helium atoms, it is useful to represent this dependence in terms of the so-called many-body expansion:
\begin{equation}\label{polar_total}
\boldsymbol{\alpha}(\mathbf{r}_1,\mathbf{r}_2,\mathbf{r}_3)= 
 3\,\alpha_0 \, \mathbf{e}  
 +\boldsymbol{\alpha}_2(\mathbf{r}_1,\mathbf{r}_2)
 +\boldsymbol{\alpha}_2(\mathbf{r}_2,\mathbf{r}_3)
 +\boldsymbol{\alpha}_2(\mathbf{r}_1,\mathbf{r}_3) 
 +\boldsymbol{\alpha}_3(\mathbf{r}_1,\mathbf{r}_2,\mathbf{r}_3),
\end{equation}
where $\alpha_0$ is the atomic polarizability, $\mathbf{e}$ is the unit tensor, $\boldsymbol{\alpha}_2(\mathbf{r}_i,\mathbf{r}_j)$ is the two-body interaction-induced polarizability tensor of the pair of atoms at $\mathbf{r}_i$ and $\mathbf{r}_j$,
\begin{equation}\label{polar_2body}
\boldsymbol{\alpha}_2(\mathbf{r}_i,\mathbf{r}_j)=
 \boldsymbol{\alpha}(\mathbf{r}_i,\mathbf{r}_j)
-2\,\alpha_0 \,\mathbf{e},
\end{equation}
and $\boldsymbol{\alpha}_3(\mathbf{r}_1,\mathbf{r}_2,\mathbf{r}_3)$ is the three-body interaction-induced polarizability tensor defined essentially by Eq.~(\ref{polar_total}). 
By $\boldsymbol{\alpha}(\mathbf{r}_i,\mathbf{r}_j)$ we denote the tensor of the total polarizability of two atoms located at $ \mathbf{r}_i$ and  $\mathbf{r}_j$. 
The tensor $\boldsymbol{\alpha}_3(\mathbf{r}_1,\mathbf{r}_2,\mathbf{r}_3)$, representing the pair-wise non-additive part of the total interaction-induced polarizability tensor $\boldsymbol{\alpha}(\mathbf{r}_1,\mathbf{r}_2,\mathbf{r}_3) -3\,\alpha_0 \,\mathbf{e}$, can be calculated from Eq.~(\ref{polar_total}) using precomputed two-body polarizability tensors  $\boldsymbol{\alpha}_2(\mathbf{r}_i,\mathbf{r}_j)$.
In this work it has been calculated directly from the %so-called supermolecular approach from the
formula
\begin{equation}\label{polar_3body}
\boldsymbol{\alpha}_3(\mathbf{r}_1,\mathbf{r}_2,\mathbf{r}_3)= 
 \boldsymbol{\alpha}(\mathbf{r}_1,\mathbf{r}_2,\mathbf{r}_3)  
-\boldsymbol{\alpha}(\mathbf{r}_1,\mathbf{r}_2)
-\boldsymbol{\alpha}(\mathbf{r}_2,\mathbf{r}_3)
-\boldsymbol{\alpha}(\mathbf{r}_1,\mathbf{r}_3)  
+3\,\alpha_0 \,\mathbf{e}.
\end{equation}
The assumption that we are dealing with closed-shell systems is essential since for three interacting open-shell atoms the quantum states of subsystems needed to determine the two-body contributions are not well defined (the simplest example is the ground, doublet state of three hydrogen atoms).
Formula (\ref{polar_total}) is particularly useful when the three-body part is much smaller than the sum of two-body polarizabilities.
This is always true when the interatomic distances are large, but in the short-range region the three-body contribution can be large or even dominant. 

For three atoms, the three-body polarizability tensor contains four independent components which can be used to define their combinations of physical interest \cite{gray1984fluids1}.  
In this work, we restrict ourselves to the mean (isotropic) polarizabilities $\alpha_n$, $n=2,3$,
\begin{equation}\label{polar_iso}
\alpha_n=
\frac{1}{3} \,{\rm Tr} \,\boldsymbol{\alpha}_n ,
\end{equation}
because of their relevance for calculations of the dielectric virial coefficients. 

While the components of the polarizability tensor depend on the spatial orientation of the triangle formed by the atoms, the isotropic polarizability $\alpha_3$ %Eq.~(\ref{polar_iso})
is invariant to rotations of the system as a whole and depends only on the interatomic distances.
When two or three interatomic distances are large one can derive asymptotically valid analytic expressions for the dependence of $\alpha_3$ on $r_{12}$, $r_{23}$, and $r_{13}$.
While only one fragmentation is possible for two-body systems, a three-body system can separate through two distinct fragmentation channels.
The first channel, that we shall refer to as the atomic channel, corresponds to the situation when all three atoms move away independently to infinity (all three interatomic distances are arbitrarily large). 
The second channel, referred to as the atom-diatom channel, describes the situation when one atom moves to infinity while the remaining two atoms stay at a limited, small distance from each other (only two interatomic distances are arbitrarily large).

The leading term of the asymptotic expansion of the isotropic three-body polarizability in the first fragmentation channel is predicted by the dipole-induced-dipole (DID) model
\cite{1991CPL...185..544B,champagne2000nonadditive} and takes the form 
\begin{equation}\label{polar_asym1}
{\alpha}_{\mathrm{DID}} = 
2\,\alpha_0^3\;\mathcal{S}_{123}\frac{P_2(\cos{\theta_3})}{r_{13}^{3}r_{23}^{3}},
\end{equation}
where $\theta_3$ is the angle between vectors $\mathbf{r}_1-\mathbf{r}_3$ and $\mathbf{r}_2-\mathbf{r}_3$, $P_n(x)$ is the Legendre polynomial of order $n$, and $\mathcal{S}_{123}$ denotes the operator that symmetrizes the expression it precedes with respect to the variables $\mathbf{r}_1$, $\mathbf{r}_2$, and $\mathbf{r}_3$. 
The classical expression for the third dielectric virial coefficient requires the integration of Eq.~(\ref{polar_asym1}) over $r_{13}$, $r_{23}$, and $\theta_3$.
It is clear that this integration is ill-defined as it is divergent both at small and at large values of $r_{13}$ or $r_{23}$.
In Sec.~\ref{sec:virial_coeff} and in Appendix~\ref{regularization} we show how this divergence is eliminated by an appropriate regularization. 
As a result, the inclusion of Eq.~(\ref{polar_asym1}) in the analytic representation of the three-body polarizability leads to a finite result.

The leading term of the asymptotic expansion in the second (atom-diatom) fragmentation channel can be obtained from the results of Hunt \emph{et al.} \cite{hunt1988transient} concerning the asymptotics of the interaction-induced polarizability of atom-diatom pairs.
It suffices to substitute the value of the polarizability anisotropy $\beta$     %=\alpha_{\parallel} - \alpha_{\perp}$
of the diatom with the polarizability anisotropy of the helium dimer and to subtract the isotropic interaction-induced polarizability of the two remaining helium pairs. 
This procedure leads to the following formula
\begin{equation}\label{polar_asym2}
{\alpha}_{\mathrm{a-d}} =
\frac{2}{3}\,\alpha_0\;\mathcal{S}_{123}\,\beta(r)\frac{P_2(\cos{\theta})}{R^3},
\end{equation}
where $r$, $R$, and $\theta$ are the Jacobi coordinates: $r$ and $R$ are the lengths of the vectors $\mathbf{r}=\mathbf{r}_2-\mathbf{r}_1$ and $\mathbf{R}=\mathbf{r}_3-(\mathbf{r}_1+\mathbf{r}_2)/2$, respectively, and $\theta$ is the angle between the vectors $\mathbf{r}$ and $\mathbf{R}$. 
By $\beta(r)$ we denote the polarizability anisotropy of two helium atoms separated by a distance $r$.
Since the function being symmetrized here [and also in Eq.~(\ref{polar_asym1})] is already symmetric with respect to the permutation of $\mathbf{r}_1$ and $\mathbf{r}_2$, the symmetrizer $\mathcal{S}_{123}$ can be replaced by the operator $2 + 2\,\mathcal{P}_{13} + 2\,\mathcal{P}_{23}$, where the operator $\mathcal{P}_{ij}$ permutes the vectors $\mathbf{r}_i$ and $\mathbf{r}_j$.

It is easy to see that when one of the interatomic distances is small, Eq.~(\ref{polar_asym1}) cannot represent the exact long-range behavior of the three-body polarizability in the atom-diatom fragmentation channel given by Eq.~(\ref{polar_asym2}). 
If the long-range terms included in the analytical representation of the three-body polarizability were given only by Eq.~(\ref{polar_asym1}) (and the higher-order terms derived in Ref.~\onlinecite{champagne2000nonadditive}), then the long-range behavior exhibited in Eq.~(\ref{polar_asym2}) would have to be modelled by fitting with short-range functions.
This is a very impractical, if not impossible, procedure. 
Therefore, the asymptotics of Eq.~(\ref{polar_asym2}) must be included in the analytic function used to represent the three-body polarizability.
It must be realized, however, that there are regions in the configuration space where Eqs.~(\ref{polar_asym1}) and (\ref{polar_asym2}) give the same, or very close, results. 
Indeed, since for a pair of rare gas atoms, the anisotropy $\beta(r)$ behaves at large $r$ as $6\,\alpha_0^2/r^3$~\cite{Certain1971}, Eqs.~(\ref{polar_asym1}) and (\ref{polar_asym2}) give the same values when both $r$ and $R$ become large. 
We thus conclude that the inclusion of both Eq.~(\ref{polar_asym1}) and Eq.~(\ref{polar_asym2}) in an analytic representation of the three-body polarizability would lead to a double counting in some parts of the configuration space.
A solution of this difficulty is presented in Sec.~\ref{sec:fit}. 

%%%%%%%%%%%%%%%%%%%%%%%%%%%%%%%%%%%%%%%%%%%%%%%%%%%%%%%%%%%%%%%%%%%%%%
%%%%%%%%%%%%%%%%%%%%%%%%%%%%%%%%%%%%%%%%%%%%%%%%%%%%%%%%%%%%%%%%%%%%%%
%%%%%%%%%%%%%%%%%%%%%%%%%%%%%%%%%%%%%%%%%%%%%%%%%%%%%%%%%%%%%%%%%%%%%%
\section{Computational details and results}
%%%%%%%%%%%%%%%%%%%%%%%%%%%%%%%%%%%%%%%%%%%%%%%%%%%%%%%%%%%%%%%%%%%%%%
%%%%%%%%%%%%%%%%%%%%%%%%%%%%%%%%%%%%%%%%%%%%%%%%%%%%%%%%%%%%%%%%%%%%%%
%%%%%%%%%%%%%%%%%%%%%%%%%%%%%%%%%%%%%%%%%%%%%%%%%%%%%%%%%%%%%%%%%%%%%%

%%%%%%%%%%%%%%%%%%%%%%%%%%%%%%%%%%%%%%%%%%%%%%%%%%%%%%%%%%%%%%%%%%%%%%
\subsection{Polarizability calculations}
%%%%%%%%%%%%%%%%%%%%%%%%%%%%%%%%%%%%%%%%%%%%%%%%%%%%%%%%%%%%%%%%%%%%%%

We performed \emph{ab initio} calculations of the three-body isotropic polarizability $\alpha_3$ 
%(r_{12},r_{23},r_{13})$ 
of helium for 748 triangles parameterized by grid points in the three-dimensional region defined by the triangle conditions for the coordinates $r_{12}$, $r_{23}$, and $r_{13}$.  
The procedure by which the set of grid points was generated is described in detail in Sec.~\ref{sec:fit}.
The calculated polarizabilities were obtained using the CC3 method \cite{hald2003calculation} as implemented in the Dalton 2018 package \cite{dalton,dalton2018}. 
For any given atomic configuration, individual components of the three-body polarizability tensor were calculated using Eq.~(\ref{polar_3body}).
All quantities in this formula were obtained using the basis set of the trimer, which is equivalent to applying the so-called counterpoise scheme to remove the basis set superposition error (BSSE) \cite{Boys1970,2009TChA..122...127H}.
Note that within this approach the polarizability tensor of a single helium atom is no longer diagonal and depends slightly on the position of the atom in the triangle. 
Therefore, the atomic term $3\,\alpha_{\mu\nu}$ in Eq.~(\ref{polar_3body}) must be replaced by the sum of three, in general different, terms coming from each atom. 
%Similarly, the two-body polarizabilities depend slightly on the position of the third atom in the triangle. 
Similarly, the two-body polarizabilities depend slightly on the shape of the triangle. 

All calculations were performed using a family of doubly-augmented correlation-consistent Gaussian basis sets, referred further as d$X$Z, where $X$ is the cardinal number, developed in Ref.~\onlinecite{cencek2012effects} specifically for interactions of helium atoms in their ground state. 
The most important difference between the d$X$Z basis sets and the standard Dunning's d-aug-cc-pV$X$Z basis sets for helium \cite{Dunning1989cc,Dunning1994he} is that in the former the augmenting functions, optimized for the coefficients in the asymptotic expansion of the pair potential, are more diffuse (have smaller exponents) than in the latter, especially for angular momenta $l\ge2$.
Therefore, the d$X$Z basis sets are more suitable for calculations of molecular properties, such as polarizability, that require proper description of the electronic wave function at large distances from the nuclei. The highest cardinal number of the employed basis sets was $X=5$.
Unfortunately, the polarizability calculations using the d5Z basis set were computationally too demanding to be routinely used for all atomic configurations. 
This is especially true for general geometries where the plane determined by positions of three non-colinear helium atoms is the sole symmetry element of the system. 
Therefore, we calculated only 45 points within the d5Z basis set, mostly for geometries with a higher symmetry where three atoms either form an isosceles (including an equilateral) triangle or are located on a straight line.
% It is also noteworthy that our d$X$Z basis sets indeed perform much better in the calculations of the polarizability than the standard d-aug-cc-pV$X$Z basis set family.
% In most cases we observed an improvement by two cardinal numbers, i.e., for each $X$ the d$X$Z basis set provides results comparable to the ones obtained with d-aug-cc-pV$(X+2)$Z.

\begin{table}
\caption{Basis set convergence of the isotropic three-body polarizability $\alpha_3$ of helium calculated at the CC3 level of theory using the d$X$Z family of basis sets. $\alpha_3$[TQ] is the extrapolated polarizability defined in Eq.~(\ref{extrapol}). 
Interatomic distances, $r_{ij}$, are in a.u. and the polarizability unit is in $10^{-3}$ a.u. 
The estimated absolute percentage uncertainty, $100\%\!\times\!|(\alpha_3[\mathrm{TQ}]-\alpha_3 [\mathrm{Q}])/ \alpha_3[\mathrm{TQ}]|$, is given in parentheses.}
\label{tab:basis_conv}
\begin{ruledtabular}
\begin{tabular}{d{1.2}d{1.2}d{1.2}d{2.5}d{2.5}d{2.5}d{2.13}}
\multicolumn{1}{c}{$r_{12}$} &
\multicolumn{1}{c}{$r_{23}$} &
\multicolumn{1}{c}{$r_{13}$} & 
\multicolumn{1}{c}{$X = \mathrm{T}$} & 
\multicolumn{1}{c}{$X = \mathrm{Q}$} &
\multicolumn{1}{c}{$X = 5$} &
\multicolumn{1}{c}{$ {\alpha}_3$[TQ]} \\ 
\hline
5.73 & 5.73 &  5.73 & -0.04065 & -0.04649 & -0.04683 & -0.05076\;(8.40\%) \\
5.87 & 5.87 &  5.87 & -0.05941 & -0.06395 & -0.06416 & -0.06726\;(4.93\%) \\
6.00 & 6.00 &  6.00 & -0.06747 & -0.07108 & -0.07122 & -0.07372\;(3.57\%) \\ 
6.50 & 6.50 &  6.50 & -0.06180 & -0.06339 & -0.06341 & -0.06456\;(1.80\%) \\
3.60 & 3.87 &  6.89 & -0.22858 & -0.21722 & -0.20985 & -0.20893\;(3.96\%) \\ 
4.15 & 4.15 &  7.00 &  0.11362 &  0.10895 &  0.10930 &  0.10554\;(3.23\%) \\
8.00 & 8.00 & 10.00 & -0.01548 & -0.01549 & -0.01547 & -0.01550\;(0.05\%) \\ 
7.95 & 8.23 & 10.10 & -0.01451 & -0.01452 & -0.01450 & -0.01453\;(0.04\%) \\
5.95 & 6.30 & 10.21 & -0.00014 & -0.00016 & -0.00022 & -0.00017\;(5.88\%) \\
\end{tabular}
\end{ruledtabular}
\end{table}

In Table~\ref{tab:basis_conv}, we present results obtained for several representative geometries.
For a vast majority of atomic configurations, the dQZ values are essentially converged and are very close to the d5Z results.
For example, for the equilateral triangle with sides lengths equal to 5.73 a.u. (a~geometry close to the minimum of the interaction potential of the helium trimer) we observe a difference of only 0.7\% between the two basis sets. 
A similar behavior was found for other atomic configurations.

The results obtained with finite basis sets were extrapolated to the complete basis set (CBS) limit. 
We assumed that the basis set truncation error of the calculated polarizability vanishes with the increasing value of the basis sets cardinal number $X$ as $1/X^n$. 
For the value of the exponent $n$ we adopted $n=3$, as recommended for extrapolation of the correlation energies \cite{Halkier1998,Helgaker2008}.
This choice can be justified by the fact that the polarizability is defined as the energy derivative, see Eq.~(\ref{polar_definition}), and its calculation involves no singular operators that could modify the value of $n$.
Such modifications appear, for example, in the extrapolation of relativistic corrections \cite{Halkier2000,Przybytek:17,cencek2012effects}.
With this choice of the exponent $n$, the CBS limit can be obtained using the following two-point formula
\begin{equation}\label{extrapol}
\alpha_3[X\!-\!1,X] = \alpha_3[X] +  (X\!-\!1)^3 \ \  \frac{\alpha_3[X]-\alpha_3[X\!-\!1]}{X^3-(X\!-\!1)^3},
\end{equation}
where $\alpha_3[X\!-\!1]$ and $\alpha_3[X]$ are polarizabilities calculated using basis sets with cardinal numbers $X\!-\!1$ and $X$, respectively, and $\alpha_3[X\!-\!1,X]$ is the extrapolated value. 
As the $X=5$ results were not available for all considered geometries, we consistently assumed the $\alpha_3[\mathrm{TQ}]$ extrapolants as our recommended values in the CBS limit. 

To each geometry of the trimer, we assigned a theoretical uncertainty of the recommended value of the polarizability. 
This uncertainty was estimated as the magnitude of the difference between the extrapolated value and the value computed with the largest basis set used in the extrapolation, i.e., $\sigma[\mathrm{TQ}]=|\alpha_3[\mathrm{TQ}]-\alpha_3[\mathrm{Q}]|$.
The results obtained using this approach are shown in Table~\ref{tab:basis_conv}.
In some cases, however, this simple procedure may lead to an underestimation of the uncertainty. 
This occurs when the dTZ value is accidentally close to the dQZ one, as can be observed, for example, for the triangles with side lengths $(8.0, 8.0, 10.0)$ a.u.\ and $(7.95, 8.23, 10.10)$ a.u.
In these two cases, the estimated uncertainty is only about 0.05\% of the extrapolated polarizability, which is half of the difference between the $X=4$ and $X=5$ results. 
To avoid the underestimation, any uncertainty estimated to be lower than 0.1\% of the extrapolated value was scaled up to a conservative value of 1.0\%. 
From now on, we use the notation $\sigma[\mathrm{TQ}]$ to refer to the modified uncertainty.
On average, the modified estimated uncertainty is 4.7\%.

To obtain a more reliable estimation of the theoretical uncertainty, we assessed the importance of the error of the CC3 model that results from an approximate treatment of triple excitations, as well as the neglect of quadruple and higher excitations.   
To this end, the full configuration interaction (FCI) calculations of the polarizability were performed for 71 grid points using a program written for the purposes of this project \cite{przybytekFCI}. 
Due to the high computational demands of the FCI method, the calculations were restricted to geometries of high symmetry and were performed using a single basis set of triple-zeta quality. 
As the dTZ basis set from Ref.~\onlinecite{cencek2012effects} was still too large to accomplish the calculations, we prepared its singly augmented version, further referred to as aTZ.
The results for six selected geometries are presented in Table~\ref{tab:fci_polar}.
\begin{table}
\caption{Comparison of the three-body polarizability of helium calculated at the CC3 ($ \alpha_3^\mathrm{CC3}$) and FCI ($\alpha_3^\mathrm{FCI}$) levels of theory using the aTZ basis set. 
Interatomic distances, $r_{ij}$, are in a.u.\ and the polarizability unit is in $ 10^{-3}$ a.u.
Two last two rows are configurations for which the DID term in Eq.~(\ref{polar_asym1}) is zero.}
\label{tab:fci_polar}
\begin{ruledtabular}
\begin{tabular}{ccd{2.2}d{2.5}d{2.5}c}
\multicolumn{1}{c}{$r_{12}$} &
\multicolumn{1}{c}{$r_{23}$} &
\multicolumn{1}{c}{$r_{13}$} &
\multicolumn{1}{c}{$\alpha_3^\mathrm{CC3}$} &
\multicolumn{1}{c}{$\alpha_3^\mathrm{FCI}$} & \multicolumn{1}{c}{$100\%\!\times\!|(\alpha^\mathrm{FCI}_3 - \alpha^\mathrm{CC3}_3)/\alpha_3^\mathrm{CC3}|$} \\[1ex]
 \hline
    7.56 &     7.56 &     5.43 & -0.07477 & -0.07745 &  3.58\% \\
    6.20 &     6.20 &     5.50 & -0.07047 & -0.07538 &  6.97\% \\     
    6.50 &     6.50 &     6.50 & -0.05993 & -0.06246 &  4.21\% \\
    2.60 &     5.40 &     8.00 &  0.87654 &  0.92274 &  5.27\% \\
5.341516 & 5.341516 &     9.00 &  0.01213 &  0.01284 &  5.84\% \\
6.528520 & 6.528520 &    11.00 &  0.00528 &  0.00550 &  4.06\% \\
\end{tabular} 
\end{ruledtabular}
\end{table} 
We found that the error resulting from an approximate character of the CC3 model is almost negligible for small triangles, only about 0.03\% of $\alpha_3$ (not shown in Table~\ref{tab:fci_polar}).
By contrast, for larger triangles it becomes comparable to the extrapolation errors, $\sigma[\mathrm{TQ}]$. The average absolute FCI contribution across all 71 grid points is 5.7\% of $\alpha_3[\mathrm{TQ}]$.
Therefore, the value 5.7\% was taken as a global, geometry independent uncertainty due to the error of the CC3 model. The total uncertainty $\sigma$ was obtained by multiplying $\alpha_3[\mathrm{TQ}]$ by this value of the relative error and adding the result in squares to $\sigma[\mathrm{TQ}]$, namely
\begin{equation}\label{uncertainty}
\sigma = \sqrt{
\sigma[\mathrm{TQ}]^2
+\left(0.057\times\alpha_3[\mathrm{TQ}]\right)^2
}.
\end{equation}
The final average total uncertainty of our polarizability surface is 9\% and its median is 6\%.
Using the local, three-dimensional analytical fits of $\alpha_3[\mathrm{TQ}]$ and $\sigma[\mathrm{TQ}]$, see Sec.~\ref{sec:fit}, the total uncertainty $\sigma$ obtained from Eq.~(\ref{uncertainty}) was employed in Sec.~\ref{sec:virial_coeff} to calculate the uncertainty of the thermophysical quantities such as the third dielectric virial coefficient.

To the best of our knowledge, the only theoretical determination of the three-body polarizability of helium has been reported by P\'erez \emph{et al.} \cite{perez1984three} using the SCEP method. 
Our results differ substantially from P\'erez \emph{et al.}, especially for small internuclear distances.
For example, the SCEP result for the equilateral triangle with side lengths equal to 2 a.u.\ is 3.5 times larger than our CC3 value. For larger triangles, the differences between the CC3 and SCEP results decrease, especially for linear configurations. 

\begin{figure}
\includegraphics[width=1\linewidth]{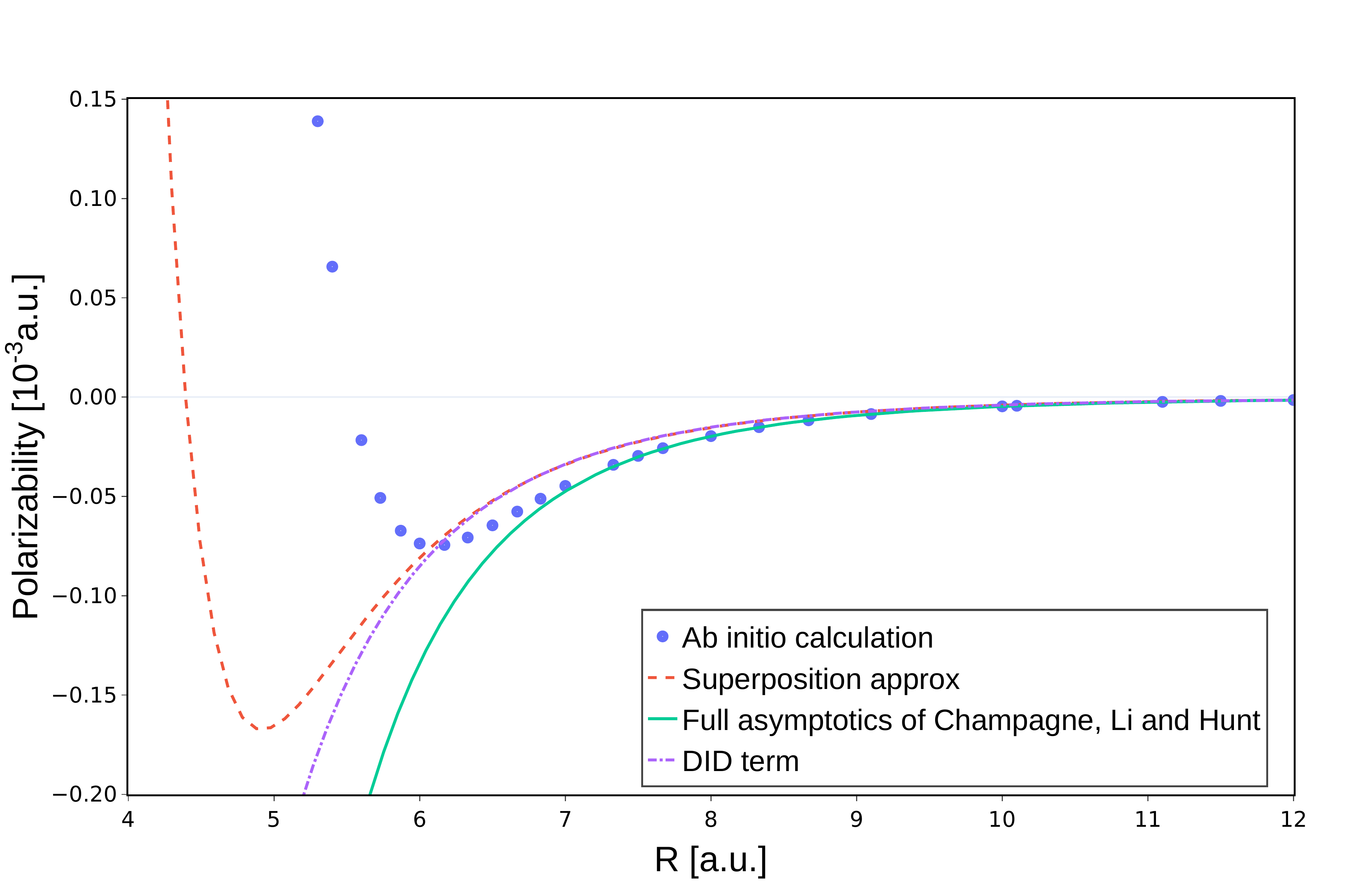}
\caption{Comparison of asymptotic models and extrapolated $\alpha_3[\mathrm{TQ}]$ \emph{ab initio} data for the equilateral triangles with side $R$.}
\label{fig:equil_approx}
\end{figure}

\begin{figure}
\begin{subfigure}[b]{.9\linewidth}
 \includegraphics[width=1\linewidth]{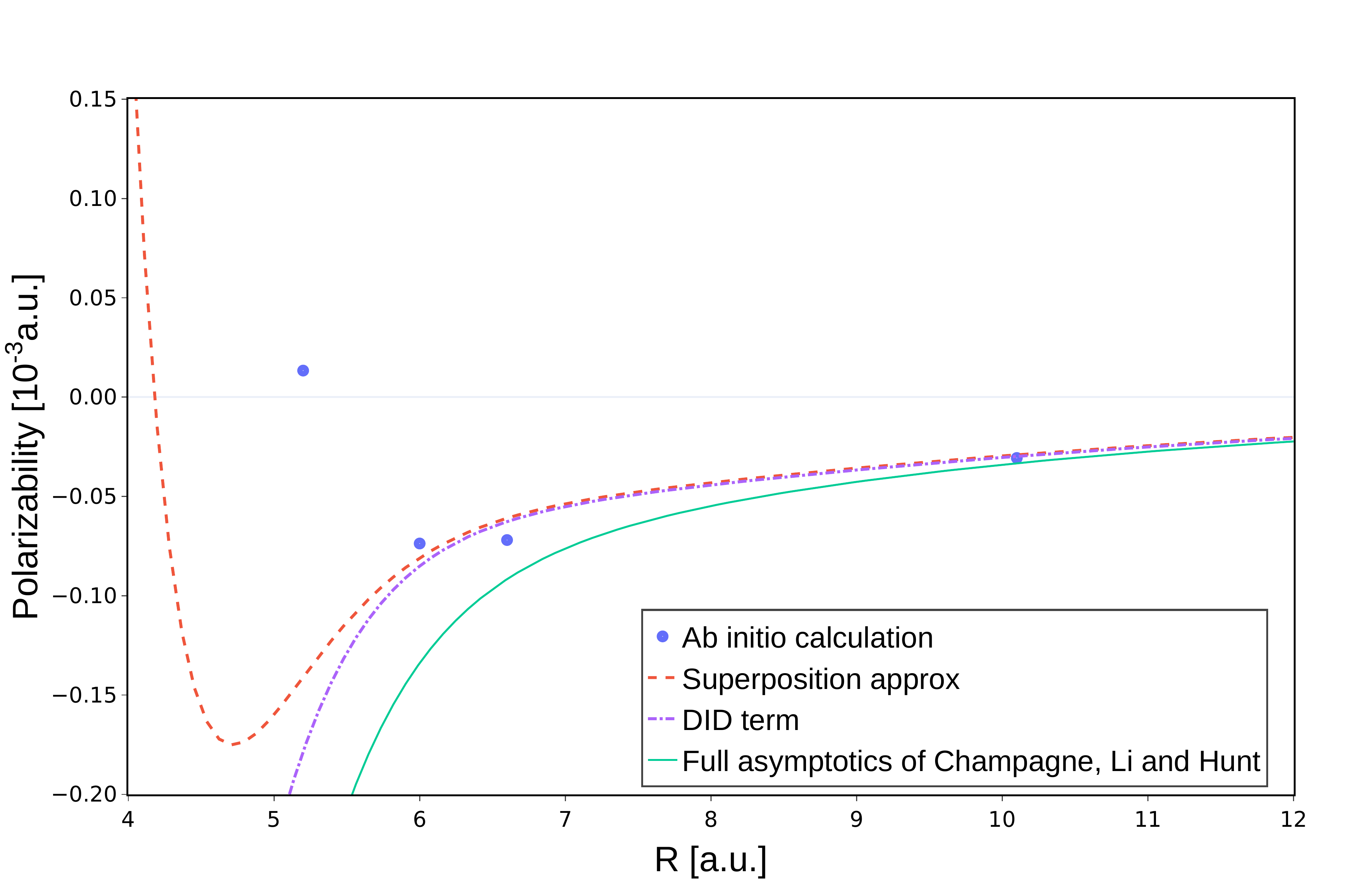}
 \caption{$R_\mathrm{base} = 6$~a.u.}
 \label{fig:iso_a}
\end{subfigure}
\begin{subfigure}[b]{.9\linewidth}
 \includegraphics[width=1\linewidth]{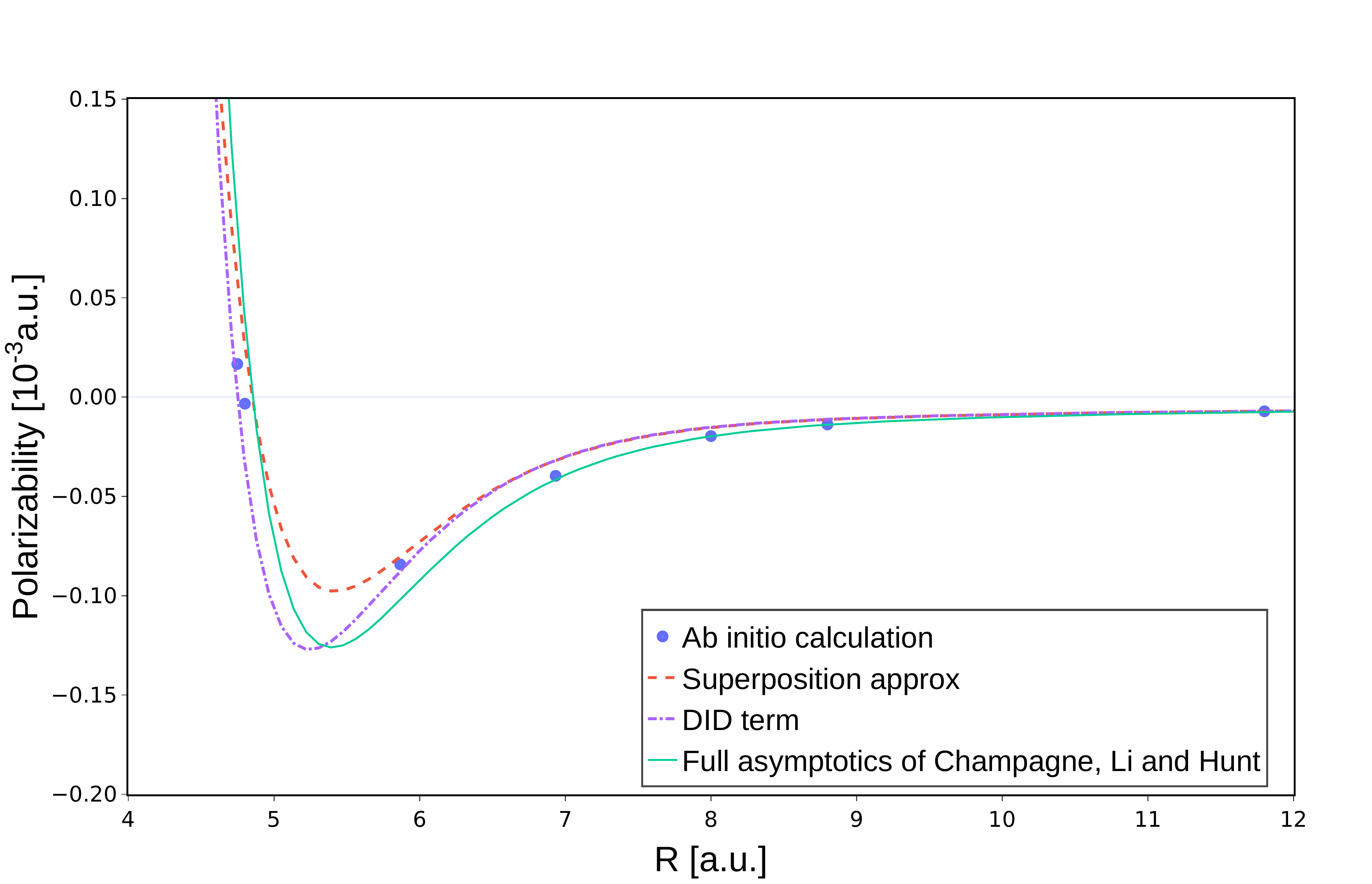}
 \caption{$R_\mathrm{base} = 8$~a.u.}
 \label{fig:iso_b}
\end{subfigure}
\caption{Comparison of asymptotic models and extrapolated $\alpha_3[\mathrm{TQ}]$ \emph{ab initio} data for the isosceles triangles with base side $R_\textrm{base}$ and leg side $R$.}
\label{fig:iso}
\end{figure}

In Fig.~\ref{fig:equil_approx}, we compare our \emph{ab initio} data for equilateral triangles with the theoretical asymptotics of Champagne \emph{et al.} \cite{champagne2000nonadditive} and the superposition approximation \cite{heller1974superpos,garberoglio20213dielec}. It is known that the superposition approximation includes the DID term from Eq.~(\ref{polar_asym1}) at long range. This is confirmed in Fig.~\ref{fig:equil_approx} where the two models give identical results for large triangles.
The DID term and the superposition approximation start to diverge from each other for triangles with side lengths below 6.5 a.u., when exponentially-decaying terms in the two-body polarizabilities start contributing significantly to the superposition approximation of the polarizability. It is worth noting that, in contrast to the pure DID term, the superposition approximation predicts a minimum of the three-body polarizability. The differences between the positions of the minima obtained from the \emph{ab initio} calculations and from the superposition approximation are about 1~a.u.\ for equilateral triangles. 
When the interatomic distances are not sufficiently large, the DID model and, consequently, the superposition approximation differ significantly from the results of the \emph{ab initio} calculations. 
For example, a reasonable agreement for equilateral triangles is found only when the side lengths are above 9~a.u., see Fig.~\ref{fig:equil_approx}. Clearly, additional asymptotic terms derived in Ref.~\onlinecite{champagne2000nonadditive} are required to correctly describe the polarizability down to 7~a.u or so. For equilateral triangles with side lengths below 7~a.u., the asymptotic expansion fails completely due to the increasing importance of the exchange and overlap effects. We observed a similar behavior also for isosceles configurations. The superposition approximation fails for isosceles triangles with the base side shorter than 7~a.u. However, it is surprisingly accurate for isosceles triangles with the base longer than 7~a.u., even if the leg sides are shorter than 7~a.u., see Fig.~\ref{fig:iso} for representative examples.

%%%%%%%%%%%%%%%%%%%%%%%%%%%%%%%%%%%%%%%%%%%%%%%%%%%%%%%%%%%%%%%%%%%%%%
\subsection{Analytic fit of the  three-body polarizability surface}
\label{sec:fit}
%%%%%%%%%%%%%%%%%%%%%%%%%%%%%%%%%%%%%%%%%%%%%%%%%%%%%%%%%%%%%%%%%%%%%%

The three-body polarizability surface $\alpha_3 (r_{12},r_{23},r_{13})$ was fitted with an analytic function $ \alpha_{\mathrm{\rm fit}}(r_{12},r_{23},r_{13})$ expressed as the sum of three components, one of a short-range and two of a long-range character:
\begin{equation}\label{polar_fit}
\alpha_{\mathrm{\rm fit}}(r_{12},r_{23},r_{13}) =  
  \alpha_{\mathrm{sr}}(r_{12},r_{23},r_{13})  
+ \alpha_{\mathrm{3a}}(r_{12},r_{23},r_{13})  
+ \alpha_{\mathrm{a-d}}(r_{12},r_{23},r_{13}).
\end{equation}
The short-range function $\alpha_\mathrm{sr}(r_{12},r_{23},r_{13})$, representing the exponentially decaying contributions to $\alpha_3(r_{12},r_{23},r_{13})$, has the form 
\begin{equation}\label{polar_SR}
\alpha_{\mathrm{sr}}(r_{12},r_{23},r_{13}) = 
\sum_{l=1}^{4} \sum_{ijk} \, A_{ijk,l}\:\,
\mathcal{S}_{123}\: e^{-\alpha_l\, r_{12}-\beta_l\, r_{23}-\gamma_l\, r_{13}} 
\: r_{12}^{i}\: r_{23}^{j}\,r_{13}^{k},
\end{equation} 
where $A_{ijk,l}$ are linear and $\alpha_l$, $\beta_l$, $\gamma_l$ nonlinear parameters to be  optimized. 
Possible values of the indices $i$, $j$, and $k$ are restricted by the inequalities $0\le i\le j \le k\le 5$ and the additional condition imposed on their sum, namely $i+j+k\le 7$. 
Overall, $\alpha_\mathrm{sr}$ includes 112 linear parameters and 12 nonlinear ones.
The form of Eq.~(\ref{polar_SR}) differs from the exponentially decaying terms employed in Refs.~\onlinecite{cencek2009threehe,cencek2013three}, where the perimeter of the triangle was used as the main argument of the exponential function. The resulting fitting function incorrectly describes linear geometries, since it remains constant when one atom moves between two atoms that have fixed positions.
By contrast, the exponential terms used in Eq.~(\ref{polar_SR}) are able to describe this movement correctly. Another difference is that we use the interatomic distances and their powers to describe the angular dependence of the potential, rather than the Legendre polynomials of the angles between triangle sides used in Refs.~\onlinecite{cencek2009threehe,cencek2013three}.

The $\alpha_{\mathrm{3a}}$ term in Eq.~(\ref{polar_fit}) provides the correct asymptotics of $\alpha_{3}$ in the three-atom fragmentation channel. It is given by the damped asymptotic expansion derived in Ref.~\onlinecite{champagne2000nonadditive},
\begin{equation}\label{longfit}
\begin{split}
\alpha_{\mathrm{3a}}(r_{12},r_{23},r_{13}) \:
&= 2 \, Z_1 \, \mathcal{S}_{123} \,
\frac{P_2(\cos{\theta_1})}{r_{12}^{3}r_{13}^{3}}
\,f_4(\eta_1  r_{12}) \,f_4(\eta_1  r_{13})
\\[0.7ex]
&+ 10\, Z_2 \, \mathcal{S}_{123} \,
\frac{P_3(\cos{\theta_1})}{r_{12}^{4}r_{13}^{4}}
\,f_5(\eta_2 r_{12}) \,f_5(\eta_2 r_{13})
\\[0.7ex]
&- 6 \, Z_3 \,
\frac{(1+3\cos\theta_1 \cos\theta_2 \cos\theta_3)}{r_{12}^{3}r_{23}^{3}r_{13}^{3}}
\,f_4(\eta_3 r_{12}) \,f_4(\eta_3 r_{23}) \,f_4(\eta_3 r_{13})
\\[0.7ex]
&+ 4 \, Z_4 \, \mathcal{S}_{123} \,
\frac{P_2(\cos{\theta_1})}{r_{12}^{3}r_{13}^{6}}
\,f_4(\eta_4 r_{12}) \,f_7(\eta_4 r_{13})
\\[0.7ex]
&- \frac{6}{\pi} \, Z_5 \, 
\frac{(1+3\cos\theta_1 \cos\theta_2 \cos\theta_3)}{r_{12}^{3}r_{23}^{3}r_{13}^{3}} 
\,f_4(\eta_5 r_{12}) \,f_4(\eta_5 r_{23}) \,f_4(\eta_5 r_{13}) 
\\[0.7ex] 
&+ \frac{4}{\pi} \, Z_6 \, \mathcal{S}_{123} \,
\frac{P_2(\cos{\theta_1})}{r_{12}^{3}r_{13}^{6}}
\,f_4(\eta_6 r_{12}) \,f_7(\eta_6 r_{13})
,
\end{split}
\end{equation}
where $f_n$ are the Tang-Toennies damping functions \cite{tang1984improved}
\begin{equation}\label{TTdamp}
f_n(x) = 1 - e^{-x } \left(1 + x+\frac{x^2}{2!} +\dots +\frac{x^n}{n!}\right).
\end{equation}
The $Z_n$ coefficients can be calculated with the knowledge of several atomic properties: the static dipole and quadrupole polarizabilities ($Z_1,\ldots ,Z_4$), the dynamic dipole polarizabilities ($Z_5$ and $Z_6$), and the second dipole hyperpolarizabilities ($Z_5$ and $Z_6$), see Ref.~\onlinecite{champagne2000nonadditive}. 
Their values, calculated using the data published in  Refs.~\onlinecite{champagne2000nonadditive,bishop1995static}, were fixed during the fitting procedure, while the nonlinear parameters $\eta_1,\ldots , \eta_6$ were fully optimized. 
The individual terms in Eq.~(\ref{longfit}) have a clear physical interpretation \cite{champagne2000nonadditive}. Their arise from the second-order induction interaction (the first two terms), the third-order induction interaction (the next two terms), the dispersion interaction (the fifth term), and the induction-dispersion interaction (the sixth term).

The term $\alpha_{\mathrm{a-d}}$ in Eq.~(\ref{polar_fit}) is an adaptation of Eq.~(\ref{polar_asym2}) representing the asymptotics of $\alpha_3(r_{12},r_{23},r_{13})$ in the atom-diatom fragmentation channel.
To avoid the problem of possible double-counting mentioned in Sec.~\ref{theory}, we substituted the exact diatomic polarizability anisotropy $\beta(r)$ in Eq.~(\ref{polar_asym2}) with a short-range exponentially decaying polynomial expansion 
\begin{equation}\label{atomdiatomfit}
 \alpha_{\mathrm{a-d}}(r_{12},r_{23},r_{13}) = 
	\widetilde{\mathcal{S}}_{123}\,
	\frac{P_2(\cos\theta)(C_0 + C_1 r+C_2 r^2)e^{-\eta_7 r}}{R^3}f_4(\eta_8R),
\end{equation}
where $\widetilde{\mathcal{S}}_{123}= 1 + \mathcal{P}_{13} + \mathcal{P}_{23}$ generates the sum over three possible ways of defining the Jacobi coordinates $r$, $R$, and $\theta$ for a triangle with sides $r_{12}$, $r_{23}$, and $r_{13}$. 
When $r$ is small, Eq.~(\ref{longfit}) gives a negligible contribution due to the strong Tang-Toennies damping and Eq.~(\ref{atomdiatomfit}) provides the correct atom-diatom asymptotics.
Conversely, when $r$ is large, Eq.~(\ref{atomdiatomfit}) gives a negligible contribution and Eq.~(\ref{longfit}) guarantees the correct asymptotics in the three-atom fragmentation channel. It may be noted that the right-hand-side of Eq.~(\ref{atomdiatomfit}) has a singularity at   $r=0$. 
This singularity is irrelevant from the physical point of view and we found that keeping $C_0$ and $C_1$ different form zero improves the quality of the fit. 
The parameters $C_0$, $C_1$, $C_2$, $\eta_7$, $\eta_8$, adjusted in the fitting procedure, are controlling the switching between Eqs.~(\ref{longfit}) and (\ref{atomdiatomfit}).
The total number of fitted parameters in Eqs.~(\ref{polar_SR}), (\ref{longfit}), and (\ref{atomdiatomfit}) is 135, including 115 linear parameters and 20 nonlinear parameters.

Effective selection of grid points is an important problem in fitting potential energy surfaces \cite{Metz:2020}. 
In our work the set of grid points was generated using an iterative approach.
The initial set of 250 points, satisfying the condition $\min(r_{12},r_{23},r_{13}) > 1.2$~a.u., was created through the Sobol sequence. 
In each step of the iterative procedure, $\sim\!\!10^6$ sets of 20 nonlinear parameters were randomly generated in a suitably chosen 20\,D box.
Subsequently, the current set of grid points was used in a local optimization of the nonlinear parameters to obtain $\sim\!\!10^6$ independent fitting functions. 
The optimization was performed using the trust-region dogleg procedure \cite{Powell1970} as implemented in the GSL library \cite{gsl_lib}.
During the optimization, the linear parameters (for fixed values of the nonlinear ones) were obtained using the weighted linear least-squares minimization with weights equal to the inverse square of the calculated uncertainties, $1/\sigma[\mathrm{TQ}]^2$. 
The fits prepared this way were sorted according to their average relative error and maximum relative error, and the top $20$ fits were picked for further analysis. 
The analysis consisted of evaluating the polarizability for $\sim\!\!10^4$ randomly distributed trimer configurations and finding regions of the configuration space where the largest differences between the preselected $20$ fits were observed. 
A new batch of $50-100$ grid points covering the problematic regions were constructed, the corresponding polarizabilities were computed, and the whole procedure was repeated with the enlarged set of grid points. 
The generation process was finished after six iterations when the differences between trial fits were acceptably small. 
The final batch of points was added to probe two particular regions: isosceles triangles for which the DID term in Eq.~(\ref{polar_asym1}) is zero and configurations where the atom-diatom contribution is important, i.e. triangles with one short side.
Altogether, the set of 748 grid points was generated. 
The final fit was prepared using a training set of 672 points ($\sim$90\% of the total number of points). 
The remaining 76 points ($\sim$10\% of the total number of points) formed a testing set and were used to assess the accuracy of the fit for points outside of the training set. 

\begin{figure}
\includegraphics[width=\textwidth]{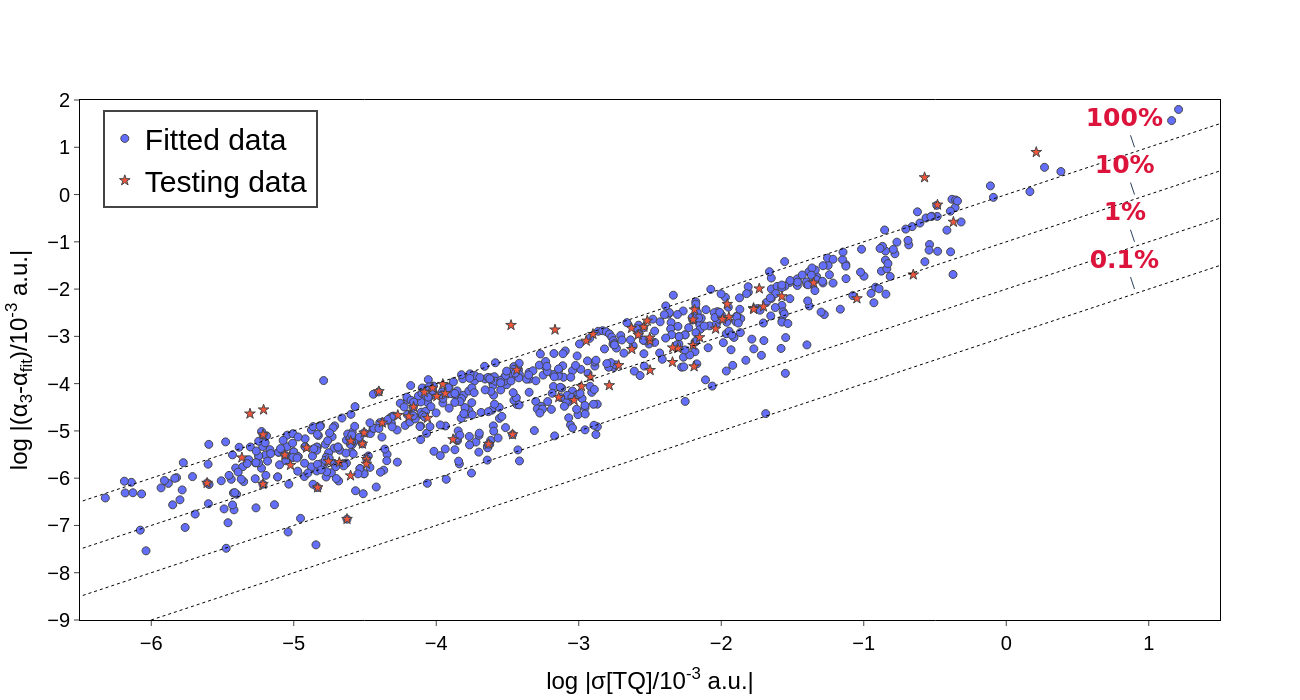}
\caption{Absolute differences between calculated and fitted three-body polarizabilities ${\alpha}_3$ vs.\ the estimated uncertainties $\sigma[\mathrm{TQ}]$. 
The straight dotted lines correspond to the relative errors.}
\label{fig:fit_plot}
\end{figure}

The final fit has a mean absolute relative error of $0.38\,\sigma[\mathrm{TQ}]$ and its median is $0.28\,\sigma[\mathrm{TQ}]$. The average relative error with respect to the values of the polarizability is 1.67\% and its median is 0.17\%. A majority of the configurations have a relative error smaller than $\sigma[\mathrm{TQ}]$, see Fig.~\ref{fig:fit_plot}. Most of the points with relative errors larger than $\sigma[\mathrm{TQ}]$ are encountered for triangles with at least one short side ($<2$~a.u.). 
In such cases, the polarizability is several orders of magnitude larger than in other regions. 
For example, the polarizability calculated for the triangle $(1.4, 4.6, 6.0)$ is $-91.48\times10^{-3}$~a.u. and the relative error is $4.8\,\sigma[\mathrm{TQ}]$. On the other hand, for a similar triangle $(2.0, 5.0, 6.9)$ the polarizability is $-0.73\times10^{-3}$~a.u. Fortunately, these configurations were found to have a minimal influence on the quality of the final results. For example, in the calculation of the third dielectric virial coefficient (see Sec.~\ref{sec:virial_coeff}), triangles with one side shorter than 2~a.u.\ were not important even at the temperature of 2400~K. 

The second group of triangles with a relative error larger than $\sigma[\mathrm{TQ}]$ are triangles with all sides longer than 16~a.u. For such large interatomic distances the values of the polarizability are tiny and hence are plagued by a large cancellation of significant digits when the supermolecular approach of Eq.~(\ref{polar_3body}) is used. However, these regions of the surface are completely determined by the asymptotic expansion. Therefore, the analytical expansion coefficients used in our fit are expected to provide accurate results.

In order to estimate uncertainties of the thermophysical properties of gaseous helium, which can be obtained using our potential, we provide a fit of $\sigma(r_{12},r_{23},r_{13})$ reflecting the estimated local uncertainties of our \emph{ab initio} calculations. Except for a small number of outliers, the exact values of the polarizability are expected to be contained within the range ${\alpha}_3(r_{12},r_{23},r_{13}) \pm \sigma(r_{12},r_{23},r_{13})$. 
The function $\sigma(r_{12},r_{23},r_{13})$ is not intended to precisely fit the uncertainties, but rather to follow the trend with respect to the size of the triangles and provide upper bounds to the calculated uncertainties for the thermophysical properties. The analytical function used to obtain the fit of the uncertainties is analogous to the short-range part of Eq.~(\ref{polar_SR}), where now the summation over $l$ goes from 1 to 2, and the allowed values of the indices $i$, $j$, and $k$ are restricted by the conditions $1 \le i\le j\le k \le 4$ and $i+j+k\le8$. The fit contains 26 linear and 6 nonlinear parameters. 

The fitting was performed for the total uncertainties as defined in Eq.~(\ref{uncertainty}).
Due to a large variation of the estimated uncertainties across the three-dimensional space of the interatomic distances, we selected a smaller subset of the uncertainties for the fitting purposes.
First, we discarded configurations with the uncertainties significantly smaller than for the neighboring ones. This is justified by the observation that the resulting fit then bounds these uncertainties from above which is sufficient for our purposes. Next, we discarded triangles with interatomic distances $\min(r_{12},r_{23},r_{13}) > 16$~a.u.\ as the asymptotic expansion is expected to provide the correct results in this case. Finally, we discarded the smallest triangles with $\min(r_{12},r_{23},r_{13}) < 2$~a.u. which are of little importance in calculations of the physical properties.
The final set of points used in the construction of $\sigma(r_{12},r_{23},r_{13})$ comprised almost $65\%$ of the total number of configurations. The average ratio of the values of the fitting function $\sigma (r_{12},r_{23},r_{13})$ to the estimated uncertainties for the whole set of points is about 1.3. 
The values of all linear and nonlinear parameters of $\alpha_3(r_{12},r_{23},r_{13})$ and $\sigma(r_{12},r_{23},r_{13})$, along with a Fortran~2003 implementation of the fitting functions, are included in the Supplementary Material. 

\subsection{Classical calculations of the third dielectric virial coefficient}
\label{sec:virial_coeff}
%%%%%%%%%%%%%%%%%%%%%%%%%%%%%%%%%%%%%%%%%%%%%%%%%%%%%%%%%%%%%%%%%%%%%%

As mentioned in the introduction, the third dielectric coefficient $C_{\varepsilon}$ plays an important role in precise determination of various metrological quantities. It is defined by the expansion of the Clausius-Mossotti function for the dielectric constant $\varepsilon$ in powers of the gas density $\rho$ \cite{1958JChPh..28...61H,gray2011fluids2}
\begin{equation}\label{clausius}
\frac{\varepsilon-1}{\varepsilon +2} 
= \rho   \, A_{\varepsilon}
+ \rho^2 \, B_{\varepsilon}
+ \rho^3 \, C_{\varepsilon} 
+ \cdots,
\end{equation}
where $A_\varepsilon$ is proportional to the atomic polarizability and $B_\varepsilon$ is the second dielectric virial coefficient.

The classical expression for the third dielectric virial coefficient $C_\varepsilon(T)$ was first published in Ref.~\onlinecite{gray2011fluids2}. In this work we use a factorized expression analogous to the one used in Refs.~\onlinecite{garberoglio2009firstprincip,kusalik1995calculation,mas1999third} for the third pressure virial calculation. Within this framework, the third virial dielectric coefficient is given by the sum of three terms,
\begin{equation}\label{dielectric_integral}
C_\varepsilon(T) 
= 4 B(T) \, B_\varepsilon(T)
+ C_\varepsilon^{\mathrm{2body}}(T) 
+ C_\varepsilon^{\mathrm{3body}}(T),
\end{equation}
where $B(T)$ is the second pressure virial coefficient, 
\begin{equation}\label{dielectric_integral1}
C_\varepsilon^{\mathrm{2body}}(T) = 
\frac{2\pi}{9V}\! \iiint \! \widetilde{\mathcal{S}}_{123} \{
\alpha_2(\textbf{r}_1,\textbf{r}_2) 
\, [ e^{-\beta U(\textbf{r}_1,\textbf{r}_2,\textbf{r}_3)}
-%\Delta\bar\alpha(\textbf{r}_1,\textbf{r}_2)
\,e^{-\beta U_2(\textbf{r}_1,\textbf{r}_2)} ]
 \}d\textbf{r}_1 d\textbf{r}_2 d\textbf{r}_3,
\end{equation}
and
\begin{equation}\label{dielectric_integral2}
C_\varepsilon^{\mathrm{3body}}(T) = 
\frac{2\pi}{9V} \iiint 
\alpha_3(\textbf{r}_1,\textbf{r}_2,\textbf{r}_3)
\,e^{-\beta U(\textbf{r}_1,\textbf{r}_2,\textbf{r}_3)}
\,d\textbf{r}_1 d\textbf{r}_2 d\textbf{r}_3. 
\end{equation}
In the above formulas, $U(\textbf{r}_1,\textbf{r}_2,\textbf{r}_3)=U_3(\textbf{r}_1,\textbf{r}_2,\textbf{r}_3)+U_2(\textbf{r}_1,\textbf{r}_2)+U_2(\textbf{r}_2,\textbf{r}_3)+U_2(\textbf{r}_1,\textbf{r}_3)$ is the sum of the two-body, $U_2$, and three-body,  $U_3$, interaction potentials, and $\beta\!=\!1/k_\mathrm{B}T$.  
%where , $U(\textbf{r}_1,\textbf{r}_2)$ and %$U(\textbf{r}_1,\textbf{r}_2,\textbf{r}_3)$ are the .
Note that the integral $C_\varepsilon^{\mathrm{2body}}(T)$ involves not only the two-body quantities (polarizability and interaction potential) but also the three-body interaction potential $U_3$.  
The calculation of the first term in Eq.~(\ref{dielectric_integral}) is trivial since accurate values of $B(T)$ and $B_{\varepsilon}(T)$ are available in the literature \cite{garberoglio2020path}. 
The computation of the second and third terms is somewhat more involved. 
The translational and rotational invariance can be used to reduce the nine-dimensional integrals in Eqs.~(\ref{dielectric_integral1}) and (\ref{dielectric_integral2}) to three-dimensional integrals over two sides of the triangle and the angle between them using the substitution $V^{-1}d\textbf{r}_{1}\,d\textbf{r}_{2}\,d\textbf{r}_{3} \rightarrow 8\pi\,r_{13}^2\,r_{23}^2\, \,dr_{13}\,dr_{23}\,d\big(\cos{\theta_3}\big)$. Once this exchange of variables is performed,
evaluation of Eq.~(\ref{dielectric_integral2}) reduces to the integration of the right-hand-side of Eq.~(\ref{polar_asym1}) over $r_{13}$, $r_{23}$, and $\theta_3$. However, the resulting integral is ill-defined as the integrals over $r_{13}$ and $r_{23}$ are divergent at infinity, even if the singularities at $r_{13}=0$ and $r_{23}=0$ are removed using suitable damping factors. In Appendix~\ref{regularization} we show that, after an appropriate regularization, this divergence is eliminated and the required integral is actually equal to zero. This does not mean that the DID contribution %${\alpha}_{\mathrm{DID}}$ 
to $C_{\epsilon}(T)$ vanishes. The inclusion of Eq.~(\ref{polar_asym1}) in the fitting function is necessary to obtain an accurate fit at intermediate distances, which are important in evaluation of $C_{\epsilon}(T)$. Moreover, Eq.~(\ref{polar_asym1}) contributes directly to $C_{\epsilon}(T)$ when multiplied by Boltzmann factors dependent on $r_{12}$. As the classical integration is valid only for higher temperatures, we also calculated the semiclassical results using the Feynman-Hibbs modified two-body potential and polarizabilities \cite{feynman1965pathintegrals,garberoglio2009firstprincip,2011JChPh.134m4106G}.
The results of the semiclassical calculations and the details of the calculations are presented in Sec.~\ref{scl_appendix}.

The integrations were performed for all possible triangle configurations with sides up to 8~nm using the adaptive Gauss-Kronrod quadrature of degree 7 and 15 \cite{kronrod1965nodes}.
%, implemented by one of the authors (J.L.). 
In the calculations of the two-body contributions we used the two-body interaction potential reported by Czachorowski \emph{et al.} \cite{czachorowski2020second} and the two-body polarizability of Cencek \emph{et al.} \cite{2011JChPh.135a4301C}. For the three-body contribution we used the three-body polarizability presented in this work and the most recent three-body interaction potential of Ref.~\onlinecite{lang2022density}. Note that all of these potentials have known local uncertainties.
We also tested the three-body potential from Ref.~\onlinecite{cencek2009threehe}, but the difference of the computed $C_\varepsilon^{\mathrm{3body}}(T)$ using the alternative three-body interaction potential was several orders of magnitude smaller than the uncertainty originating from the accuracy of the three-body polarizability. The error of the numerical integration was assessed using the Gauss-Kronrod method and was much smaller than the effect of the uncertainties of the polarizability and the potential energy surfaces. 

The uncertainties of the third dielectric virial coefficient were estimated through the propagation of errors for both the two-body and three-body polarizabilities, and the three-body interaction potential.
The errors of the two-body potential are negligible \cite{czachorowski2020second} and were not included in the propagation. The uncertainties due to the errors in the polarizabilities and the three-body potential were computed using the approach developed by Garberoglio and Harvey \cite{Garberoglio2021a} for calculations of the pressure viral coefficients. The uncertainties due to the polarizabilities were estimated employing the formula
\begin{equation}\label{uncert}
    |C_{\varepsilon}(\alpha_n^{\rm acc}) - C_{\varepsilon}(\alpha_n)| =
     \bigg|   \int  (\alpha_n^{\rm acc} - \alpha_n) 
     \frac{\delta C_{\varepsilon}}{\delta\alpha_n} dX \bigg| 
     \le  \int \sigma(\alpha_n) \bigg| 
     \frac{\delta C_{\varepsilon}}{\delta \alpha_n} \bigg| dX ,
     \end{equation}
where $\alpha_n^{\rm acc}$ and $\alpha_n$ are the accurate and approximate $n$-body polarizability functions, respectively, $\delta C_{\varepsilon} / \delta\alpha_n $ is the functional derivative of $C_{\varepsilon}$ with respect to $\alpha_n$, $\sigma(\alpha_n)$ is our estimate of the $k\!=\!2$ level uncertainty of $\alpha_n$, and $dX$ is the volume element in the space of coordinates on which $\alpha_n$ depends. The equality in Eq.~(\ref{uncert}) is exact since $C_{\varepsilon}$ depends linearly on $\alpha_n$. The inequality in Eq.~(\ref{uncert}) holds at the 95\% confidence level since $|C_{\varepsilon}(\alpha_n^{\rm acc}) - C_{\varepsilon}(\alpha_n)| \le \sigma(\alpha_n)$ and since $\sigma(\alpha_n)$ is interpreted as the $k\!=\!2$ level uncertainty of $\alpha_n$.
The rightmost side of Eq.~(\ref{uncert}) was also used to estimate a minor source of uncertainty due to the approximate nature of the three-body potential energy $U_3$. In this case the equality in Eq.~(\ref{uncert}) is valid only approximately with an error quadratic in $\sigma(U_n)$. 
%ince  $C_{\varepsilon}$ is a nonlinear function  of   $U_$.
The total uncertainty was computed as the square root of the sum of squares of the uncertainties of $\alpha_2$, $\alpha_3$, and $U_3$.

%First, we evaluated Eq.~(\ref{dielectric_integral}) using perturbed potentials, $U_\pm = U\pm\sigma_U$, with a fixed value of the polarizability $\Delta\bar{\alpha}$. 
%The uncertainty due to errors in the potential is estimated as half of the absolute difference between $C_\varepsilon(T)$ obtained with $U_+$ and $U_-$. 
%Analogously, we evaluated the uncertainty due to the polarizability as the half of the difference between $C_\varepsilon(T)$ computed with $\Delta\bar{\alpha}_\pm = \Delta\bar{\alpha}\pm \sigma_{\Delta\bar{\alpha}}$ with fixed interaction potential. 
%These estimations of the uncertainties were performed separately for all three terms appearing in Eq.~(\ref{dielectric_integral}). 

\begin{figure}
\includegraphics[width=\linewidth]{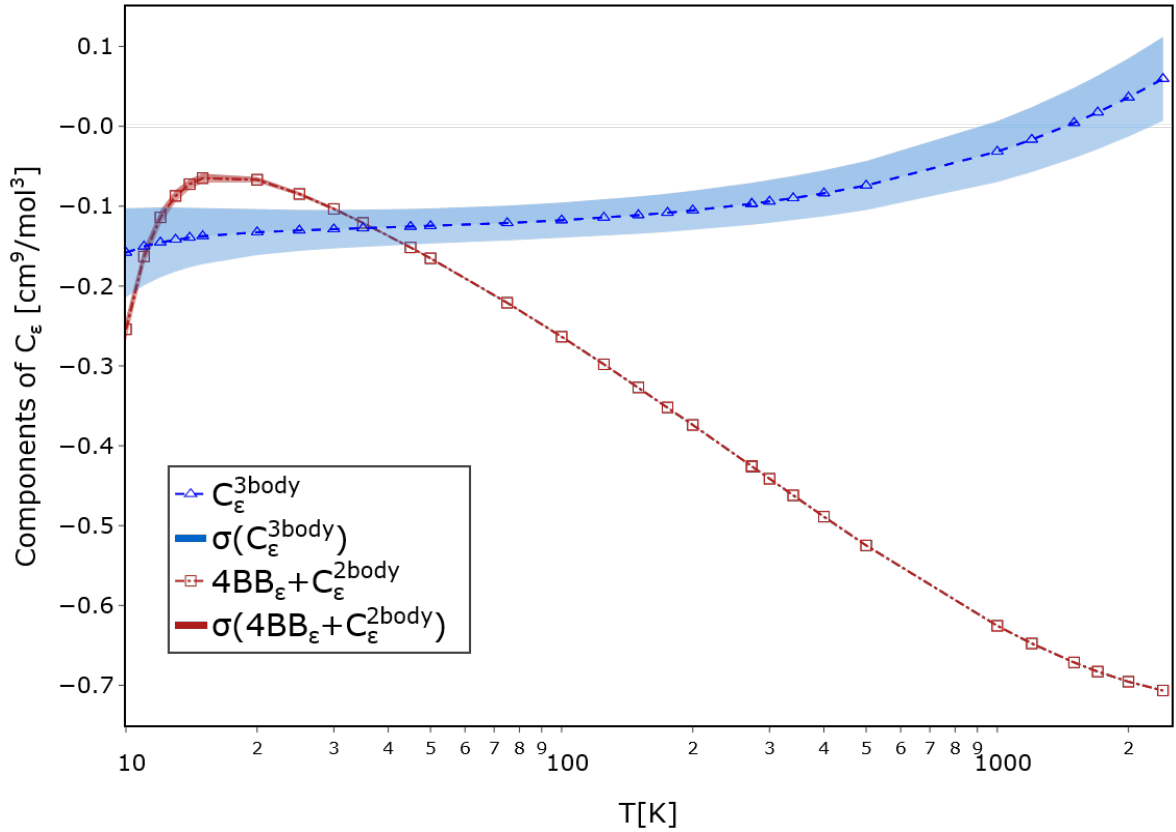}
\caption{Comparison of the two-body and three-body contributions  to $C_\varepsilon$. 
%The light blue area around the plots corresponds to the calculated uncertainty of the each contribution.
}
\label{fig:virial_contrib}
\end{figure}

The results of the classical calculations of $C_\varepsilon(T)$ and of the contributions identified in Eq.~(\ref{dielectric_integral}) are presented in Fig.~\ref{fig:virial_contrib}.
In contrast to the third pressure virial coefficient \cite{garberoglio2009firstprincip}, the three-body contribution $C_\varepsilon^{\mathrm{3body}}(T)$ to the third dielectric virial coefficient has a comparable magnitude to the combined two-body terms. For temperatures of about 1270~K, the values of $C_\varepsilon^{\mathrm{3body}}(T)$ cross zero and the two-body terms dominate. However, this trend suggests an increasing importance of $C_\varepsilon^{\mathrm{3body}}(T)$ for temperatures higher than 2500~K.  

From the inspection of the contributions to $C_\varepsilon^{\mathrm{3body}}(T)$ originating from the three terms included in the fitting function, see Eq.~(\ref{polar_fit}), we conclude that for higher temperatures, the short-range contributions dominate. This is an expected behavior since the helium atoms have enough kinetic energy to penetrate the repulsive wall of the potential. As these short-range contributions are mostly positive, they compensate the (typically negative) long-range contributions.
Regarding the uncertainties, the uncertainty of $C_\varepsilon^{\mathrm{3body}}(T)$ dominates for the entire temperature range. 
% An exception occurs for very low temperatures (around 10~K), where the combined two-body and three-body uncertainties are of a similar magnitude. 
The largest contribution to $C_\varepsilon^{\mathrm{3body}}(T)$ then comes from the uncertainty of the three-body polarizability.

\begin{table}
\caption{The third dielectric virial coefficient $C_\varepsilon(T)$ for helium and its many-body contributions.  The uncertainties are shown in parentheses. The total uncertainty is calculated as the sum of squares.}
\label{tab:c_epsilon}
\begin{ruledtabular}
\begin{tabular}{d{4.2}d{2.7}d{2.9}d{2.7}}
\multicolumn{1}{c}{$T$ [K]} &
\multicolumn{1}{c}{$C_\varepsilon^{\mathrm{3body}}$} &
\multicolumn{1}{c}{$4BB_\varepsilon + C_\varepsilon^{\mathrm{2body}}$} &
\multicolumn{1}{c}{$C_\varepsilon$} \\
\hline
10 & -0.158(56) & -0.254(14) & -0.412(57) \\
11 & -0.150(49) & -0.163(11) & -0.313(50) \\
12 & -0.145(44) & -0.1140(90) & -0.259(45) \\
13 & -0.142(40) & -0.0872(76) & -0.229(41) \\
14 & -0.139(37) & -0.0725(65) & -0.212(38) \\
15 & -0.138(35) & -0.0649(57) & -0.202(35) \\
20 & -0.133(29) & -0.0668(34) & -0.199(29) \\
25 & -0.130(26) & -0.0847(24) & -0.215(26) \\
30 & -0.129(24) & -0.1034(19) & -0.232(24) \\
35 & -0.128(23) & -0.1209(16) & -0.249(23) \\
45 & -0.126(22) & -0.1517(12) & -0.277(22) \\
50 & -0.125(22) & -0.1652(11) & -0.290(22) \\
75 & -0.121(22) & -0.22095(73) & -0.342(22) \\
100 & -0.118(22) & -0.26349(58) & -0.381(22) \\
125 & -0.114(23) & -0.29800(50) & -0.412(23) \\
150 & -0.111(23) & -0.32702(44) & -0.438(23) \\
175 & -0.108(24) & -0.35202(40) & -0.460(24) \\
200 & -0.105(25) & -0.37396(37) & -0.479(25) \\
273.15 & -0.097(26) & -0.42573(32) & -0.523(26) \\
273.16 & -0.097(26) & -0.42573(32) & -0.523(26) \\
300 & -0.094(27) & -0.44134(30) & -0.535(27) \\
340 & -0.090(28) & -0.46211(28) & -0.552(28) \\
400 & -0.084(29) & -0.48886(26) & -0.573(29) \\
500 & -0.074(31) & -0.52483(24) & -0.599(31) \\
1000 & -0.032(38) & -0.62547(18) & -0.657(38) \\
1200 & -0.017(41) & -0.64761(16) & -0.664(41) \\
1500 & 0.004(44) & -0.67134(15) & -0.667(44) \\
1700 & 0.018(46) & -0.68281(14) & -0.665(46) \\
2000 & 0.036(49) & -0.69551(14) & -0.659(49) \\
2400 & 0.059(53) & -0.70657(13) & -0.647(53) \\
\end{tabular}
\end{ruledtabular}
\end{table}

\begin{figure}
\includegraphics[width=0.9\textwidth]{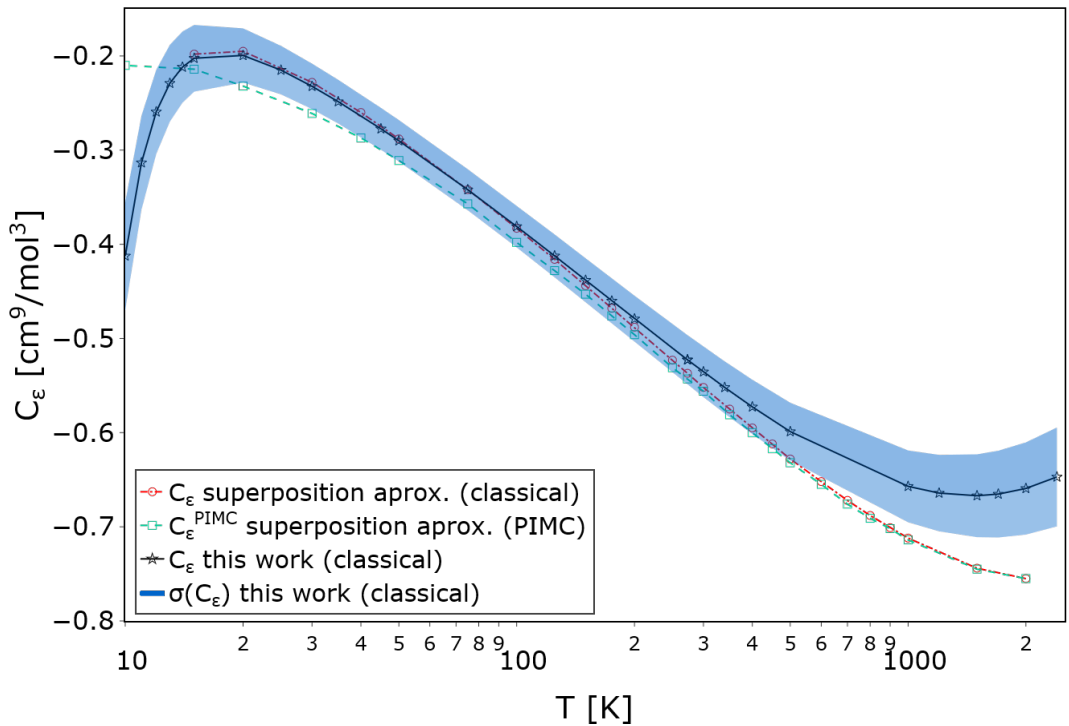}
\caption{Comparison of the classical values of $C_\varepsilon(T)$ calculated in this work with the results obtained using the superposition approximation by Garberoglio \emph{et al.} \cite{garberoglio20213dielec}.
% For experiments of White and Gugan \cite{1992Metro..29...37W} only upper error bars are shown in the small plot.
% The light blue area around  black solid  line corresponds to the uncertainty of theoretical values of $C_\varepsilon(T)$.
}
\label{fig:virial_total_teo}
\end{figure}

\begin{figure}
\includegraphics[width=0.9\textwidth]{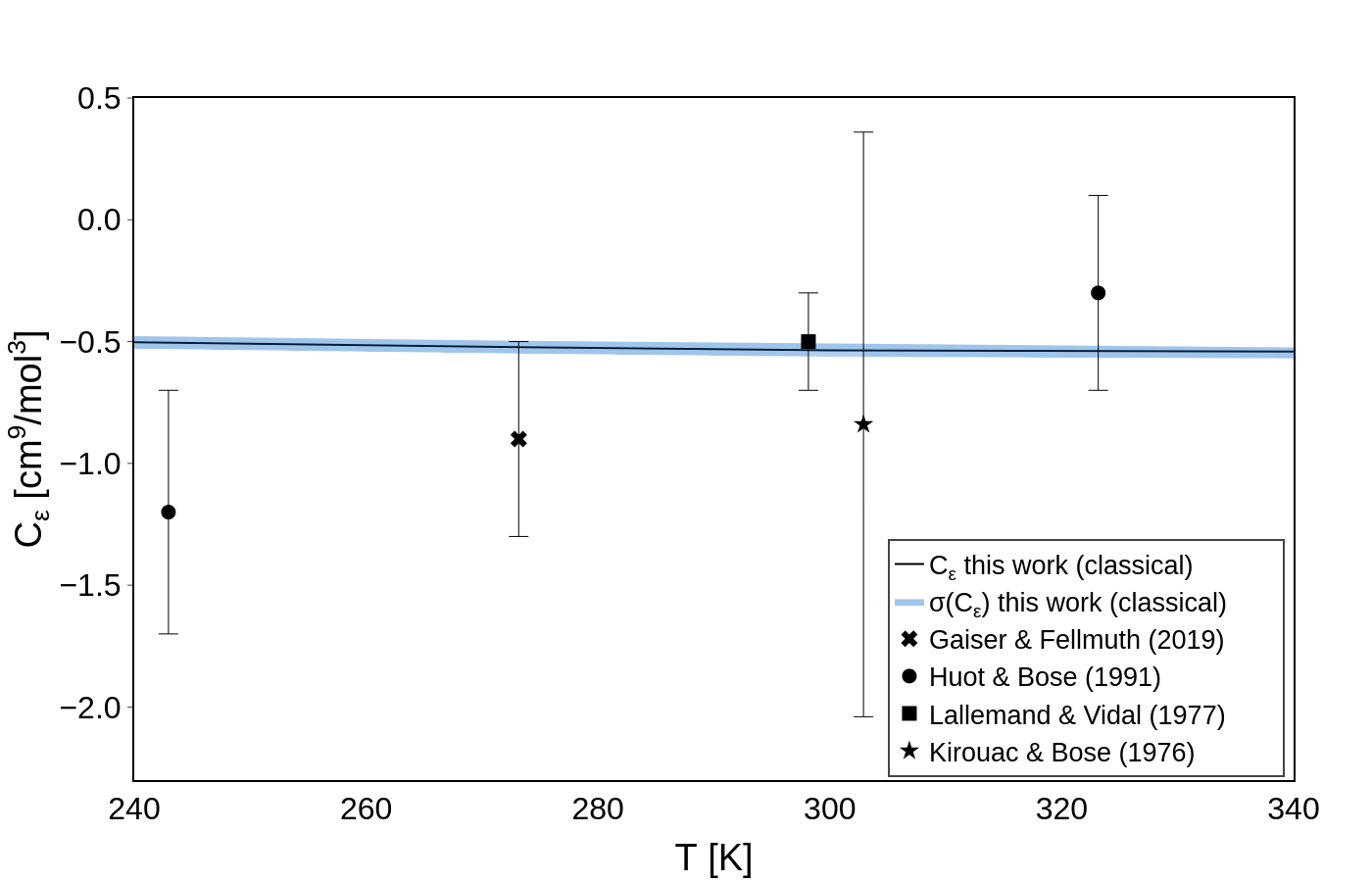}
\caption{Comparison of the classical values of $C_\varepsilon(T)$ calculated in this work with the experimental data \cite{2019JChPh.150m4303G,1977JChPh..66.4776L,1991JChPh..95.2683H,1976JChPh..64.1580K}.
% For experiments of White and Gugan \cite{1992Metro..29...37W} only upper error bars are shown in the small plot.
% The light blue area around  black solid  line corresponds to the uncertainty of theoretical values of $C_\varepsilon(T)$.
}
\label{fig:virial_total_exp}
\end{figure}

The calculated values of $C_\varepsilon^{\mathrm{3body}}(T)$, as well as of the individual contributions defined in Eq.~(\ref{dielectric_integral}), are presented in Table~\ref{tab:c_epsilon} for a wide range of temperatures. In Figs.~\ref{fig:virial_total_teo} and \ref{fig:virial_total_exp} we compare our calculations with theoretical results and experimental values found in the literature. To the best of our knowledge, there are only two relevant theoretical calculations of the third dielectric virial coefficient available in the literature. Both of them use the superposition approximation to obtain the three-body polarizability. Heller and Gelbart provided only one value of $C_\varepsilon(300\,\textrm{K}) = -0.719\;\text{cm}^9\: \text{mol}^{-3}$ \cite{heller1974superpos}. In their work, they used relatively simple models for both the two-body potential and the two-body polarizability. More recently, Garberoglio \emph{et al.} \cite{garberoglio20213dielec} used the superposition approximation for the three-body polarizability and the same two-body potential and two-body polarizability as in the present work. 
Their PIMC value of $C_\varepsilon(T)$ at 300~K is $-0.556\; \text{cm}^9\:\text{mol}^{-3}$ \cite{garberoglio20213dielec} which is close to the value $C_\varepsilon(300\,\textrm{K}) = -0.535\; \text{cm}^9\: \text{mol}^{-3}$ obtained by us. Overall, our values of $C_\varepsilon(T)$ are negative over the whole range of temperatures, in agreement with the results of Garberoglio \emph{et al.} \cite{garberoglio20213dielec}. The classical results are similar to their values up to $T = 400$~K, but the two curves start to diverge for higher temperatures. According to our results, a local minimum around 1500~K is present, see Fig.~\ref{fig:virial_total_teo}, while this feature is absent in the results of Garberoglio \emph{et al.} \cite{garberoglio20213dielec}. However, note that the rate of decrease of $C_\varepsilon(T)$ with the temperature computed using the superposition approximation slows down around 2000~K. Therefore, it is possible that this may result in a minimum for higher temperatures. One can conclude that the difference between the results of Ref.~\onlinecite{garberoglio20213dielec} and our work is due to the incorrect description of the short-range polarizability within the superposition approximation. 

Regarding the experimental values, only a few of them can be found in the literature \cite{2019JChPh.150m4303G,1977JChPh..66.4776L,1991JChPh..95.2683H,1976JChPh..64.1580K}.
The most recent value of Gaiser and Fellmuth comes from the dielectric-constant gas thermometry \cite{2019JChPh.150m4303G}. They obtained the third pressure virial coefficient from the measured isotherms and converted this quantity to the third dielectric virial coefficient with the use of the best available pressure virials \cite{cencek2012effects,garberoglio2011improved,bich2007ab} and the second dielectric virial coefficient \cite{rizzo2002effect}. Other experiments performed at the room temperature are also consistent with our data. However, for low temperatures the difference between the experiments and our theoretical calculations is substantial, even if we take large experimental errors into account.
This suggests that more accurate measurements are needed to resolve this discrepancy.

It is also important to mention the missing contribution to $C_\varepsilon(T)$ due to the three-body dipole moment. To date, no dipole moment surface has been published and Garberoglio \emph{et al.} \cite{garberoglio20213dielec} used an approximate model containing only the long-range asymptotic description of Hunt and Li \cite{li1997threedipoles}. They determined that the contribution due to the dipole moment is negligible. Our preliminary study of the three-body dipole moment confirms their findings.
When expressed in atomic units, the three-body dipole moments computed by us using the CC3 theory are one to two orders of magnitude smaller than the corresponding three-body polarizabilities. Note that the dipole moments contribute to $C_\varepsilon(T)$ through squares of their length divided by $k_{\rm B}T$. 
Therefore, the contribution of the three-body dipoles to $C_\varepsilon(T)$ is expected to be smaller than the present uncertainty of $C_\varepsilon(T)$, except possibly for very low temperatures. 

%%%%%%%%%%%%%%%%%%%%%%%%%%%%%%%%%%%%%%%%%%%%%%%%%%%%%%%%%%%%%%%%%%%%%%
%%%%%%%%%%%%%%%%%%%%%%%%%%%%%%%%%%%%%%%%%%%%%%%%%%%%%%%%%%%%%%%%%%%%%%
%%%%%%%%%%%%%%%%%%%%%%%%%%%%%%%%%%%%%%%%%%%%%%%%%%%%%%%%%%%%%%%%%%%%%%
\section{Semiclassical quantum correction}
\label{scl_appendix}
%%%%%%%%%%%%%%%%%%%%%%%%%%%%%%%%%%%%%%%%%%%%%%%%%%%%%%%%%%%%%%%%%%%%%%
%%%%%%%%%%%%%%%%%%%%%%%%%%%%%%%%%%%%%%%%%%%%%%%%%%%%%%%%%%%%%%%%%%%%%%
%%%%%%%%%%%%%%%%%%%%%%%%%%%%%%%%%%%%%%%%%%%%%%%%%%%%%%%%%%%%%%%%%%%%%%

It is well known that the classical calculations of the virial coefficients fail to properly describe properties of the gas at low temperatures when quantum effects play an important role \cite{garberoglio20213dielec,garberoglio2009firstprincip,shaul2012semiclassical,ram1973quantum,1995CPL...247..440M,song2020accurate,hirschfelder1954molecular}. 
For this reason, we also consider the leading-order quantum correction and provide a semiclassical estimation of the third dielectric virial coefficient. The Wigner-Kirkwood expansion is widely used for this purpose  \cite{1995CPL...247..440M,dewitt1962analytic,kihara1953virial}, but the Feynman-Hibbs effective potentials have been recently used as well \cite{shaul2012semiclassical}. As noted by Guillot and Guissani~\cite{guillot1998quantum}, the Feynman-Hibbs effective potentials are more accurate and have better convergence properties than the Wigner-Kirkwood expansion. Therefore, we used the Feynman-Hibbs effective potential in this work and modified the two-body interaction potentials with the quadratic Feynman-Hibbs (QFH) correction \cite{feynman1965pathintegrals,feynman1972statistical,guillot1998quantum}
\begin{equation}\label{qfh_pot}
U^{QFH}(r_{ij})
= U(r_{ij})
+ \frac{\beta}{12m} \left(
    \frac{\partial^2 U(r_{ij})}{\partial r_{ij}^2}
   +\frac{2}{r_{ij}}\frac{\partial U(r_{ij})}{\partial r_{ij}}
 \right),
\end{equation}
where $m$ is the mass of helium atom. The modified two-body potential is used instead of the original two-body potential in calculation of the virial coefficients. The same modification was used by Shaul and collaborators to calculate semiclassical values of the pressure virial coefficients \cite{shaul2012semiclassical}.

\begin{figure}
\includegraphics[width=\linewidth]{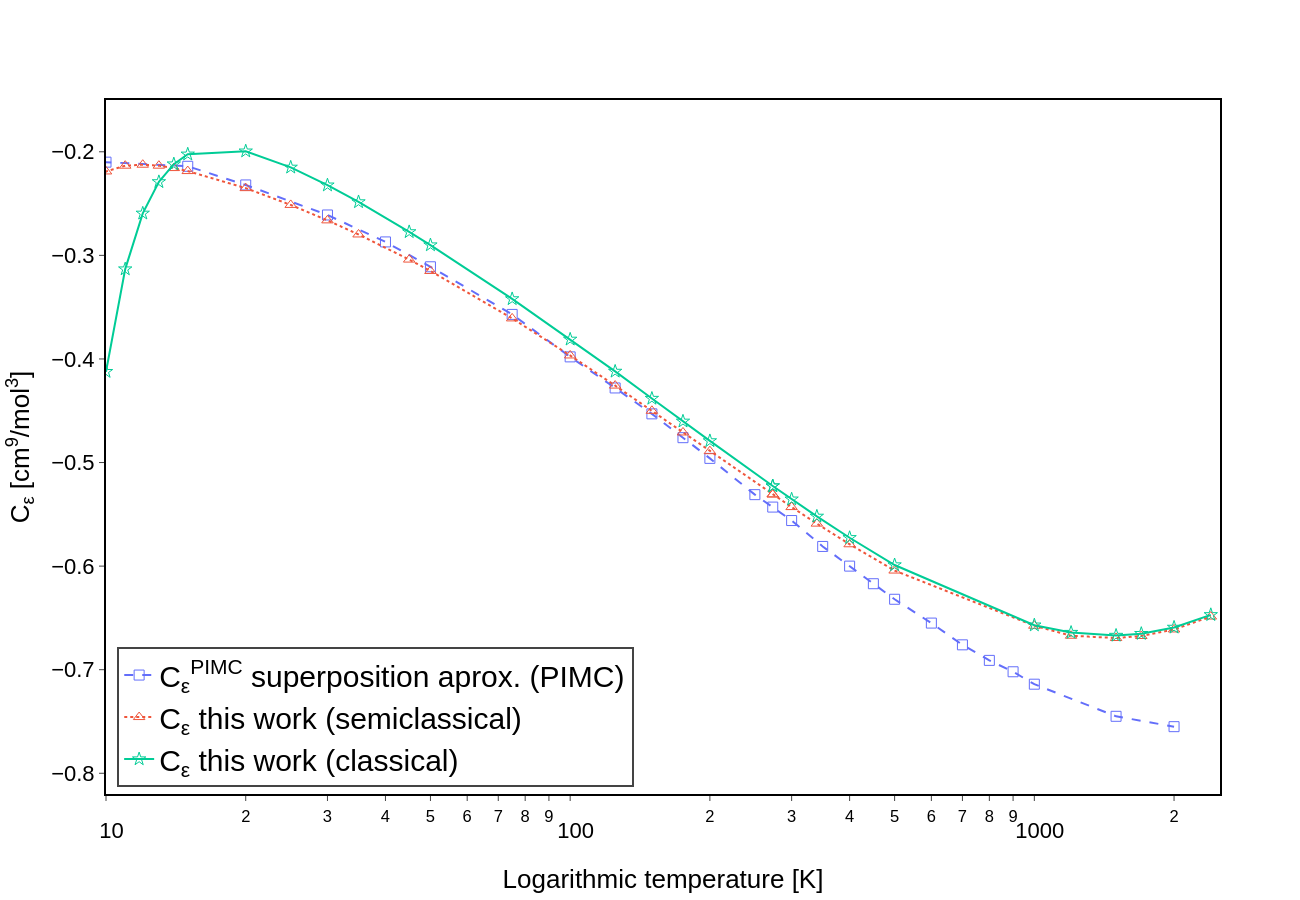}
\caption{
Comparison of the classical and semiclassical values of $C_\varepsilon(T)$ calculated in this work with the results from the quantum calculation (PIMC) of Garberoglio \emph{et al.} \cite{garberoglio20213dielec} employing the superposition approximation.}
\label{fig:virial_total_scl}
\end{figure}

Similarly, it is possible to introduce the same modification to the two-body polarizability as the polarizability is the second derivative of the potential with respect to the electric field. 
The QFH modified polarizability reads
\begin{equation}\label{qfh_polar}
\alpha^{QFH}_2(r_{ij}) 
= \alpha_2 (r_{ij}) 
+ \frac{\beta}{12m} \left(
    \frac{\partial^2 \alpha_2(r_{ij})}{\partial r_{ij}^2}
   +\frac{2}{r_{ij}}\frac{\partial \alpha_2 (r_{ij})}{\partial r_{ij}}
 \right).
\end{equation}
Since there is no three-body Feynman-Hibbs effective potential available, we modified only the two-body potentials and polarizabilities. Using the modified two-body potential and polarizability from Eqs.~(\ref{qfh_pot}) and (\ref{qfh_polar}) we calculated semiclassical results for the two-body terms ($BB_\varepsilon$, $C_\varepsilon^\mathrm{2body}$) in Eq.~(\ref{dielectric_integral}). The resulting semiclassical values of the third dielectric virial coefficients are presented in Fig.~\ref{fig:virial_total_scl}.

Note that this procedure is easily justified for the third pressure virial coefficient where the two-body contribution accounts for over 95\% of its value. The ratio of the two- and three-body contributions to the third dielectric virial coefficient is closer to the unity (see Fig.~\ref{fig:virial_contrib}) and hence our semiclassical values may provide an incomplete information. Nevertheless, this crude semiclassical approximation to the two-body terms of $C_\varepsilon(T)$ only, gives results that agree with the PIMC values of the Garberoglio \emph{et al.} \cite{garberoglio20213dielec} down to 10~K. It is possible that this is a consequence of a fortuitous error cancellation, but it may also indicate that the semiclassical quantum correction to the three-body polarizability surface is small. This requires a further study employing the quadratic Feynmann-Hibbs potential for three particles.

%%%%%%%%%%%%%%%%%%%%%%%%%%%%%%%%%%%%%%%%%%%%%%%%%%%%%%%%%%%%%%%%%%%%%%
%%%%%%%%%%%%%%%%%%%%%%%%%%%%%%%%%%%%%%%%%%%%%%%%%%%%%%%%%%%%%%%%%%%%%%
%%%%%%%%%%%%%%%%%%%%%%%%%%%%%%%%%%%%%%%%%%%%%%%%%%%%%%%%%%%%%%%%%%%%%%
\section{Conclusions}
%%%%%%%%%%%%%%%%%%%%%%%%%%%%%%%%%%%%%%%%%%%%%%%%%%%%%%%%%%%%%%%%%%%%%%
%%%%%%%%%%%%%%%%%%%%%%%%%%%%%%%%%%%%%%%%%%%%%%%%%%%%%%%%%%%%%%%%%%%%%%
%%%%%%%%%%%%%%%%%%%%%%%%%%%%%%%%%%%%%%%%%%%%%%%%%%%%%%%%%%%%%%%%%%%%%%

We have presented the first complete \emph{ab initio} calculations of the three-body polarizability of helium and constructed an analytical three-dimensional fit of its isotropic part. The fit is based on \emph{ab initio} calculations at the CC3 level of theory extrapolated to the complete basis set limit.   
We have taken into account the exact asymptotic behavior of the three-body polarizability for atomic configurations corresponding to two possible fragmentation channels: three free atoms and atom-diatom.
Our uncertainty budget takes into account the uncertainty of the complete basis set extrapolation as well as
the approximate nature of the CC3 model. The analytical fit obtained by us has a mean absolute relative error of $0.4\, \sigma$ which corresponds to the mean absolute percentage error of 1.7\% with respect to the computed \emph{ab initio} polarizabilities. Furthermore, we present a three-dimensional fit of local uncertainties of our calculations, which is important from the point of view of applications, e.g., in metrology. 

Employing the fits developed in this work and the best available two-body potentials and polarizabilities, we performed classical and semiclassical calculations of the third dielectric virial coefficient $C_\varepsilon(T)$ for gaseous helium. In contrast to the third pressure virial coefficient, the three-body effects give a substantial contribution to the total value of $C_\varepsilon(T)$---as large as 50\% for temperature range from 30~K to 75~K.

Our classical results agree with the classical calculations of Garberoglio \emph{et al.} \cite{garberoglio20213dielec} from 10~K up to 400~K, but start to differ at higher temperatures.
This difference is attributed to the increasing importance of short-range contributions to the three-body polarizability, treated in Ref.~\onlinecite{garberoglio20213dielec} using the superposition approximation. 
For temperatures below 100~K we observed significant differences in comparison with the results of quantum calculations of Garberoglio \emph{et al.} \cite{garberoglio20213dielec}. Nonetheless, our semiclassical calculations employing the \emph{ab initio} three-body polarizability agree very well with the fully quantum PIMC calculations \cite{garberoglio20213dielec} down to 10~K. Above 273~K our results are consistent with the available experimental data but have much smaller uncertainties.  For lower temperatures we observed differences with the results of White and Gugan \cite{1992Metro..29...37W} %obtained  near 11~K and 18 K  
and of Hout and Bose~\cite{1991JChPh..95.2683H}, which have uncertainties more than an order of magnitude larger than the uncertainties of our semiclassical calculations. These differences are significant in view of the excellent agreement between the semiclassical and fully quantum PIMC calculations at the relevant temperature range.

%Additionally, we have performed semiclassical %calculations using the quadratic Feyn\-man-Hibbs %effective two-body potential and polarizability. 
%The resulting quantum corrections to $C_\varepsilon(T)$ substantially improve the results of classical calculations and recover the quantum PIMC values for temperatures down to 10~K. However, this approach requires further study, focused on solving the problem of missing three-particle Feynman-Hibbs effective potential.

\begin{acknowledgments}
We thank Giovanni Garberoglio and Krzysztof Szalewicz for reading and commenting on the manuscript.
Professor Robert Moszy\'nski is thanked for his involvement at the early stage of this work. 
The authors acknowledge the support from QuantumPascal project 18SIB04, which has received funding from the EMPIR programme co-financed by the Participating States and from the European Union’s Horizon 2020 research and innovation programme. The authors thank also for the support from the National Science Center, Poland, Project No. 2017/27/B/ST4/02739. This research was supported in part by PL-Grid infrastructure.
\end{acknowledgments}

\appendix
%\section{Appendix}
%\setcounter{equation}{0}
%\numberwithin{equation}{subsection}

%%%%%%%%%%%%%%%%%%%%%%%%%%%%%%%%%%%%%%%%%%%%%%%%%%%%%%%%%%%%%%%%%%%%%%
%%%%%%%%%%%%%%%%%%%%%%%%%%%%%%%%%%%%%%%%%%%%%%%%%%%%%%%%%%%%%%%%%%%%%%
%%%%%%%%%%%%%%%%%%%%%%%%%%%%%%%%%%%%%%%%%%%%%%%%%%%%%%%%%%%%%%%%%%%%%%
\section{Regularization of integrals contributing to $C_\varepsilon(T)$}
\label{regularization}
%%%%%%%%%%%%%%%%%%%%%%%%%%%%%%%%%%%%%%%%%%%%%%%%%%%%%%%%%%%%%%%%%%%%%%
%%%%%%%%%%%%%%%%%%%%%%%%%%%%%%%%%%%%%%%%%%%%%%%%%%%%%%%%%%%%%%%%%%%%%%
%%%%%%%%%%%%%%%%%%%%%%%%%%%%%%%%%%%%%%%%%%%%%%%%%%%%%%%%%%%%%%%%%%%%%%

In evaluating the three-body contribution to $C_{\varepsilon}$ one encounters ill-defined integrals of the form 
\begin{equation} \label{A1}
\iint 
\frac{f_4(\eta r_1)\,f_{n+1}(\eta r_2)}{ r^3_1\,r_2^{n}} 
\,P_2(\cos\gamma)
\,d\mathbf{r}_1 d\mathbf{r}_2,
\end{equation}
where the $\mathbf{r}_1$ and $\mathbf{r}_2$ integrations extend over the whole 6\,D space, $\gamma$ is the angle between the vectors $\mathbf{r}_1$ and $\mathbf{r}_2$, and $\eta$ is the fitting parameter specifying the strength of the short-range damping effected by the Tang-Toennies functions (\ref{TTdamp}). 
The integrals with $n=3$ and $n=6$ are generated by the asymptotics of the three-body polarizability, see Eq.~(\ref{longfit}), whereas  when $n \ge 9$, they appear due to the long-range behavior of the two-body Mayer functions $e^{-\beta U(r_{ij}) }-1$, see Eq.~(\ref{dielectric_integral2}).
The integrals (\ref{A1}) are ill-defined since they are divergent when the integration is performed first over $r_1$ or $r_2$ and   vanish if the angular integration over $\gamma$ is executed first. 
To eliminate this ambiguity, we note that the  integrals (\ref{A1}) are derived from the well defined three-atom,  9\,D  integrals in Eq.~(\ref{dielectric_integral2}) when the integration over  center of mass is performed, which leads to the factorization of volume $V$.
This procedure is mathematically justified only if the integral~(\ref{A1}) is convergent.
We thus have to go back to the original three-particle integral over the finite volume $V$.
This leads to the following 9\,D integral 
\begin{equation} \label{A2}
\mathcal{I}_n =
\lim_{V\rightarrow \infty }\frac{1}{V}
\iiint
\frac{f_4(\eta |\mathbf{r}_1 - \mathbf{a}|) }{|\mathbf{r}_1 - \mathbf{a}|^3}
\frac{f_{n+1}(\eta |\mathbf{r}_2 - \mathbf{a}|) }{|\mathbf{r}_2 - \mathbf{a}|^n }
\, P_2(\cos\gamma)
\,d\mathbf{a}\,d\mathbf{r}_1d\mathbf{r}_2,
\end{equation}
where each of the three 3\,D integrations is over the interior of the sphere of radius $R$ with the center at the origin of the coordinate system, and $\gamma$ denotes the angle between the vectors 
\mbox{$\mathbf{r}_1 - \mathbf{a}$} and $\mathbf{r}_2 - \mathbf{a}$.
Actually the parameter $\eta > 0$ can be different for $f_4$ and $f_{n+1}$, but this complication is irrelevant for further reasoning. 
Unlike in the conventional statistical mechanics treatment, no center of mass is introduced here and the integration is over positions of all three particles, the position of the particle 3 being denoted by $\mathbf{a}$.

Employing the spherical harmonics addition theorem and the spherical symmetry of the integrand in Eq.~(\ref{A2}), one can see that $\mathcal{I}_n$ is equal to the $R\rightarrow \infty$ limit of the one-dimensional integral $\mathcal{I}_n(R)$ 
\begin{equation} \label{A3}
\mathcal{I}_n(R)= \frac{3}{R^3} \int_0^R a^2 \, I_3(R,a) \, I_n(R,a) \, da,
\end{equation}  
over the products of one-particle (3D) integrals $I_n(R,a)$
\begin{equation}
I_n(R,a) =
\int \frac{f_{n+1}(\eta |\mathbf{r} - \mathbf{a}|)}{|\mathbf{r} - \mathbf{a}|^{n}}
\,P_2(\cos \theta_a ) 
\,d\mathbf{r},
\end{equation}
where the integration is over the interior of the sphere of radius $R$ located at the origin of the coordinate system, the vector $\mathbf{a}$ is placed on the $z$ axis, and $\theta_a$ is the angle between the vector $\mathbf{r} - \mathbf{a}$ and the $z$ axis.
Note that due to cylindrical symmetry of the integration over $\textbf{r}_1$ and $\textbf{r}_2$, the spherical harmonics $Y_{2m}(\theta_a,\phi_a)$, with $m=\pm1$ and $m=\pm2$ do not contribute to $\mathcal{I}_n$.
Representing $P_2(\cos\theta_a)$ in terms of Cartesian coordinates we can write 
\begin{equation}
I_n(R,a) =
\frac{1}{2} \int 
\frac{f_{n+1}(\eta |\mathbf{r} - \mathbf{a}|)}{|\mathbf{r} - \mathbf{a}|^{n+2}}
\,(3z^2-r^2-4az+2a^2) 
\,d\mathbf{r},
\end{equation}   
where the integration is again over the interior of the sphere $|\mathbf{r} | \le R$.
Performing this integration in spherical coordinates $r$, $\theta$, $\phi$, and making the substitution $u=\cos\theta$ we find  
\begin{equation}
\begin{split}
I_n(R,a)&
= \pi \int_0^R r^2 \int_{-1}^{+1}
\frac{f_{n+1} [\eta(r^2-2aru+a^2 )^{1/2}]}{\big(r^2-2aru+a^2\big)^{n/2 +1}}\\
&\times (3r^2u^2 -r^2 -4aru +2a^2)
\,du\,dr.
\end{split}
\end{equation}

Let us first  consider the case $n=3$.
It is then possible to express the integration over $u$ and  $r$ in closed form.
The result reads
\begin{equation}
\begin{split}
I_3(R,a) &
= \frac{2\pi R^3}{3\,a^3} 
  \left[ f_4(\eta R+ \eta a) -f_4 (\eta  R- \eta  a)) \right]\\
&+ \frac{\pi\,\eta^5 R^8}{576\,a^3}
\bigg[ 
                                                             g_7(\eta R,a)
 - 3\left(                  \frac{a^2}{R^2} +3 \right)       g_5(\eta R,a) \\
&+ 3\left(\frac{a^4}{R^4} +2\frac{a^2}{R^2} -3 \right)       g_3(\eta R,a) \\
&-  \left(                  \frac{a^2}{R^2} -1 \right)^{\!3} g_1(\eta R,a)
\bigg],
\end{split}
\end{equation}
where
\begin{equation}
g_0(w,a) = \frac{e^{-w}}{w} (e^{aw} -e^{-aw})   
\end{equation}
and $g_n(w,a)$ is the $n$-th derivative of $g_0(w,a)$ with respect to $w$ multiplied by the factor of $(-1)^n$.
The evaluation of the final integral over the variable $a$ in Eq.~(\ref{A3}) is now elementary, although rather tedious and cannot be performed by hand.
However, it can by accomplished by the {\sc Mathematica} \cite{mathematica} software using symbolic integration methods.
The result is lengthy, but we are interested only in the leading terms in the large $R$ asymptotic expansion which are
\begin{equation} \label{A9}
\mathcal{I}_3(R)
= \frac{89 \pi^2}{192\eta}   \,\frac{1}{R}
- \frac{25\pi^2}{8\eta^2}    \,\frac{1}{R^2}
+ \frac{763\pi^2}{576\eta^3} \,\frac{1}{R^3} 
+ \mathcal{O}\left(\frac{1}{R^4}\right).
\end{equation}
It is clear that the ${R\rightarrow\infty}$ limit of $\mathcal{I}_3(R)$ equals to zero.
This means that the regularized integral $\mathcal{I}_3$ vanishes and the integral (\ref{A1}) can be assumed to be zero for $n=3$.

Let us now consider the general case of arbitrary $n \ge 4$.
Using the obvious inequality $|ab|\le (a^2 +b^2)/2$, valid for any real numbers $a$ and $b$, we can make the following estimate 
\begin{equation} \label{A10}
|\mathcal{I}_n(R)| \,\le\, \frac{1}{2}\,\mathcal{I}_3(R) 
+ \frac{3}{2R^3} \int_0^R a^2\,I^2_n(R,a)\,da .
\end{equation}  
When $n \ge 4$ the second term on right-hand-side of the inequality (\ref{A10}) can be rigorously estimated without analytic calculation.
For this  purpose we  define a "small" ball $|\mathbf{r} -\mathbf{a}|\le\rho$ of a radius
\begin{equation} \label{A11}
\rho = \left( \frac{4\pi \sqrt{R}}{ n-3} \right)^{1/(n-3)}
\end{equation}
centered around the point $\mathbf{a}$.
Note that this "small" ball does not have to be entirely included in the "large" ball, $|\mathbf{r}|\le R$.  
Since $|f_{n+1}(x)| < 1$ and $P_2(\cos \theta_a) \le 1$ one can easily show that 
\begin{equation} \label{A12}
\int_{|\mathbf{r} - \mathbf{a}|\ge \rho\,\wedge \,|\mathbf{r}| \le R}   \frac{f_{n+1}(\eta |\mathbf{r} - \mathbf{a}|)}{|\mathbf{r} - \mathbf{a}|^{n}}  
\,P_2(\cos \theta_a) 
\,d\mathbf{r} \,\le\, \frac{1}{\sqrt{R}},
\end{equation}
where the integration extends now over the intersection of the exterior of the "small" ball and the interior of the "large" ball. 
For $a \le R-\rho$ the integral $I_n(R,a)$ can be estimated as
\begin{equation} \label{A13}
\begin{split}
I_n(R,a)&
=\int_{|\mathbf{r} - \mathbf{a}|< \rho\,\wedge \,|\mathbf{r}| \le R}
 \frac{f_{n+1}(\eta |\mathbf{r} - \mathbf{a}|)}{|\mathbf{r} - \mathbf{a}|^{n}}
 \,P_2(\cos \theta_a ) 
 \,d\mathbf{r} \\
&+\int_{|\mathbf{r} - \mathbf{a}|\ge \rho\,\wedge \,|\mathbf{r}| \le R}
  \frac{f_{n+1}(\eta |\mathbf{r} - \mathbf{a}|)}{|\mathbf{r} - \mathbf{a}|^{n}}
  \,P_2(\cos \theta_a )
  \,d\mathbf{r} \;\le\,  \frac{1}{\sqrt{R}},
\end{split}
\end{equation}
because the first integral vanishes by spherical symmetry.
For $a > R-\rho$ we can use the following estimate 
\begin{equation} \label{A14}
I_n(R,a) \,\le\,
\int \frac{f_{n+1}(\eta |\mathbf{r} - \mathbf{a}|)}{|\mathbf{r} - \mathbf{a}|^{n}}
\,d\mathbf{r} \,=C_n,
\end{equation}
where the integration extends over the whole 3D space, so that $C_n$ does not depend on $R$. 
Splitting the $a$ integration into $a \le R-\rho$ and $a\ge R-\rho$ and using Eqs.~(\ref{A13}) and (\ref{A14}), the second term on the right-hand-side of the inequality (\ref{A10}) can be estimated as follows 
\begin{equation} \label{A15}  
\frac{3}{2R^3} \int_0^R a^2 \,I^2_n(R,a)\,da
\,\le\, 
 \frac{1}{2R}\left(1-\frac{\rho}{R}\right)^3
+\frac{1}{2}\,C_n^2 \left[\left(1-\frac{\rho}{R}\right)^3-1\right]
=\mathcal{O}\left(\frac{\rho}{R}\right).
\end{equation}
Since $\rho \sim R^{1/(2n-6)}$ we see that for $n \ge 4$ the second term in the inequality in Eq.~(\ref{A10}) goes to zero when $R\rightarrow\infty$.
In view of Eq.~(\ref{A9}) the same applies to $\mathcal{I}_n(R)$.
Therefore, the regularized integral of Eq.~(\ref{A2}) is equal to zero and the ill-defined integrals of Eq.~(\ref{A1}) can be neglected in calculations of $C_{\varepsilon}$.

Although in the short-range damping of the asymptotics three-body energy, as given by Eq.~(\ref{longfit}), we used only the $f_4(x)$ and $f_{n+1}(x) $damping functions, the reasoning similar to that of Eqs.~(\ref{A10})-(\ref{A15}) shows that the regularized integral (\ref{A1}) vanishes also when the damping function $f_4(x)$ is replaced by any $f_k(x)$, with $k\ge 2$ or $f_{n+1}(x)$ is replaced by any $f_k(x)$, $k \ge n-1$.

%\setcounter{equation}{0}
%\numberwithin{equation}{subsection} 

% \bibliographystyle{short}
\bibliography{c_epsilon_paper}

\begin{thebibliography}{89}%
\makeatletter
\providecommand \@ifxundefined [1]{%
 \@ifx{#1\undefined}
}%
\providecommand \@ifnum [1]{%
 \ifnum #1\expandafter \@firstoftwo
 \else \expandafter \@secondoftwo
 \fi
}%
\providecommand \@ifx [1]{%
 \ifx #1\expandafter \@firstoftwo
 \else \expandafter \@secondoftwo
 \fi
}%
\providecommand \natexlab [1]{#1}%
\providecommand \enquote  [1]{``#1''}%
\providecommand \bibnamefont  [1]{#1}%
\providecommand \bibfnamefont [1]{#1}%
\providecommand \citenamefont [1]{#1}%
\providecommand \href@noop [0]{\@secondoftwo}%
\providecommand \href [0]{\begingroup \@sanitize@url \@href}%
\providecommand \@href[1]{\@@startlink{#1}\@@href}%
\providecommand \@@href[1]{\endgroup#1\@@endlink}%
\providecommand \@sanitize@url [0]{\catcode `\\12\catcode `\$12\catcode
  `\&12\catcode `\#12\catcode `\^12\catcode `\_12\catcode `\%12\relax}%
\providecommand \@@startlink[1]{}%
\providecommand \@@endlink[0]{}%
\providecommand \url  [0]{\begingroup\@sanitize@url \@url }%
\providecommand \@url [1]{\endgroup\@href {#1}{\urlprefix }}%
\providecommand \urlprefix  [0]{URL }%
\providecommand \Eprint [0]{\href }%
\providecommand \doibase [0]{https://doi.org/}%
\providecommand \selectlanguage [0]{\@gobble}%
\providecommand \bibinfo  [0]{\@secondoftwo}%
\providecommand \bibfield  [0]{\@secondoftwo}%
\providecommand \translation [1]{[#1]}%
\providecommand \BibitemOpen [0]{}%
\providecommand \bibitemStop [0]{}%
\providecommand \bibitemNoStop [0]{.\EOS\space}%
\providecommand \EOS [0]{\spacefactor3000\relax}%
\providecommand \BibitemShut  [1]{\csname bibitem#1\endcsname}%
\let\auto@bib@innerbib\@empty
%</preamble>
\bibitem [{\citenamefont {Borysow}\ and\ \citenamefont
  {Frommhold}(1989)}]{1989borysowcollision}%
  \BibitemOpen
  \bibfield  {author} {\bibinfo {author} {\bibfnamefont {A.}~\bibnamefont
  {Borysow}}\ and\ \bibinfo {author} {\bibfnamefont {L.}~\bibnamefont
  {Frommhold}},\ }\bibfield  {title} {\bibinfo {title} {{Collision-induced
  light scattering: a bibliography}},\ }\href@noop {} {\bibfield  {journal}
  {\bibinfo  {journal} {Adv. Chem. Phys.}\ }\textbf {\bibinfo {volume} {75}},\
  \bibinfo {pages} {439} (\bibinfo {year} {1989})}\BibitemShut {NoStop}%
\bibitem [{\citenamefont {{Moszynski}}\ \emph {et~al.}(1995)\citenamefont
  {{Moszynski}}, \citenamefont {{Heijmen}},\ and\ \citenamefont {{Van der
  Avoird}}}]{1995CPL...247..440M}%
  \BibitemOpen
  \bibfield  {author} {\bibinfo {author} {\bibfnamefont {R.}~\bibnamefont
  {{Moszynski}}}, \bibinfo {author} {\bibfnamefont {T.~G.~A.}\ \bibnamefont
  {{Heijmen}}},\ and\ \bibinfo {author} {\bibfnamefont {A.}~\bibnamefont {{Van
  der Avoird}}},\ }\bibfield  {title} {\bibinfo {title} {{Second dielectric
  virial coefficient of helium gas: quantum-statistical calculations from an ab
  initio interaction-induced polarizability}},\ }\href
  {https://doi.org/10.1016/S0009-2614(95)01271-0} {\bibfield  {journal}
  {\bibinfo  {journal} {Chem. Phys. Lett.}\ }\textbf {\bibinfo {volume}
  {247}},\ \bibinfo {pages} {440} (\bibinfo {year} {1995})}\BibitemShut
  {NoStop}%
\bibitem [{\citenamefont {McTague}\ and\ \citenamefont
  {Birnbaum}(1968)}]{1968mctaguecollision}%
  \BibitemOpen
  \bibfield  {author} {\bibinfo {author} {\bibfnamefont {J.~P.}\ \bibnamefont
  {McTague}}\ and\ \bibinfo {author} {\bibfnamefont {G.}~\bibnamefont
  {Birnbaum}},\ }\bibfield  {title} {\bibinfo {title} {{Collision-induced light
  scattering in gaseous Ar and Kr}},\ }\href@noop {} {\bibfield  {journal}
  {\bibinfo  {journal} {Phys. Rev. Lett.}\ }\textbf {\bibinfo {volume} {21}},\
  \bibinfo {pages} {661} (\bibinfo {year} {1968})}\BibitemShut {NoStop}%
\bibitem [{\citenamefont {Moszynski}\ \emph {et~al.}(1996)\citenamefont
  {Moszynski}, \citenamefont {Heijmen}, \citenamefont {Wormer},\ and\
  \citenamefont {{Van der Avoird}}}]{Moszynski1996raman}%
  \BibitemOpen
  \bibfield  {author} {\bibinfo {author} {\bibfnamefont {R.}~\bibnamefont
  {Moszynski}}, \bibinfo {author} {\bibfnamefont {T.~G.~A.}\ \bibnamefont
  {Heijmen}}, \bibinfo {author} {\bibfnamefont {P.~E.~S.}\ \bibnamefont
  {Wormer}},\ and\ \bibinfo {author} {\bibfnamefont {A.}~\bibnamefont {{Van der
  Avoird}}},\ }\bibfield  {title} {\bibinfo {title} {{Ab initio
  collision‐induced polarizability, polarized and depolarized Raman spectra,
  and second dielectric virial coefficient of the helium diatom}},\ }\href
  {https://doi.org/10.1063/1.471416} {\bibfield  {journal} {\bibinfo  {journal}
  {J. Chem. Phys.}\ }\textbf {\bibinfo {volume} {104}},\ \bibinfo {pages}
  {6997} (\bibinfo {year} {1996})}\BibitemShut {NoStop}%
\bibitem [{\citenamefont {Skomorowski}\ and\ \citenamefont
  {Moszynski}(2013)}]{Skomorowski2013}%
  \BibitemOpen
  \bibfield  {author} {\bibinfo {author} {\bibfnamefont {W.}~\bibnamefont
  {Skomorowski}}\ and\ \bibinfo {author} {\bibfnamefont {R.}~\bibnamefont
  {Moszynski}},\ }\bibfield  {title} {\bibinfo {title} {{Kerr and
  Cotton–Mouton effects in atomic gases: a quantum-statistical study}},\
  }\href@noop {} {\bibfield  {journal} {\bibinfo  {journal} {Mol. Phys.}\
  }\textbf {\bibinfo {volume} {111}},\ \bibinfo {pages} {1414} (\bibinfo {year}
  {2013})}\BibitemShut {NoStop}%
\bibitem [{\citenamefont {Buckingham}\ and\ \citenamefont
  {Pople}(1955)}]{1955buckinghamstatistical}%
  \BibitemOpen
  \bibfield  {author} {\bibinfo {author} {\bibfnamefont {A.~D.}\ \bibnamefont
  {Buckingham}}\ and\ \bibinfo {author} {\bibfnamefont {J.~A.}\ \bibnamefont
  {Pople}},\ }\bibfield  {title} {\bibinfo {title} {{The statistical mechanics
  of imperfect polar gases. Part 2.—Dielectric polarization}},\ }\href@noop
  {} {\bibfield  {journal} {\bibinfo  {journal} {Trans.~Faraday Soc.}\ }\textbf
  {\bibinfo {volume} {51}},\ \bibinfo {pages} {1179} (\bibinfo {year}
  {1955})}\BibitemShut {NoStop}%
\bibitem [{\citenamefont {{Gaiser}}\ \emph
  {et~al.}(2017{\natexlab{a}})\citenamefont {{Gaiser}}, \citenamefont
  {{Fellmuth}}, \citenamefont {{Haft}}, \citenamefont {{Kuhn}}, \citenamefont
  {{Thiele-Krivoi}}, \citenamefont {{Zandt}}, \citenamefont {{Fischer}},
  \citenamefont {{Jusko}},\ and\ \citenamefont
  {{Sabuga}}}]{2017Metro..54..280G}%
  \BibitemOpen
  \bibfield  {author} {\bibinfo {author} {\bibfnamefont {C.}~\bibnamefont
  {{Gaiser}}}, \bibinfo {author} {\bibfnamefont {B.}~\bibnamefont
  {{Fellmuth}}}, \bibinfo {author} {\bibfnamefont {N.}~\bibnamefont {{Haft}}},
  \bibinfo {author} {\bibfnamefont {A.}~\bibnamefont {{Kuhn}}}, \bibinfo
  {author} {\bibfnamefont {B.}~\bibnamefont {{Thiele-Krivoi}}}, \bibinfo
  {author} {\bibfnamefont {T.}~\bibnamefont {{Zandt}}}, \bibinfo {author}
  {\bibfnamefont {J.}~\bibnamefont {{Fischer}}}, \bibinfo {author}
  {\bibfnamefont {O.}~\bibnamefont {{Jusko}}},\ and\ \bibinfo {author}
  {\bibfnamefont {W.}~\bibnamefont {{Sabuga}}},\ }\bibfield  {title} {\bibinfo
  {title} {{Final determination of the Boltzmann constant by
  dielectric-constant gas thermometry}},\ }\href
  {https://doi.org/10.1088/1681-7575/aa62e3} {\bibfield  {journal} {\bibinfo
  {journal} {Metrologia}\ }\textbf {\bibinfo {volume} {54}},\ \bibinfo {pages}
  {280} (\bibinfo {year} {2017}{\natexlab{a}})}\BibitemShut {NoStop}%
\bibitem [{\citenamefont {Gaiser}\ \emph {et~al.}(2015)\citenamefont {Gaiser},
  \citenamefont {Zandt},\ and\ \citenamefont
  {Fellmuth}}]{gaiser2015dielectric}%
  \BibitemOpen
  \bibfield  {author} {\bibinfo {author} {\bibfnamefont {C.}~\bibnamefont
  {Gaiser}}, \bibinfo {author} {\bibfnamefont {T.}~\bibnamefont {Zandt}},\ and\
  \bibinfo {author} {\bibfnamefont {B.}~\bibnamefont {Fellmuth}},\ }\bibfield
  {title} {\bibinfo {title} {{Dielectric-constant gas thermometry}},\
  }\href@noop {} {\bibfield  {journal} {\bibinfo  {journal} {Metrologia}\
  }\textbf {\bibinfo {volume} {52}},\ \bibinfo {pages} {S217} (\bibinfo {year}
  {2015})}\BibitemShut {NoStop}%
\bibitem [{\citenamefont {{Gaiser}}\ \emph
  {et~al.}(2017{\natexlab{b}})\citenamefont {{Gaiser}}, \citenamefont
  {{Fellmuth}},\ and\ \citenamefont {{Haft}}}]{2017Metro..54..141G}%
  \BibitemOpen
  \bibfield  {author} {\bibinfo {author} {\bibfnamefont {C.}~\bibnamefont
  {{Gaiser}}}, \bibinfo {author} {\bibfnamefont {B.}~\bibnamefont
  {{Fellmuth}}},\ and\ \bibinfo {author} {\bibfnamefont {N.}~\bibnamefont
  {{Haft}}},\ }\bibfield  {title} {\bibinfo {title} {{{Primary thermometry from
  2.5{\,}K to 140{\,}K applying dielectric-constant gas thermometry}}},\ }\href
  {https://doi.org/10.1088/1681-7575/aa5389} {\bibfield  {journal} {\bibinfo
  {journal} {Metrologia}\ }\textbf {\bibinfo {volume} {54}},\ \bibinfo {pages}
  {141} (\bibinfo {year} {2017}{\natexlab{b}})}\BibitemShut {NoStop}%
\bibitem [{\citenamefont {Gaiser}\ \emph {et~al.}(2020)\citenamefont {Gaiser},
  \citenamefont {Fellmuth},\ and\ \citenamefont
  {Haft}}]{gaiser2020thermodynamic}%
  \BibitemOpen
  \bibfield  {author} {\bibinfo {author} {\bibfnamefont {C.}~\bibnamefont
  {Gaiser}}, \bibinfo {author} {\bibfnamefont {B.}~\bibnamefont {Fellmuth}},\
  and\ \bibinfo {author} {\bibfnamefont {N.}~\bibnamefont {Haft}},\ }\bibfield
  {title} {\bibinfo {title} {{Thermodynamic-temperature data from 30 K to 200
  K}},\ }\href@noop {} {\bibfield  {journal} {\bibinfo  {journal} {Metrologia}\
  }\textbf {\bibinfo {volume} {57}},\ \bibinfo {pages} {055003} (\bibinfo
  {year} {2020})}\BibitemShut {NoStop}%
\bibitem [{\citenamefont {{Ripa}}\ \emph {et~al.}(2021)\citenamefont {{Ripa}},
  \citenamefont {{Imbraguglio}}, \citenamefont {{Gaiser}}, \citenamefont
  {{Steur}}, \citenamefont {{Giraudi}}, \citenamefont {{Fogliati}},
  \citenamefont {{Bertinetti}}, \citenamefont {{Lopardo}}, \citenamefont
  {{Dematteis}},\ and\ \citenamefont {{Gavioso}}}]{2021Metro..58b5008R}%
  \BibitemOpen
  \bibfield  {author} {\bibinfo {author} {\bibfnamefont {D.~M.}\ \bibnamefont
  {{Ripa}}}, \bibinfo {author} {\bibfnamefont {D.}~\bibnamefont
  {{Imbraguglio}}}, \bibinfo {author} {\bibfnamefont {C.}~\bibnamefont
  {{Gaiser}}}, \bibinfo {author} {\bibfnamefont {P.~P.~M.}\ \bibnamefont
  {{Steur}}}, \bibinfo {author} {\bibfnamefont {D.}~\bibnamefont {{Giraudi}}},
  \bibinfo {author} {\bibfnamefont {M.}~\bibnamefont {{Fogliati}}}, \bibinfo
  {author} {\bibfnamefont {M.}~\bibnamefont {{Bertinetti}}}, \bibinfo {author}
  {\bibfnamefont {G.}~\bibnamefont {{Lopardo}}}, \bibinfo {author}
  {\bibfnamefont {R.}~\bibnamefont {{Dematteis}}},\ and\ \bibinfo {author}
  {\bibfnamefont {R.~M.}\ \bibnamefont {{Gavioso}}},\ }\bibfield  {title}
  {\bibinfo {title} {{Refractive index gas thermometry between 13.8 K and 161.4
  K}},\ }\href {https://doi.org/10.1088/1681-7575/abe249} {\bibfield  {journal}
  {\bibinfo  {journal} {Metrologia}\ }\textbf {\bibinfo {volume} {58}},\
  \bibinfo {eid} {025008} (\bibinfo {year} {2021})}\BibitemShut {NoStop}%
\bibitem [{\citenamefont {Schmidt}\ \emph {et~al.}(2007)\citenamefont
  {Schmidt}, \citenamefont {Gavioso}, \citenamefont {May},\ and\ \citenamefont
  {Moldover}}]{schmidt2007polarizability}%
  \BibitemOpen
  \bibfield  {author} {\bibinfo {author} {\bibfnamefont {J.~W.}\ \bibnamefont
  {Schmidt}}, \bibinfo {author} {\bibfnamefont {R.}~\bibnamefont {Gavioso}},
  \bibinfo {author} {\bibfnamefont {E.}~\bibnamefont {May}},\ and\ \bibinfo
  {author} {\bibfnamefont {M.~R.}\ \bibnamefont {Moldover}},\ }\bibfield
  {title} {\bibinfo {title} {Polarizability of helium and gas metrology},\
  }\href@noop {} {\bibfield  {journal} {\bibinfo  {journal} {Phys. Rev. Lett.}\
  }\textbf {\bibinfo {volume} {98}},\ \bibinfo {pages} {254504} (\bibinfo
  {year} {2007})}\BibitemShut {NoStop}%
\bibitem [{\citenamefont {{Gaiser}}\ \emph {et~al.}(2020)\citenamefont
  {{Gaiser}}, \citenamefont {{Fellmuth}},\ and\ \citenamefont
  {{Sabuga}}}]{2020NatPh..16..177G}%
  \BibitemOpen
  \bibfield  {author} {\bibinfo {author} {\bibfnamefont {C.}~\bibnamefont
  {{Gaiser}}}, \bibinfo {author} {\bibfnamefont {B.}~\bibnamefont
  {{Fellmuth}}},\ and\ \bibinfo {author} {\bibfnamefont {W.}~\bibnamefont
  {{Sabuga}}},\ }\bibfield  {title} {\bibinfo {title} {{Primary gas-pressure
  standard from electrical measurements and thermophysical ab initio
  calculations}},\ }\href {https://doi.org/10.1038/s41567-019-0722-2}
  {\bibfield  {journal} {\bibinfo  {journal} {Nat.~Phys.}\ }\textbf {\bibinfo
  {volume} {16}},\ \bibinfo {pages} {177} (\bibinfo {year} {2020})}\BibitemShut
  {NoStop}%
\bibitem [{\citenamefont {{Piszczatowski}}\ \emph {et~al.}(2015)\citenamefont
  {{Piszczatowski}}, \citenamefont {{Puchalski}}, \citenamefont {{Komasa}},
  \citenamefont {{Jeziorski}},\ and\ \citenamefont
  {{Szalewicz}}}]{2015PhRvL.114q3004P}%
  \BibitemOpen
  \bibfield  {author} {\bibinfo {author} {\bibfnamefont {K.}~\bibnamefont
  {{Piszczatowski}}}, \bibinfo {author} {\bibfnamefont {M.}~\bibnamefont
  {{Puchalski}}}, \bibinfo {author} {\bibfnamefont {J.}~\bibnamefont
  {{Komasa}}}, \bibinfo {author} {\bibfnamefont {B.}~\bibnamefont
  {{Jeziorski}}},\ and\ \bibinfo {author} {\bibfnamefont {K.}~\bibnamefont
  {{Szalewicz}}},\ }\bibfield  {title} {\bibinfo {title} {{Frequency-Dependent
  Polarizability of Helium Including Relativistic Effects with Nuclear Recoil
  Terms}},\ }\href {https://doi.org/10.1103/PhysRevLett.114.173004} {\bibfield
  {journal} {\bibinfo  {journal} {Phys.~Rev.~Lett.}\ }\textbf {\bibinfo
  {volume} {114}},\ \bibinfo {eid} {173004} (\bibinfo {year}
  {2015})}\BibitemShut {NoStop}%
\bibitem [{\citenamefont {{Puchalski}}\ \emph {et~al.}(2020)\citenamefont
  {{Puchalski}}, \citenamefont {{Szalewicz}}, \citenamefont {{Lesiuk}},\ and\
  \citenamefont {{Jeziorski}}}]{2020PhRvA.101b2505P}%
  \BibitemOpen
  \bibfield  {author} {\bibinfo {author} {\bibfnamefont {M.}~\bibnamefont
  {{Puchalski}}}, \bibinfo {author} {\bibfnamefont {K.}~\bibnamefont
  {{Szalewicz}}}, \bibinfo {author} {\bibfnamefont {M.}~\bibnamefont
  {{Lesiuk}}},\ and\ \bibinfo {author} {\bibfnamefont {B.}~\bibnamefont
  {{Jeziorski}}},\ }\bibfield  {title} {\bibinfo {title} {{QED calculation of
  the dipole polarizability of helium atom}},\ }\href
  {https://doi.org/10.1103/PhysRevA.101.022505} {\bibfield  {journal} {\bibinfo
   {journal} {Phys.~Rev.~A}\ }\textbf {\bibinfo {volume} {101}},\ \bibinfo
  {eid} {022505} (\bibinfo {year} {2020})}\BibitemShut {NoStop}%
\bibitem [{\citenamefont {{Czachorowski}}\ \emph {et~al.}(2020)\citenamefont
  {{Czachorowski}}, \citenamefont {{Przybytek}}, \citenamefont {{Lesiuk}},
  \citenamefont {{Puchalski}},\ and\ \citenamefont
  {{Jeziorski}}}]{czachorowski2020second}%
  \BibitemOpen
  \bibfield  {author} {\bibinfo {author} {\bibfnamefont {P.}~\bibnamefont
  {{Czachorowski}}}, \bibinfo {author} {\bibfnamefont {M.}~\bibnamefont
  {{Przybytek}}}, \bibinfo {author} {\bibfnamefont {M.}~\bibnamefont
  {{Lesiuk}}}, \bibinfo {author} {\bibfnamefont {M.}~\bibnamefont
  {{Puchalski}}},\ and\ \bibinfo {author} {\bibfnamefont {B.}~\bibnamefont
  {{Jeziorski}}},\ }\bibfield  {title} {\bibinfo {title} {{Second virial
  coefficients for $^{4}$He and $^{3}$He from an accurate relativistic
  interaction potential}},\ }\href
  {https://doi.org/10.1103/PhysRevA.102.042810} {\bibfield  {journal} {\bibinfo
   {journal} {Phys.~Rev.~A}\ }\textbf {\bibinfo {volume} {102}},\ \bibinfo
  {eid} {042810} (\bibinfo {year} {2020})}\BibitemShut {NoStop}%
\bibitem [{\citenamefont {{Cencek}}\ \emph {et~al.}(2009)\citenamefont
  {{Cencek}}, \citenamefont {{Patkowski}},\ and\ \citenamefont
  {{Szalewicz}}}]{cencek2009threehe}%
  \BibitemOpen
  \bibfield  {author} {\bibinfo {author} {\bibfnamefont {W.}~\bibnamefont
  {{Cencek}}}, \bibinfo {author} {\bibfnamefont {K.}~\bibnamefont
  {{Patkowski}}},\ and\ \bibinfo {author} {\bibfnamefont {K.}~\bibnamefont
  {{Szalewicz}}},\ }\bibfield  {title} {\bibinfo {title}
  {{Full-configuration-interaction calculation of three-body nonadditive
  contribution to helium interaction potential}},\ }\href
  {https://doi.org/10.1063/1.3204319} {\bibfield  {journal} {\bibinfo
  {journal} {J.~Chem.~Phys.}\ }\textbf {\bibinfo {volume} {131}},\ \bibinfo
  {pages} {064105} (\bibinfo {year} {2009})}\BibitemShut {NoStop}%
\bibitem [{\citenamefont {{Cencek}}\ \emph {et~al.}(2011)\citenamefont
  {{Cencek}}, \citenamefont {{Komasa}},\ and\ \citenamefont
  {{Szalewicz}}}]{2011JChPh.135a4301C}%
  \BibitemOpen
  \bibfield  {author} {\bibinfo {author} {\bibfnamefont {W.}~\bibnamefont
  {{Cencek}}}, \bibinfo {author} {\bibfnamefont {J.}~\bibnamefont {{Komasa}}},\
  and\ \bibinfo {author} {\bibfnamefont {K.}~\bibnamefont {{Szalewicz}}},\
  }\bibfield  {title} {\bibinfo {title} {{Collision-induced dipole
  polarizability of helium dimer from explicitly correlated calculations}},\
  }\href {https://doi.org/10.1063/1.3603968} {\bibfield  {journal} {\bibinfo
  {journal} {J.~Chem.~Phys.}\ }\textbf {\bibinfo {volume} {135}},\ \bibinfo
  {pages} {014301} (\bibinfo {year} {2011})}\BibitemShut {NoStop}%
\bibitem [{\citenamefont {{Vogel}}\ \emph {et~al.}(2010)\citenamefont
  {{Vogel}}, \citenamefont {{J{\"a}ger}}, \citenamefont {{Hellmann}},\ and\
  \citenamefont {{Bich}}}]{2010MolPh.108.3335V}%
  \BibitemOpen
  \bibfield  {author} {\bibinfo {author} {\bibfnamefont {E.}~\bibnamefont
  {{Vogel}}}, \bibinfo {author} {\bibfnamefont {B.}~\bibnamefont
  {{J{\"a}ger}}}, \bibinfo {author} {\bibfnamefont {R.}~\bibnamefont
  {{Hellmann}}},\ and\ \bibinfo {author} {\bibfnamefont {E.}~\bibnamefont
  {{Bich}}},\ }\bibfield  {title} {\bibinfo {title} {{Ab initio pair potential
  energy curve for the argon atom pair and thermophysical properties for the
  dilute argon gas. II. Thermophysical properties for low-density argon}},\
  }\href {https://doi.org/10.1080/00268976.2010.507557} {\bibfield  {journal}
  {\bibinfo  {journal} {Mol. Phys.}\ }\textbf {\bibinfo {volume} {108}},\
  \bibinfo {pages} {3335} (\bibinfo {year} {2010})}\BibitemShut {NoStop}%
\bibitem [{\citenamefont {{Patkowski}}\ and\ \citenamefont
  {{Szalewicz}}(2010)}]{2010JChPh.133i4304P}%
  \BibitemOpen
  \bibfield  {author} {\bibinfo {author} {\bibfnamefont {K.}~\bibnamefont
  {{Patkowski}}}\ and\ \bibinfo {author} {\bibfnamefont {K.}~\bibnamefont
  {{Szalewicz}}},\ }\bibfield  {title} {\bibinfo {title} {{Argon pair potential
  at basis set and excitation limits}},\ }\href
  {https://doi.org/10.1063/1.3478513} {\bibfield  {journal} {\bibinfo
  {journal} {J.~Chem.~Phys.}\ }\textbf {\bibinfo {volume} {133}},\ \bibinfo
  {pages} {094304} (\bibinfo {year} {2010})}\BibitemShut {NoStop}%
\bibitem [{\citenamefont {Cencek}\ \emph {et~al.}(2007)\citenamefont {Cencek},
  \citenamefont {Jeziorska}, \citenamefont {Akin-Ojo},\ and\ \citenamefont
  {Szalewicz}}]{cencek2007three}%
  \BibitemOpen
  \bibfield  {author} {\bibinfo {author} {\bibfnamefont {W.}~\bibnamefont
  {Cencek}}, \bibinfo {author} {\bibfnamefont {M.}~\bibnamefont {Jeziorska}},
  \bibinfo {author} {\bibfnamefont {O.}~\bibnamefont {Akin-Ojo}},\ and\
  \bibinfo {author} {\bibfnamefont {K.}~\bibnamefont {Szalewicz}},\ }\bibfield
  {title} {\bibinfo {title} {Three-body contribution to the helium interaction
  potential},\ }\href@noop {} {\bibfield  {journal} {\bibinfo  {journal} {J.
  Phys. Chem. A}\ }\textbf {\bibinfo {volume} {111}},\ \bibinfo {pages} {11311}
  (\bibinfo {year} {2007})}\BibitemShut {NoStop}%
\bibitem [{\citenamefont {Cencek}\ \emph {et~al.}(2013)\citenamefont {Cencek},
  \citenamefont {Garberoglio}, \citenamefont {Harvey}, \citenamefont
  {McLinden},\ and\ \citenamefont {Szalewicz}}]{cencek2013three}%
  \BibitemOpen
  \bibfield  {author} {\bibinfo {author} {\bibfnamefont {W.}~\bibnamefont
  {Cencek}}, \bibinfo {author} {\bibfnamefont {G.}~\bibnamefont {Garberoglio}},
  \bibinfo {author} {\bibfnamefont {A.~H.}\ \bibnamefont {Harvey}}, \bibinfo
  {author} {\bibfnamefont {M.~O.}\ \bibnamefont {McLinden}},\ and\ \bibinfo
  {author} {\bibfnamefont {K.}~\bibnamefont {Szalewicz}},\ }\bibfield  {title}
  {\bibinfo {title} {{Three-body Nonadditive Potential for Argon with Estimated
  Uncertainties and Third Virial Coefficient}},\ }\href@noop {} {\bibfield
  {journal} {\bibinfo  {journal} {J.~Phys.~Chem.~A}\ }\textbf {\bibinfo
  {volume} {117}},\ \bibinfo {pages} {7542} (\bibinfo {year}
  {2013})}\BibitemShut {NoStop}%
\bibitem [{\citenamefont {Bafile}\ \emph {et~al.}(1987)\citenamefont {Bafile},
  \citenamefont {Ulivi}, \citenamefont {Zoppi},\ and\ \citenamefont
  {Barocchi}}]{bafile1987three}%
  \BibitemOpen
  \bibfield  {author} {\bibinfo {author} {\bibfnamefont {U.}~\bibnamefont
  {Bafile}}, \bibinfo {author} {\bibfnamefont {L.}~\bibnamefont {Ulivi}},
  \bibinfo {author} {\bibfnamefont {M.}~\bibnamefont {Zoppi}},\ and\ \bibinfo
  {author} {\bibfnamefont {F.}~\bibnamefont {Barocchi}},\ }\bibfield  {title}
  {\bibinfo {title} {{The three-body correlation spectral moments in
  depolarized interaction-induced light scattering of H$_{2}$ at 297 K}},\
  }\href@noop {} {\bibfield  {journal} {\bibinfo  {journal} {Chem. Phys.
  Lett.}\ }\textbf {\bibinfo {volume} {138}},\ \bibinfo {pages} {559} (\bibinfo
  {year} {1987})}\BibitemShut {NoStop}%
\bibitem [{\citenamefont {Bafile}\ \emph {et~al.}(1991)\citenamefont {Bafile},
  \citenamefont {Ulivi}, \citenamefont {Zoppi}, \citenamefont {Moraldi},\ and\
  \citenamefont {Frommhold}}]{bafile1991third}%
  \BibitemOpen
  \bibfield  {author} {\bibinfo {author} {\bibfnamefont {U.}~\bibnamefont
  {Bafile}}, \bibinfo {author} {\bibfnamefont {L.}~\bibnamefont {Ulivi}},
  \bibinfo {author} {\bibfnamefont {M.}~\bibnamefont {Zoppi}}, \bibinfo
  {author} {\bibfnamefont {M.}~\bibnamefont {Moraldi}},\ and\ \bibinfo {author}
  {\bibfnamefont {L.}~\bibnamefont {Frommhold}},\ }\bibfield  {title} {\bibinfo
  {title} {{Third virial coefficients of collision-induced, depolarized light
  scattering of hydrogen}},\ }\href@noop {} {\bibfield  {journal} {\bibinfo
  {journal} {Phys.~Rev.~A}\ }\textbf {\bibinfo {volume} {44}},\ \bibinfo
  {pages} {4450} (\bibinfo {year} {1991})}\BibitemShut {NoStop}%
\bibitem [{\citenamefont {Pestelli}\ \emph {et~al.}(1994)\citenamefont
  {Pestelli}, \citenamefont {Bafile}, \citenamefont {Ulivi},\ and\
  \citenamefont {Zoppi}}]{pestelli1994three}%
  \BibitemOpen
  \bibfield  {author} {\bibinfo {author} {\bibfnamefont {S.}~\bibnamefont
  {Pestelli}}, \bibinfo {author} {\bibfnamefont {U.}~\bibnamefont {Bafile}},
  \bibinfo {author} {\bibfnamefont {L.}~\bibnamefont {Ulivi}},\ and\ \bibinfo
  {author} {\bibfnamefont {M.}~\bibnamefont {Zoppi}},\ }\bibfield  {title}
  {\bibinfo {title} {{Three-body depolarized interaction-induced
  light-scattering spectrum of neon}},\ }\href@noop {} {\bibfield  {journal}
  {\bibinfo  {journal} {Phys.~Rev.~A}\ }\textbf {\bibinfo {volume} {49}},\
  \bibinfo {pages} {4602} (\bibinfo {year} {1994})}\BibitemShut {NoStop}%
\bibitem [{\citenamefont {Barocchi}\ \emph {et~al.}(1988)\citenamefont
  {Barocchi}, \citenamefont {Celli},\ and\ \citenamefont
  {Zoppi}}]{barocchi1988interaction}%
  \BibitemOpen
  \bibfield  {author} {\bibinfo {author} {\bibfnamefont {F.}~\bibnamefont
  {Barocchi}}, \bibinfo {author} {\bibfnamefont {M.}~\bibnamefont {Celli}},\
  and\ \bibinfo {author} {\bibfnamefont {M.}~\bibnamefont {Zoppi}},\ }\bibfield
   {title} {\bibinfo {title} {{Interaction-induced translational Raman
  scattering in dense krypton gas: Evidence of irreducible many-body
  effects}},\ }\href@noop {} {\bibfield  {journal} {\bibinfo  {journal}
  {Phys.~Rev.~A}\ }\textbf {\bibinfo {volume} {38}},\ \bibinfo {pages} {3984}
  (\bibinfo {year} {1988})}\BibitemShut {NoStop}%
\bibitem [{\citenamefont {{Van der Elsken}}\ and\ \citenamefont
  {Huijts}(1988)}]{van1988density}%
  \BibitemOpen
  \bibfield  {author} {\bibinfo {author} {\bibfnamefont {J.}~\bibnamefont {{Van
  der Elsken}}}\ and\ \bibinfo {author} {\bibfnamefont {R.}~\bibnamefont
  {Huijts}},\ }\bibfield  {title} {\bibinfo {title} {{Density dependence of the
  depolarized light scattering spectrum of xenon}},\ }\href@noop {} {\bibfield
  {journal} {\bibinfo  {journal} {J.~Chem.~Phys.}\ }\textbf {\bibinfo {volume}
  {88}},\ \bibinfo {pages} {3007} (\bibinfo {year} {1988})}\BibitemShut
  {NoStop}%
\bibitem [{\citenamefont {{Buckingham}}\ and\ \citenamefont
  {{Hands}}(1991)}]{1991CPL...185..544B}%
  \BibitemOpen
  \bibfield  {author} {\bibinfo {author} {\bibfnamefont {A.~D.}\ \bibnamefont
  {{Buckingham}}}\ and\ \bibinfo {author} {\bibfnamefont {I.~D.}\ \bibnamefont
  {{Hands}}},\ }\bibfield  {title} {\bibinfo {title} {{The three-body
  contribution to the polarizability of a trimer of inert gas atoms using a
  dipole{\textemdash}induced-dipole model}},\ }\href
  {https://doi.org/10.1016/0009-2614(91)80257-X} {\bibfield  {journal}
  {\bibinfo  {journal} {Chem. Phys. Lett.}\ }\textbf {\bibinfo {volume}
  {185}},\ \bibinfo {pages} {544} (\bibinfo {year} {1991})}\BibitemShut
  {NoStop}%
\bibitem [{\citenamefont {P{\'e}rez}\ \emph {et~al.}(1984)\citenamefont
  {P{\'e}rez}, \citenamefont {Clarke},\ and\ \citenamefont
  {Hinchliffe}}]{perez1984three}%
  \BibitemOpen
  \bibfield  {author} {\bibinfo {author} {\bibfnamefont {J.~J.}\ \bibnamefont
  {P{\'e}rez}}, \bibinfo {author} {\bibfnamefont {J.~H.~R.}\ \bibnamefont
  {Clarke}},\ and\ \bibinfo {author} {\bibfnamefont {A.}~\bibnamefont
  {Hinchliffe}},\ }\bibfield  {title} {\bibinfo {title} {{Three-body
  contributions to the dipole polarizability of He$_{3}$ clusters}},\
  }\href@noop {} {\bibfield  {journal} {\bibinfo  {journal} {Chem. Phys.
  Lett.}\ }\textbf {\bibinfo {volume} {104}},\ \bibinfo {pages} {583} (\bibinfo
  {year} {1984})}\BibitemShut {NoStop}%
\bibitem [{\citenamefont {{Champagne}}\ \emph {et~al.}(2000)\citenamefont
  {{Champagne}}, \citenamefont {{Li}},\ and\ \citenamefont
  {{Hunt}}}]{champagne2000nonadditive}%
  \BibitemOpen
  \bibfield  {author} {\bibinfo {author} {\bibfnamefont {M.~H.}\ \bibnamefont
  {{Champagne}}}, \bibinfo {author} {\bibfnamefont {X.}~\bibnamefont {{Li}}},\
  and\ \bibinfo {author} {\bibfnamefont {K.~L.~C.}\ \bibnamefont {{Hunt}}},\
  }\bibfield  {title} {\bibinfo {title} {{Nonadditive three-body
  polarizabilities of molecules interacting at long range: Theory and numerical
  results for the inert gases, H$_{2}$, N$_{2}$, CO$_{2}$, and CH$_{4}$}},\
  }\href {https://doi.org/10.1063/1.480753} {\bibfield  {journal} {\bibinfo
  {journal} {J.~Chem.~Phys.}\ }\textbf {\bibinfo {volume} {112}},\ \bibinfo
  {pages} {1893} (\bibinfo {year} {2000})}\BibitemShut {NoStop}%
\bibitem [{\citenamefont {Li}\ and\ \citenamefont {Hunt}(1996)}]{Li96nonlocal}%
  \BibitemOpen
  \bibfield  {author} {\bibinfo {author} {\bibfnamefont {X.}~\bibnamefont
  {Li}}\ and\ \bibinfo {author} {\bibfnamefont {K.~L.~C.}\ \bibnamefont
  {Hunt}},\ }\bibfield  {title} {\bibinfo {title} {{Nonadditive, three‐body
  dipoles and forces on nuclei: New interrelations and an electrostatic
  interpretation}},\ }\href {https://doi.org/10.1063/1.472280} {\bibfield
  {journal} {\bibinfo  {journal} {J. Chem. Phys.}\ }\textbf {\bibinfo {volume}
  {105}},\ \bibinfo {pages} {4076} (\bibinfo {year} {1996})}\BibitemShut
  {NoStop}%
\bibitem [{\citenamefont {Li}\ and\ \citenamefont {Hunt}(1997)}]{Li97nonlocal}%
  \BibitemOpen
  \bibfield  {author} {\bibinfo {author} {\bibfnamefont {X.}~\bibnamefont
  {Li}}\ and\ \bibinfo {author} {\bibfnamefont {K.~L.~C.}\ \bibnamefont
  {Hunt}},\ }\bibfield  {title} {\bibinfo {title} {{Nonadditive three-body
  dipoles of inert gas trimers and $\text{H}_2\cdots \text{H}_2\cdots
  \text{H}_2$: Long-range effects in far infrared absorption and triple
  vibrational transitions}},\ }\href {https://doi.org/10.1063/1.474790}
  {\bibfield  {journal} {\bibinfo  {journal} {J. Chem. Phys.}\ }\textbf
  {\bibinfo {volume} {107}},\ \bibinfo {pages} {4133} (\bibinfo {year}
  {1997})}\BibitemShut {NoStop}%
\bibitem [{\citenamefont {Aleman}\ \emph {et~al.}(1992)\citenamefont {Aleman},
  \citenamefont {Perez},\ and\ \citenamefont {Hinchliffe}}]{aleman1992ab}%
  \BibitemOpen
  \bibfield  {author} {\bibinfo {author} {\bibfnamefont {C.}~\bibnamefont
  {Aleman}}, \bibinfo {author} {\bibfnamefont {J.~J.}\ \bibnamefont {Perez}},\
  and\ \bibinfo {author} {\bibfnamefont {A.}~\bibnamefont {Hinchliffe}},\
  }\bibfield  {title} {\bibinfo {title} {{Ab initio SCF-MO incremental triplet
  polarizabilities of neon clusters}},\ }\href@noop {} {\bibfield  {journal}
  {\bibinfo  {journal} {Int.~J.~Mass Spectrom.~Ion Processes}\ }\textbf
  {\bibinfo {volume} {122}},\ \bibinfo {pages} {331} (\bibinfo {year}
  {1992})}\BibitemShut {NoStop}%
\bibitem [{\citenamefont {{Heller}}\ and\ \citenamefont
  {{Gelbart}}(1974)}]{heller1974superpos}%
  \BibitemOpen
  \bibfield  {author} {\bibinfo {author} {\bibfnamefont {D.~F.}\ \bibnamefont
  {{Heller}}}\ and\ \bibinfo {author} {\bibfnamefont {W.~M.}\ \bibnamefont
  {{Gelbart}}},\ }\bibfield  {title} {\bibinfo {title} {{Short range electronic
  distortion and the density dependent dielectric function of simple gases}},\
  }\href {https://doi.org/10.1016/0009-2614(74)90241-3} {\bibfield  {journal}
  {\bibinfo  {journal} {Chem. Phys. Lett.}\ }\textbf {\bibinfo {volume} {27}},\
  \bibinfo {pages} {359} (\bibinfo {year} {1974})}\BibitemShut {NoStop}%
\bibitem [{\citenamefont {Garberoglio}\ \emph {et~al.}(2021)\citenamefont
  {Garberoglio}, \citenamefont {Harvey},\ and\ \citenamefont
  {Jeziorski}}]{garberoglio20213dielec}%
  \BibitemOpen
  \bibfield  {author} {\bibinfo {author} {\bibfnamefont {G.}~\bibnamefont
  {Garberoglio}}, \bibinfo {author} {\bibfnamefont {A.~H.}\ \bibnamefont
  {Harvey}},\ and\ \bibinfo {author} {\bibfnamefont {B.}~\bibnamefont
  {Jeziorski}},\ }\bibfield  {title} {\bibinfo {title} {{Path-integral
  calculation of the third dielectric virial coefficient of noble gases}},\
  }\href {https://doi.org/10.1063/5.0077684} {\bibfield  {journal} {\bibinfo
  {journal} {J.~Chem.~Phys.}\ }\textbf {\bibinfo {volume} {155}},\ \bibinfo
  {pages} {234103} (\bibinfo {year} {2021})}\BibitemShut {NoStop}%
\bibitem [{\citenamefont {Feynman}\ and\ \citenamefont
  {Hibbs}(1965)}]{feynman1965pathintegrals}%
  \BibitemOpen
  \bibfield  {author} {\bibinfo {author} {\bibfnamefont {R.~P.}\ \bibnamefont
  {Feynman}}\ and\ \bibinfo {author} {\bibfnamefont {A.~R.}\ \bibnamefont
  {Hibbs}},\ }\href@noop {} {\emph {\bibinfo {title} {{Quantum Mechanics and
  Path Integrals}}}}\ (\bibinfo  {publisher} {McGraw-Hill, New York},\ \bibinfo
  {year} {1965})\BibitemShut {NoStop}%
\bibitem [{\citenamefont {Garberoglio}\ and\ \citenamefont
  {Harvey}(2009)}]{garberoglio2009firstprincip}%
  \BibitemOpen
  \bibfield  {author} {\bibinfo {author} {\bibfnamefont {G.}~\bibnamefont
  {Garberoglio}}\ and\ \bibinfo {author} {\bibfnamefont {A.~H.}\ \bibnamefont
  {Harvey}},\ }\bibfield  {title} {\bibinfo {title} {{First-principles
  calculation of the third virial coefficient of helium}},\ }\href@noop {}
  {\bibfield  {journal} {\bibinfo  {journal} {J. Res. Natl. Inst. Stand.
  Technol.}\ }\textbf {\bibinfo {volume} {114}},\ \bibinfo {pages} {249}
  (\bibinfo {year} {2009})}\BibitemShut {NoStop}%
\bibitem [{\citenamefont {{Garberoglio}}\ and\ \citenamefont
  {{Harvey}}(2011)}]{2011JChPh.134m4106G}%
  \BibitemOpen
  \bibfield  {author} {\bibinfo {author} {\bibfnamefont {G.}~\bibnamefont
  {{Garberoglio}}}\ and\ \bibinfo {author} {\bibfnamefont {A.~H.}\ \bibnamefont
  {{Harvey}}},\ }\bibfield  {title} {\bibinfo {title} {{Path-integral
  calculation of the third virial coefficient of quantum gases at low
  temperatures}},\ }\href {https://doi.org/10.1063/1.3573564} {\bibfield
  {journal} {\bibinfo  {journal} {J.~Chem.~Phys.}\ }\textbf {\bibinfo {volume}
  {134}},\ \bibinfo {pages} {134106} (\bibinfo {year} {2011})}\BibitemShut
  {NoStop}%
\bibitem [{\citenamefont {Shaul}\ \emph
  {et~al.}(2012{\natexlab{a}})\citenamefont {Shaul}, \citenamefont {Schultz},\
  and\ \citenamefont {Kofke}}]{shaul2012path}%
  \BibitemOpen
  \bibfield  {author} {\bibinfo {author} {\bibfnamefont {K.~R.~S.}\
  \bibnamefont {Shaul}}, \bibinfo {author} {\bibfnamefont {A.~J.}\ \bibnamefont
  {Schultz}},\ and\ \bibinfo {author} {\bibfnamefont {D.~A.}\ \bibnamefont
  {Kofke}},\ }\bibfield  {title} {\bibinfo {title} {{Path-integral
  Mayer-sampling calculations of the quantum Boltzmann contribution to virial
  coefficients of helium-4}},\ }\href@noop {} {\bibfield  {journal} {\bibinfo
  {journal} {J.~Chem.~Phys.}\ }\textbf {\bibinfo {volume} {137}},\ \bibinfo
  {pages} {184101} (\bibinfo {year} {2012}{\natexlab{a}})}\BibitemShut
  {NoStop}%
\bibitem [{\citenamefont {{Alder}}\ \emph {et~al.}(1980)\citenamefont
  {{Alder}}, \citenamefont {{Beers}}, \citenamefont {{Strauss}},\ and\
  \citenamefont {{Weis}}}]{1980PNAS...77.3098A}%
  \BibitemOpen
  \bibfield  {author} {\bibinfo {author} {\bibfnamefont {B.~J.}\ \bibnamefont
  {{Alder}}}, \bibinfo {author} {\bibfnamefont {J.~C.}\ \bibnamefont {{Beers}},
  \bibfnamefont {II}}, \bibinfo {author} {\bibfnamefont {H.~L.}\ \bibnamefont
  {{Strauss}}},\ and\ \bibinfo {author} {\bibfnamefont {J.~J.}\ \bibnamefont
  {{Weis}}},\ }\bibfield  {title} {\bibinfo {title} {{Dielectric Constant of
  Atomic Fluids with Variable Polarizability}},\ }\href
  {https://doi.org/10.1073/pnas.77.6.3098} {\bibfield  {journal} {\bibinfo
  {journal} {Proc. Natl. Acad. Sci. U.S.A.}\ }\textbf {\bibinfo {volume}
  {77}},\ \bibinfo {pages} {3098} (\bibinfo {year} {1980})}\BibitemShut
  {NoStop}%
\bibitem [{\citenamefont {Hald}\ \emph {et~al.}(2003)\citenamefont {Hald},
  \citenamefont {Paw{\l}owski}, \citenamefont {J{\o}rgensen},\ and\
  \citenamefont {H{\"a}ttig}}]{hald2003calculation}%
  \BibitemOpen
  \bibfield  {author} {\bibinfo {author} {\bibfnamefont {K.}~\bibnamefont
  {Hald}}, \bibinfo {author} {\bibfnamefont {F.}~\bibnamefont {Paw{\l}owski}},
  \bibinfo {author} {\bibfnamefont {P.}~\bibnamefont {J{\o}rgensen}},\ and\
  \bibinfo {author} {\bibfnamefont {C.}~\bibnamefont {H{\"a}ttig}},\ }\bibfield
   {title} {\bibinfo {title} {{Calculation of frequency-dependent
  polarizabilities using the approximate coupled-cluster triples model CC3}},\
  }\href@noop {} {\bibfield  {journal} {\bibinfo  {journal} {J. Chem. Phys.}\
  }\textbf {\bibinfo {volume} {118}},\ \bibinfo {pages} {1292} (\bibinfo {year}
  {2003})}\BibitemShut {NoStop}%
\bibitem [{\citenamefont {Garberoglio}\ \emph {et~al.}(2022)\citenamefont
  {Garberoglio}, \citenamefont {Harvey}, \citenamefont {Lang},\ and\
  \citenamefont {Jeziorski}}]{Garberoglio_at_al_2022}%
  \BibitemOpen
  \bibfield  {author} {\bibinfo {author} {\bibfnamefont {G.}~\bibnamefont
  {Garberoglio}}, \bibinfo {author} {\bibfnamefont {A.~H.}\ \bibnamefont
  {Harvey}}, \bibinfo {author} {\bibfnamefont {J.}~\bibnamefont {Lang}},\ and\
  \bibinfo {author} {\bibfnamefont {B.}~\bibnamefont {Jeziorski}},\ }\href@noop
  {} {\bibinfo {title} {{Path-integral calculation of the third dielectric
  virial coefficient of helium employing ab initio three-body dipole and
  plarizability surfaces}}} (\bibinfo {year} {2022}),\ \bibinfo {note} {to be
  published}\BibitemShut {NoStop}%
\bibitem [{\citenamefont {{Gaiser}}\ \emph {et~al.}(2022)\citenamefont
  {{Gaiser}}, \citenamefont {{Fellmuth}},\ and\ \citenamefont
  {{Sabuga}}}]{Geiser_Annalen_2022}%
  \BibitemOpen
  \bibfield  {author} {\bibinfo {author} {\bibfnamefont {C.}~\bibnamefont
  {{Gaiser}}}, \bibinfo {author} {\bibfnamefont {B.}~\bibnamefont
  {{Fellmuth}}},\ and\ \bibinfo {author} {\bibfnamefont {W.}~\bibnamefont
  {{Sabuga}}},\ }\bibfield  {title} {\bibinfo {title} {{Primary pressure
  standard passed next stress test}},\ }\href
  {https://doi.org/10:1002/andp.202200336} {\bibfield  {journal} {\bibinfo
  {journal} {Ann. Phys. (Berlin)}\ }\textbf {\bibinfo {volume} {534}},\
  \bibinfo {pages} {2200336} (\bibinfo {year} {2022})}\BibitemShut {NoStop}%
\bibitem [{\citenamefont {Gray}\ and\ \citenamefont
  {Gubbins}(1984)}]{gray1984fluids1}%
  \BibitemOpen
  \bibfield  {author} {\bibinfo {author} {\bibfnamefont {C.~G.}\ \bibnamefont
  {Gray}}\ and\ \bibinfo {author} {\bibfnamefont {K.~E.}\ \bibnamefont
  {Gubbins}},\ }\href@noop {} {\emph {\bibinfo {title} {{Theory of Molecular
  Fluids: Volume 1: Fundamentals}}}},\ Vol.~\bibinfo {volume} {1}\ (\bibinfo
  {publisher} {Oxford University Press},\ \bibinfo {year} {1984})\BibitemShut
  {NoStop}%
\bibitem [{\citenamefont {Hunt}\ \emph {et~al.}(1988)\citenamefont {Hunt},
  \citenamefont {Liang},\ and\ \citenamefont {Sethuraman}}]{hunt1988transient}%
  \BibitemOpen
  \bibfield  {author} {\bibinfo {author} {\bibfnamefont {K.~L.~C.}\
  \bibnamefont {Hunt}}, \bibinfo {author} {\bibfnamefont {Y.~Q.}\ \bibnamefont
  {Liang}},\ and\ \bibinfo {author} {\bibfnamefont {S.}~\bibnamefont
  {Sethuraman}},\ }\bibfield  {title} {\bibinfo {title} {{Transient,
  collision-induced changes in polarizability for atoms interacting with
  linear, centrosymmetric molecules at long range}},\ }\href@noop {} {\bibfield
   {journal} {\bibinfo  {journal} {J.~Chem.~Phys.}\ }\textbf {\bibinfo {volume}
  {89}},\ \bibinfo {pages} {7126} (\bibinfo {year} {1988})}\BibitemShut
  {NoStop}%
\bibitem [{\citenamefont {Certain}\ and\ \citenamefont
  {Fortune}(1971)}]{Certain1971}%
  \BibitemOpen
  \bibfield  {author} {\bibinfo {author} {\bibfnamefont {P.~R.}\ \bibnamefont
  {Certain}}\ and\ \bibinfo {author} {\bibfnamefont {P.~J.}\ \bibnamefont
  {Fortune}},\ }\bibfield  {title} {\bibinfo {title} {{Long‐Range
  Polarizability of the Helium Diatom}},\ }\href
  {https://doi.org/10.1063/1.1675752} {\bibfield  {journal} {\bibinfo
  {journal} {J. Chem. Phys.}\ }\textbf {\bibinfo {volume} {55}},\ \bibinfo
  {pages} {5818} (\bibinfo {year} {1971})}\BibitemShut {NoStop}%
\bibitem [{\citenamefont {Aidas}\ \emph {et~al.}(2014)\citenamefont {Aidas},
  \citenamefont {Angeli}, \citenamefont {Bak}, \citenamefont {Bakken},
  \citenamefont {Bast}, \citenamefont {Boman}, \citenamefont {Christiansen},
  \citenamefont {Cimiraglia}, \citenamefont {Coriani}, \citenamefont {Dahle}
  \emph {et~al.}}]{dalton}%
  \BibitemOpen
  \bibfield  {author} {\bibinfo {author} {\bibfnamefont {K.}~\bibnamefont
  {Aidas}}, \bibinfo {author} {\bibfnamefont {C.}~\bibnamefont {Angeli}},
  \bibinfo {author} {\bibfnamefont {K.~L.}\ \bibnamefont {Bak}}, \bibinfo
  {author} {\bibfnamefont {V.}~\bibnamefont {Bakken}}, \bibinfo {author}
  {\bibfnamefont {R.}~\bibnamefont {Bast}}, \bibinfo {author} {\bibfnamefont
  {L.}~\bibnamefont {Boman}}, \bibinfo {author} {\bibfnamefont
  {O.}~\bibnamefont {Christiansen}}, \bibinfo {author} {\bibfnamefont
  {R.}~\bibnamefont {Cimiraglia}}, \bibinfo {author} {\bibfnamefont
  {S.}~\bibnamefont {Coriani}}, \bibinfo {author} {\bibfnamefont
  {P.}~\bibnamefont {Dahle}}, \emph {et~al.},\ }\bibfield  {title} {\bibinfo
  {title} {{The Dalton quantum chemistry program system}},\ }\href@noop {}
  {\bibfield  {journal} {\bibinfo  {journal} {Wiley Interdiscip. Rev.: Comput.
  Mol. Sci.}\ }\textbf {\bibinfo {volume} {4}},\ \bibinfo {pages} {269}
  (\bibinfo {year} {2014})}\BibitemShut {NoStop}%
\bibitem [{dal(2018)}]{dalton2018}%
  \BibitemOpen
  \href@noop {} {\bibinfo {title} {{Dalton, a molecular electronic structure
  program, release 2018}}} (\bibinfo {year} {2018}),\ \bibinfo {note} {see
  http://daltonprogram.org}\BibitemShut {NoStop}%
\bibitem [{\citenamefont {Boys}\ and\ \citenamefont
  {Bernardi}(1970)}]{Boys1970}%
  \BibitemOpen
  \bibfield  {author} {\bibinfo {author} {\bibfnamefont {S.~F.}\ \bibnamefont
  {Boys}}\ and\ \bibinfo {author} {\bibfnamefont {F.}~\bibnamefont
  {Bernardi}},\ }\bibfield  {title} {\bibinfo {title} {{The calculation of
  small molecular interactions by the differences of separate total energies.
  Some procedures with reduced errors}},\ }\href
  {https://doi.org/10.1080/00268977000101561} {\bibfield  {journal} {\bibinfo
  {journal} {Mol. Phys.}\ }\textbf {\bibinfo {volume} {19}},\ \bibinfo {pages}
  {553} (\bibinfo {year} {1970})}\BibitemShut {NoStop}%
\bibitem [{\citenamefont {{Skwara}}\ \emph {et~al.}(2009)\citenamefont
  {{Skwara}}, \citenamefont {{Bartkowiak}},\ and\ \citenamefont
  {{Silva}}}]{2009TChA..122...127H}%
  \BibitemOpen
  \bibfield  {author} {\bibinfo {author} {\bibfnamefont {B.}~\bibnamefont
  {{Skwara}}}, \bibinfo {author} {\bibfnamefont {W.}~\bibnamefont
  {{Bartkowiak}}},\ and\ \bibinfo {author} {\bibfnamefont {D.~L.}\ \bibnamefont
  {{Silva}}},\ }\bibfield  {title} {\bibinfo {title} {{On the Basis Set
  Superposition Error in Supermolecular Calculations of Interaction-Induced
  Electric Properties: Many-Body Components}},\ }\href
  {https://doi.org/10.1063/1.1744081} {\bibfield  {journal} {\bibinfo
  {journal} {Theor. Chim. Acta}\ }\textbf {\bibinfo {volume} {122}},\ \bibinfo
  {pages} {127} (\bibinfo {year} {2009})}\BibitemShut {NoStop}%
\bibitem [{\citenamefont {Cencek}\ \emph {et~al.}(2012)\citenamefont {Cencek},
  \citenamefont {Przybytek}, \citenamefont {Komasa}, \citenamefont {Mehl},
  \citenamefont {Jeziorski},\ and\ \citenamefont
  {Szalewicz}}]{cencek2012effects}%
  \BibitemOpen
  \bibfield  {author} {\bibinfo {author} {\bibfnamefont {W.}~\bibnamefont
  {Cencek}}, \bibinfo {author} {\bibfnamefont {M.}~\bibnamefont {Przybytek}},
  \bibinfo {author} {\bibfnamefont {J.}~\bibnamefont {Komasa}}, \bibinfo
  {author} {\bibfnamefont {J.~B.}\ \bibnamefont {Mehl}}, \bibinfo {author}
  {\bibfnamefont {B.}~\bibnamefont {Jeziorski}},\ and\ \bibinfo {author}
  {\bibfnamefont {K.}~\bibnamefont {Szalewicz}},\ }\bibfield  {title} {\bibinfo
  {title} {{Effects of adiabatic, relativistic, and quantum electrodynamics
  interactions on the pair potential and thermophysical properties of
  helium}},\ }\href@noop {} {\bibfield  {journal} {\bibinfo  {journal}
  {J.~Chem.~Phys.}\ }\textbf {\bibinfo {volume} {136}},\ \bibinfo {pages}
  {224303} (\bibinfo {year} {2012})}\BibitemShut {NoStop}%
\bibitem [{\citenamefont {Dunning}(1989)}]{Dunning1989cc}%
  \BibitemOpen
  \bibfield  {author} {\bibinfo {author} {\bibfnamefont {T.~H.}\ \bibnamefont
  {Dunning}},\ }\bibfield  {title} {\bibinfo {title} {{Gaussian basis sets for
  use in correlated molecular calculations. I. The atoms boron through neon and
  hydrogen}},\ }\href {https://doi.org/10.1063/1.456153} {\bibfield  {journal}
  {\bibinfo  {journal} {J. Chem. Phys.}\ }\textbf {\bibinfo {volume} {90}},\
  \bibinfo {pages} {1007} (\bibinfo {year} {1989})}\BibitemShut {NoStop}%
\bibitem [{\citenamefont {Woon}\ and\ \citenamefont
  {Dunning}(1994)}]{Dunning1994he}%
  \BibitemOpen
  \bibfield  {author} {\bibinfo {author} {\bibfnamefont {D.~E.}\ \bibnamefont
  {Woon}}\ and\ \bibinfo {author} {\bibfnamefont {T.~H.}\ \bibnamefont
  {Dunning}},\ }\bibfield  {title} {\bibinfo {title} {{Gaussian basis sets for
  use in correlated molecular calculations. IV. Calculation of static
  electrical response properties}},\ }\href {https://doi.org/10.1063/1.466439}
  {\bibfield  {journal} {\bibinfo  {journal} {J. Chem. Phys.}\ }\textbf
  {\bibinfo {volume} {100}},\ \bibinfo {pages} {2975} (\bibinfo {year}
  {1994})}\BibitemShut {NoStop}%
\bibitem [{\citenamefont {Halkier}\ \emph {et~al.}(1998)\citenamefont
  {Halkier}, \citenamefont {Helgaker}, \citenamefont {J{\o}rgensen},
  \citenamefont {Klopper}, \citenamefont {Koch}, \citenamefont {Olsen},\ and\
  \citenamefont {Wilson}}]{Halkier1998}%
  \BibitemOpen
  \bibfield  {author} {\bibinfo {author} {\bibfnamefont {A.}~\bibnamefont
  {Halkier}}, \bibinfo {author} {\bibfnamefont {T.}~\bibnamefont {Helgaker}},
  \bibinfo {author} {\bibfnamefont {P.}~\bibnamefont {J{\o}rgensen}}, \bibinfo
  {author} {\bibfnamefont {W.}~\bibnamefont {Klopper}}, \bibinfo {author}
  {\bibfnamefont {H.}~\bibnamefont {Koch}}, \bibinfo {author} {\bibfnamefont
  {J.}~\bibnamefont {Olsen}},\ and\ \bibinfo {author} {\bibfnamefont {A.~K.}\
  \bibnamefont {Wilson}},\ }\bibfield  {title} {\bibinfo {title} {{Basis-set
  convergence in correlated calculations on Ne, N$_2$, and H$_2$O}},\ }\href
  {https://doi.org/https://doi.org/10.1016/S0009-2614(98)00111-0} {\bibfield
  {journal} {\bibinfo  {journal} {Chem. Phys. Lett.}\ }\textbf {\bibinfo
  {volume} {286}},\ \bibinfo {pages} {243} (\bibinfo {year}
  {1998})}\BibitemShut {NoStop}%
\bibitem [{\citenamefont {Helgaker}\ \emph {et~al.}(2008)\citenamefont
  {Helgaker}, \citenamefont {Klopper},\ and\ \citenamefont
  {Tew}}]{Helgaker2008}%
  \BibitemOpen
  \bibfield  {author} {\bibinfo {author} {\bibfnamefont {T.}~\bibnamefont
  {Helgaker}}, \bibinfo {author} {\bibfnamefont {W.}~\bibnamefont {Klopper}},\
  and\ \bibinfo {author} {\bibfnamefont {D.~P.}\ \bibnamefont {Tew}},\
  }\bibfield  {title} {\bibinfo {title} {{Quantitative quantum chemistry}},\
  }\href {https://doi.org/10.1080/00268970802258591} {\bibfield  {journal}
  {\bibinfo  {journal} {Mol. Phys.}\ }\textbf {\bibinfo {volume} {106}},\
  \bibinfo {pages} {2107} (\bibinfo {year} {2008})}\BibitemShut {NoStop}%
\bibitem [{\citenamefont {Halkier}\ \emph {et~al.}(2000)\citenamefont
  {Halkier}, \citenamefont {Helgaker}, \citenamefont {Klopper},\ and\
  \citenamefont {Olsen}}]{Halkier2000}%
  \BibitemOpen
  \bibfield  {author} {\bibinfo {author} {\bibfnamefont {A.}~\bibnamefont
  {Halkier}}, \bibinfo {author} {\bibfnamefont {T.}~\bibnamefont {Helgaker}},
  \bibinfo {author} {\bibfnamefont {W.}~\bibnamefont {Klopper}},\ and\ \bibinfo
  {author} {\bibfnamefont {J.}~\bibnamefont {Olsen}},\ }\bibfield  {title}
  {\bibinfo {title} {{Basis-set convergence of the two-electron Darwin term}},\
  }\href {https://doi.org/https://doi.org/10.1016/S0009-2614(00)00161-5}
  {\bibfield  {journal} {\bibinfo  {journal} {Chem. Phys. Lett.}\ }\textbf
  {\bibinfo {volume} {319}},\ \bibinfo {pages} {287} (\bibinfo {year}
  {2000})}\BibitemShut {NoStop}%
\bibitem [{\citenamefont {Przybytek}\ \emph {et~al.}(2017)\citenamefont
  {Przybytek}, \citenamefont {Cencek}, \citenamefont {Jeziorski},\ and\
  \citenamefont {Szalewicz}}]{Przybytek:17}%
  \BibitemOpen
  \bibfield  {author} {\bibinfo {author} {\bibfnamefont {M.}~\bibnamefont
  {Przybytek}}, \bibinfo {author} {\bibfnamefont {W.}~\bibnamefont {Cencek}},
  \bibinfo {author} {\bibfnamefont {B.}~\bibnamefont {Jeziorski}},\ and\
  \bibinfo {author} {\bibfnamefont {K.}~\bibnamefont {Szalewicz}},\ }\bibfield
  {title} {\bibinfo {title} {Pair potential with submillikelvin uncertainties
  and nonadiabatic treatment of the halo state of the helium dimer},\ }\href
  {https://doi.org/10.1103/PhysRevLett.119.123401} {\bibfield  {journal}
  {\bibinfo  {journal} {Phys. Rev. Lett.}\ }\textbf {\bibinfo {volume} {119}},\
  \bibinfo {pages} {123401} (\bibinfo {year} {2017})}\BibitemShut {NoStop}%
\bibitem [{\citenamefont {Przybytek}(2014)}]{przybytekFCI}%
  \BibitemOpen
  \bibfield  {author} {\bibinfo {author} {\bibfnamefont {M.}~\bibnamefont
  {Przybytek}},\ }\href@noop {} {\bibinfo {title} {{General {FCI} program
  {Hector}.}}} (\bibinfo {year} {2014})\BibitemShut {NoStop}%
\bibitem [{\citenamefont {Tang}\ and\ \citenamefont
  {Toennies}(1984)}]{tang1984improved}%
  \BibitemOpen
  \bibfield  {author} {\bibinfo {author} {\bibfnamefont {K.~T.}\ \bibnamefont
  {Tang}}\ and\ \bibinfo {author} {\bibfnamefont {J.~P.}\ \bibnamefont
  {Toennies}},\ }\bibfield  {title} {\bibinfo {title} {{An improved simple
  model for the van der Waals potential based on universal damping functions
  for the dispersion coefficients}},\ }\href@noop {} {\bibfield  {journal}
  {\bibinfo  {journal} {J.~Chem.~Phys.}\ }\textbf {\bibinfo {volume} {80}},\
  \bibinfo {pages} {3726} (\bibinfo {year} {1984})}\BibitemShut {NoStop}%
\bibitem [{\citenamefont {Bishop}\ and\ \citenamefont
  {Pipin}(1995)}]{bishop1995static}%
  \BibitemOpen
  \bibfield  {author} {\bibinfo {author} {\bibfnamefont {D.~M.}\ \bibnamefont
  {Bishop}}\ and\ \bibinfo {author} {\bibfnamefont {J.}~\bibnamefont {Pipin}},\
  }\bibfield  {title} {\bibinfo {title} {{Static electric properties of H and
  He}},\ }\href@noop {} {\bibfield  {journal} {\bibinfo  {journal} {Chem. Phys.
  Lett.}\ }\textbf {\bibinfo {volume} {236}},\ \bibinfo {pages} {15} (\bibinfo
  {year} {1995})}\BibitemShut {NoStop}%
\bibitem [{\citenamefont {Metz}\ and\ \citenamefont
  {Szalewicz}(2020)}]{Metz:2020}%
  \BibitemOpen
  \bibfield  {author} {\bibinfo {author} {\bibfnamefont {M.~P.}\ \bibnamefont
  {Metz}}\ and\ \bibinfo {author} {\bibfnamefont {K.}~\bibnamefont
  {Szalewicz}},\ }\bibfield  {title} {\bibinfo {title} {A statistically guided
  grid generation method and its application to intermolecular potential energy
  surfaces},\ }\href@noop {} {\bibfield  {journal} {\bibinfo  {journal} {J.
  Chem. Phys.}\ }\textbf {\bibinfo {volume} {152}},\ \bibinfo {pages} {134111}
  (\bibinfo {year} {2020})}\BibitemShut {NoStop}%
\bibitem [{\citenamefont {Powell}(1970)}]{Powell1970}%
  \BibitemOpen
  \bibfield  {author} {\bibinfo {author} {\bibfnamefont {M.~J.~D.}\
  \bibnamefont {Powell}},\ }\href@noop {} {\emph {\bibinfo {title} {{Numerical
  Methods for Nonlinear Algebraic Equations}}}},\ edited by\ \bibinfo {editor}
  {\bibfnamefont {P.}~\bibnamefont {Robinowitz}}\ (\bibinfo  {publisher}
  {Gordon and Breach Science},\ \bibinfo {address} {London},\ \bibinfo {year}
  {1970})\ p.\ \bibinfo {pages} {87–144}\BibitemShut {NoStop}%
\bibitem [{\citenamefont {Galassi}\ \emph {et~al.}(2002)\citenamefont
  {Galassi}, \citenamefont {Davies}, \citenamefont {Theiler}, \citenamefont
  {Gough}, \citenamefont {Jungman}, \citenamefont {Alken}, \citenamefont
  {Booth}, \citenamefont {Rossi},\ and\ \citenamefont {Ulerich}}]{gsl_lib}%
  \BibitemOpen
  \bibfield  {author} {\bibinfo {author} {\bibfnamefont {M.}~\bibnamefont
  {Galassi}}, \bibinfo {author} {\bibfnamefont {J.}~\bibnamefont {Davies}},
  \bibinfo {author} {\bibfnamefont {J.}~\bibnamefont {Theiler}}, \bibinfo
  {author} {\bibfnamefont {B.}~\bibnamefont {Gough}}, \bibinfo {author}
  {\bibfnamefont {G.}~\bibnamefont {Jungman}}, \bibinfo {author} {\bibfnamefont
  {P.}~\bibnamefont {Alken}}, \bibinfo {author} {\bibfnamefont
  {M.}~\bibnamefont {Booth}}, \bibinfo {author} {\bibfnamefont
  {F.}~\bibnamefont {Rossi}},\ and\ \bibinfo {author} {\bibfnamefont
  {R.}~\bibnamefont {Ulerich}},\ }\href@noop {} {\emph {\bibinfo {title} {{GNU
  scientific library}}}}\ (\bibinfo  {publisher} {Network Theory Limited},\
  \bibinfo {year} {2002})\BibitemShut {NoStop}%
\bibitem [{\citenamefont {{Hill}}(1958)}]{1958JChPh..28...61H}%
  \BibitemOpen
  \bibfield  {author} {\bibinfo {author} {\bibfnamefont {T.~L.}\ \bibnamefont
  {{Hill}}},\ }\bibfield  {title} {\bibinfo {title} {{Theory of the Dielectric
  Constant of Imperfect Gases and Dilute Solutions}},\ }\href
  {https://doi.org/10.1063/1.1744081} {\bibfield  {journal} {\bibinfo
  {journal} {J.~Chem.~Phys.}\ }\textbf {\bibinfo {volume} {28}},\ \bibinfo
  {pages} {61} (\bibinfo {year} {1958})}\BibitemShut {NoStop}%
\bibitem [{\citenamefont {Gray}\ \emph {et~al.}(2011)\citenamefont {Gray},
  \citenamefont {Gubbins},\ and\ \citenamefont {Joslin}}]{gray2011fluids2}%
  \BibitemOpen
  \bibfield  {author} {\bibinfo {author} {\bibfnamefont {C.~G.}\ \bibnamefont
  {Gray}}, \bibinfo {author} {\bibfnamefont {K.~E.}\ \bibnamefont {Gubbins}},\
  and\ \bibinfo {author} {\bibfnamefont {C.~G.}\ \bibnamefont {Joslin}},\
  }\href@noop {} {\emph {\bibinfo {title} {{Theory of Molecular Fluids: Volume
  2: Applications}}}},\ Vol.~\bibinfo {volume} {2}\ (\bibinfo  {publisher}
  {Oxford University Press},\ \bibinfo {year} {2011})\BibitemShut {NoStop}%
\bibitem [{\citenamefont {Kusalik}\ \emph {et~al.}(1995)\citenamefont
  {Kusalik}, \citenamefont {Liden},\ and\ \citenamefont
  {Svishchev}}]{kusalik1995calculation}%
  \BibitemOpen
  \bibfield  {author} {\bibinfo {author} {\bibfnamefont {P.~G.}\ \bibnamefont
  {Kusalik}}, \bibinfo {author} {\bibfnamefont {F.}~\bibnamefont {Liden}},\
  and\ \bibinfo {author} {\bibfnamefont {I.~M.}\ \bibnamefont {Svishchev}},\
  }\bibfield  {title} {\bibinfo {title} {{Calculation of the third virial
  coefficient for water}},\ }\href@noop {} {\bibfield  {journal} {\bibinfo
  {journal} {J.~Chem.~Phys.}\ }\textbf {\bibinfo {volume} {103}},\ \bibinfo
  {pages} {10169} (\bibinfo {year} {1995})}\BibitemShut {NoStop}%
\bibitem [{\citenamefont {Mas}\ \emph {et~al.}(1999)\citenamefont {Mas},
  \citenamefont {Lotrich},\ and\ \citenamefont {Szalewicz}}]{mas1999third}%
  \BibitemOpen
  \bibfield  {author} {\bibinfo {author} {\bibfnamefont {E.~M.}\ \bibnamefont
  {Mas}}, \bibinfo {author} {\bibfnamefont {V.~F.}\ \bibnamefont {Lotrich}},\
  and\ \bibinfo {author} {\bibfnamefont {K.}~\bibnamefont {Szalewicz}},\
  }\bibfield  {title} {\bibinfo {title} {{Third virial coefficient of argon}},\
  }\href@noop {} {\bibfield  {journal} {\bibinfo  {journal} {J.~Chem.~Phys.}\
  }\textbf {\bibinfo {volume} {110}},\ \bibinfo {pages} {6694} (\bibinfo {year}
  {1999})}\BibitemShut {NoStop}%
\bibitem [{\citenamefont {Garberoglio}\ and\ \citenamefont
  {Harvey}(2020)}]{garberoglio2020path}%
  \BibitemOpen
  \bibfield  {author} {\bibinfo {author} {\bibfnamefont {G.}~\bibnamefont
  {Garberoglio}}\ and\ \bibinfo {author} {\bibfnamefont {A.~H.}\ \bibnamefont
  {Harvey}},\ }\bibfield  {title} {\bibinfo {title} {{Path-integral calculation
  of the second dielectric and refractivity virial coefficients of helium,
  neon, and argon}},\ }\href@noop {} {\bibfield  {journal} {\bibinfo  {journal}
  {J. Res. Natl. Inst. Stand. Technol.}\ }\textbf {\bibinfo {volume} {125}},\
  \bibinfo {pages} {022} (\bibinfo {year} {2020})}\BibitemShut {NoStop}%
\bibitem [{\citenamefont {Kronrod}(1965)}]{kronrod1965nodes}%
  \BibitemOpen
  \bibfield  {author} {\bibinfo {author} {\bibfnamefont {A.}~\bibnamefont
  {Kronrod}},\ }\href@noop {} {\bibinfo {title} {{Nodes and Weights of
  Quadrature Formulas: Sixteen-place Tables. Consultants Bureau, New York}}}
  (\bibinfo {year} {1965})\BibitemShut {NoStop}%
\bibitem [{\citenamefont {Lang}\ \emph {et~al.}(2022)\citenamefont {Lang},
  \citenamefont {Garberoglio}, \citenamefont {Przybytek}, \citenamefont
  {Jeziorska},\ and\ \citenamefont {Jeziorski}}]{lang2022density}%
  \BibitemOpen
  \bibfield  {author} {\bibinfo {author} {\bibfnamefont {J.}~\bibnamefont
  {Lang}}, \bibinfo {author} {\bibfnamefont {G.}~\bibnamefont {Garberoglio}},
  \bibinfo {author} {\bibfnamefont {M.}~\bibnamefont {Przybytek}}, \bibinfo
  {author} {\bibfnamefont {M.}~\bibnamefont {Jeziorska}},\ and\ \bibinfo
  {author} {\bibfnamefont {B.}~\bibnamefont {Jeziorski}},\ }\href@noop {}
  {\bibinfo {title} {{Three-body potential and third virial coefficients for
  helium including relativistic and the nuclear motion effects}}} (\bibinfo
  {year} {2022}),\ \bibinfo {note} {to be published}\BibitemShut {NoStop}%
\bibitem [{\citenamefont {Garberoglio}\ and\ \citenamefont
  {Harvey}(2021)}]{Garberoglio2021a}%
  \BibitemOpen
  \bibfield  {author} {\bibinfo {author} {\bibfnamefont {G.}~\bibnamefont
  {Garberoglio}}\ and\ \bibinfo {author} {\bibfnamefont {A.~H.}\ \bibnamefont
  {Harvey}},\ }\bibfield  {title} {\bibinfo {title} {Path-integral calculation
  of the fourth virial coefficient of helium isotopes},\ }\href@noop {}
  {\bibfield  {journal} {\bibinfo  {journal} {J. Chem. Phys.}\ }\textbf
  {\bibinfo {volume} {154}},\ \bibinfo {pages} {104107} (\bibinfo {year}
  {2021})}\BibitemShut {NoStop}%
\bibitem [{\citenamefont {{Gaiser}}\ and\ \citenamefont
  {{Fellmuth}}(2019)}]{2019JChPh.150m4303G}%
  \BibitemOpen
  \bibfield  {author} {\bibinfo {author} {\bibfnamefont {C.}~\bibnamefont
  {{Gaiser}}}\ and\ \bibinfo {author} {\bibfnamefont {B.}~\bibnamefont
  {{Fellmuth}}},\ }\bibfield  {title} {\bibinfo {title} {{Highly-accurate
  density-virial-coefficient values for helium, neon, and argon at 0.01 C
  determined by dielectric-constant gas thermometry}},\ }\href
  {https://doi.org/10.1063/1.5090224} {\bibfield  {journal} {\bibinfo
  {journal} {J.~Chem.~Phys.}\ }\textbf {\bibinfo {volume} {150}},\ \bibinfo
  {eid} {134303} (\bibinfo {year} {2019})}\BibitemShut {NoStop}%
\bibitem [{\citenamefont {{Lallemand}}\ and\ \citenamefont
  {{Vidal}}(1977)}]{1977JChPh..66.4776L}%
  \BibitemOpen
  \bibfield  {author} {\bibinfo {author} {\bibfnamefont {M.}~\bibnamefont
  {{Lallemand}}}\ and\ \bibinfo {author} {\bibfnamefont {D.}~\bibnamefont
  {{Vidal}}},\ }\bibfield  {title} {\bibinfo {title} {{Variation of the
  polarizability of noble gases with density}},\ }\href
  {https://doi.org/10.1063/1.433839} {\bibfield  {journal} {\bibinfo  {journal}
  {J.~Chem.~Phys.}\ }\textbf {\bibinfo {volume} {66}},\ \bibinfo {pages} {4776}
  (\bibinfo {year} {1977})}\BibitemShut {NoStop}%
\bibitem [{\citenamefont {{Huot}}\ and\ \citenamefont
  {{Bose}}(1991)}]{1991JChPh..95.2683H}%
  \BibitemOpen
  \bibfield  {author} {\bibinfo {author} {\bibfnamefont {J.}~\bibnamefont
  {{Huot}}}\ and\ \bibinfo {author} {\bibfnamefont {T.~K.}\ \bibnamefont
  {{Bose}}},\ }\bibfield  {title} {\bibinfo {title} {{Experimental
  determination of the dielectric virial coefficients of atomic gases as a
  function of temperature}},\ }\href {https://doi.org/10.1063/1.461801}
  {\bibfield  {journal} {\bibinfo  {journal} {J.~Chem.~Phys.}\ }\textbf
  {\bibinfo {volume} {95}},\ \bibinfo {pages} {2683} (\bibinfo {year}
  {1991})}\BibitemShut {NoStop}%
\bibitem [{\citenamefont {{Kirouac}}\ and\ \citenamefont
  {{Bose}}(1976)}]{1976JChPh..64.1580K}%
  \BibitemOpen
  \bibfield  {author} {\bibinfo {author} {\bibfnamefont {S.}~\bibnamefont
  {{Kirouac}}}\ and\ \bibinfo {author} {\bibfnamefont {T.~K.}\ \bibnamefont
  {{Bose}}},\ }\bibfield  {title} {\bibinfo {title} {{Polarizability and
  dielectric properties of helium}},\ }\href {https://doi.org/10.1063/1.432383}
  {\bibfield  {journal} {\bibinfo  {journal} {J.~Chem.~Phys.}\ }\textbf
  {\bibinfo {volume} {64}},\ \bibinfo {pages} {1580} (\bibinfo {year}
  {1976})}\BibitemShut {NoStop}%
\bibitem [{\citenamefont {Garberoglio}\ \emph {et~al.}(2011)\citenamefont
  {Garberoglio}, \citenamefont {Moldover},\ and\ \citenamefont
  {Harvey}}]{garberoglio2011improved}%
  \BibitemOpen
  \bibfield  {author} {\bibinfo {author} {\bibfnamefont {G.}~\bibnamefont
  {Garberoglio}}, \bibinfo {author} {\bibfnamefont {M.~R.}\ \bibnamefont
  {Moldover}},\ and\ \bibinfo {author} {\bibfnamefont {A.~H.}\ \bibnamefont
  {Harvey}},\ }\bibfield  {title} {\bibinfo {title} {{Improved first-principles
  calculation of the third virial coefficient of helium}},\ }\href@noop {}
  {\bibfield  {journal} {\bibinfo  {journal} {J. Res. Natl. Inst. Stand.
  Technol.}\ }\textbf {\bibinfo {volume} {116}},\ \bibinfo {pages} {729}
  (\bibinfo {year} {2011})}\BibitemShut {NoStop}%
\bibitem [{\citenamefont {Bich}\ \emph {et~al.}(2007)\citenamefont {Bich},
  \citenamefont {Hellmann},\ and\ \citenamefont {Vogel}}]{bich2007ab}%
  \BibitemOpen
  \bibfield  {author} {\bibinfo {author} {\bibfnamefont {E.}~\bibnamefont
  {Bich}}, \bibinfo {author} {\bibfnamefont {R.}~\bibnamefont {Hellmann}},\
  and\ \bibinfo {author} {\bibfnamefont {E.}~\bibnamefont {Vogel}},\ }\bibfield
   {title} {\bibinfo {title} {{Ab initio potential energy curve for the helium
  atom pair and thermophysical properties of the dilute helium gas. II.
  Thermophysical standard values for low-density helium}},\ }\href@noop {}
  {\bibfield  {journal} {\bibinfo  {journal} {Mol. Phys.}\ }\textbf {\bibinfo
  {volume} {105}},\ \bibinfo {pages} {3035} (\bibinfo {year}
  {2007})}\BibitemShut {NoStop}%
\bibitem [{\citenamefont {Rizzo}\ \emph {et~al.}(2002)\citenamefont {Rizzo},
  \citenamefont {H{\"a}ttig}, \citenamefont {Fern{\'a}ndez},\ and\
  \citenamefont {Koch}}]{rizzo2002effect}%
  \BibitemOpen
  \bibfield  {author} {\bibinfo {author} {\bibfnamefont {A.}~\bibnamefont
  {Rizzo}}, \bibinfo {author} {\bibfnamefont {C.}~\bibnamefont {H{\"a}ttig}},
  \bibinfo {author} {\bibfnamefont {B.}~\bibnamefont {Fern{\'a}ndez}},\ and\
  \bibinfo {author} {\bibfnamefont {H.}~\bibnamefont {Koch}},\ }\bibfield
  {title} {\bibinfo {title} {{The effect of intermolecular interactions on the
  electric properties of helium and argon. III. Quantum statistical
  calculations of the dielectric second virial coefficients}},\ }\href@noop {}
  {\bibfield  {journal} {\bibinfo  {journal} {J.~Chem.~Phys.}\ }\textbf
  {\bibinfo {volume} {117}},\ \bibinfo {pages} {2609} (\bibinfo {year}
  {2002})}\BibitemShut {NoStop}%
\bibitem [{\citenamefont {{Li}}\ and\ \citenamefont
  {{Hunt}}(1997)}]{li1997threedipoles}%
  \BibitemOpen
  \bibfield  {author} {\bibinfo {author} {\bibfnamefont {X.}~\bibnamefont
  {{Li}}}\ and\ \bibinfo {author} {\bibfnamefont {K.~L.~C.}\ \bibnamefont
  {{Hunt}}},\ }\bibfield  {title} {\bibinfo {title} {{Nonadditive three-body
  dipoles of inert gas trimers and
  H$_{2}${\ensuremath{\cdots}}H$_{2}${\ensuremath{\cdots}}H$_{2}$: Long-range
  effects in far infrared absorption and triple vibrational transitions}},\
  }\href {https://doi.org/10.1063/1.474790} {\bibfield  {journal} {\bibinfo
  {journal} {J.~Chem.~Phys.}\ }\textbf {\bibinfo {volume} {107}},\ \bibinfo
  {pages} {4133} (\bibinfo {year} {1997})}\BibitemShut {NoStop}%
\bibitem [{\citenamefont {Shaul}\ \emph
  {et~al.}(2012{\natexlab{b}})\citenamefont {Shaul}, \citenamefont {Schultz},
  \citenamefont {Kofke},\ and\ \citenamefont
  {Moldover}}]{shaul2012semiclassical}%
  \BibitemOpen
  \bibfield  {author} {\bibinfo {author} {\bibfnamefont {K.~R.~S.}\
  \bibnamefont {Shaul}}, \bibinfo {author} {\bibfnamefont {A.~J.}\ \bibnamefont
  {Schultz}}, \bibinfo {author} {\bibfnamefont {D.~A.}\ \bibnamefont {Kofke}},\
  and\ \bibinfo {author} {\bibfnamefont {M.~R.}\ \bibnamefont {Moldover}},\
  }\bibfield  {title} {\bibinfo {title} {{Semiclassical fifth virial
  coefficients for improved ab initio helium-4 standards}},\ }\href@noop {}
  {\bibfield  {journal} {\bibinfo  {journal} {Chem. Phys. Lett.}\ }\textbf
  {\bibinfo {volume} {531}},\ \bibinfo {pages} {11} (\bibinfo {year}
  {2012}{\natexlab{b}})}\BibitemShut {NoStop}%
\bibitem [{\citenamefont {Ram}\ and\ \citenamefont
  {Singh}(1973)}]{ram1973quantum}%
  \BibitemOpen
  \bibfield  {author} {\bibinfo {author} {\bibfnamefont {J.}~\bibnamefont
  {Ram}}\ and\ \bibinfo {author} {\bibfnamefont {Y.}~\bibnamefont {Singh}},\
  }\bibfield  {title} {\bibinfo {title} {{On the quantum corrections to the
  virial coefficients of the equation of state of a fluid}},\ }\href@noop {}
  {\bibfield  {journal} {\bibinfo  {journal} {Mol. Phys.}\ }\textbf {\bibinfo
  {volume} {26}},\ \bibinfo {pages} {539} (\bibinfo {year} {1973})}\BibitemShut
  {NoStop}%
\bibitem [{\citenamefont {Song}\ and\ \citenamefont
  {Luo}(2020)}]{song2020accurate}%
  \BibitemOpen
  \bibfield  {author} {\bibinfo {author} {\bibfnamefont {B.}~\bibnamefont
  {Song}}\ and\ \bibinfo {author} {\bibfnamefont {Q.-Y.}\ \bibnamefont {Luo}},\
  }\bibfield  {title} {\bibinfo {title} {{Accurate second dielectric virial
  coefficient of helium, neon, and argon from ab initio potentials and
  polarizabilities}},\ }\href@noop {} {\bibfield  {journal} {\bibinfo
  {journal} {Metrologia}\ }\textbf {\bibinfo {volume} {57}},\ \bibinfo {pages}
  {025007} (\bibinfo {year} {2020})}\BibitemShut {NoStop}%
\bibitem [{\citenamefont {{Hirschfelder}}\ \emph {et~al.}(1954)\citenamefont
  {{Hirschfelder}}, \citenamefont {{Curtiss}},\ and\ \citenamefont
  {{Bird}}}]{hirschfelder1954molecular}%
  \BibitemOpen
  \bibfield  {author} {\bibinfo {author} {\bibfnamefont {J.~O.}\ \bibnamefont
  {{Hirschfelder}}}, \bibinfo {author} {\bibfnamefont {C.~F.}\ \bibnamefont
  {{Curtiss}}},\ and\ \bibinfo {author} {\bibfnamefont {R.~B.}\ \bibnamefont
  {{Bird}}},\ }\href@noop {} {\emph {\bibinfo {title} {{Molecular theory of
  gases and liquids}}}}\ (\bibinfo  {publisher} {Wiley, New York},\ \bibinfo
  {year} {1954})\BibitemShut {NoStop}%
\bibitem [{\citenamefont {DeWitt}(1962)}]{dewitt1962analytic}%
  \BibitemOpen
  \bibfield  {author} {\bibinfo {author} {\bibfnamefont {H.~E.}\ \bibnamefont
  {DeWitt}},\ }\bibfield  {title} {\bibinfo {title} {{Analytic Properties of
  the Quantum Corrections to the Second Virial Coefficient}},\ }\href@noop {}
  {\bibfield  {journal} {\bibinfo  {journal} {J.~Math.~Phys.}\ }\textbf
  {\bibinfo {volume} {3}},\ \bibinfo {pages} {1003} (\bibinfo {year}
  {1962})}\BibitemShut {NoStop}%
\bibitem [{\citenamefont {Kihara}(1953)}]{kihara1953virial}%
  \BibitemOpen
  \bibfield  {author} {\bibinfo {author} {\bibfnamefont {T.}~\bibnamefont
  {Kihara}},\ }\bibfield  {title} {\bibinfo {title} {{Virial coefficients and
  models of molecules in gases}},\ }\href@noop {} {\bibfield  {journal}
  {\bibinfo  {journal} {Rev. Mod. Phys.}\ }\textbf {\bibinfo {volume} {25}},\
  \bibinfo {pages} {831} (\bibinfo {year} {1953})}\BibitemShut {NoStop}%
\bibitem [{\citenamefont {Guillot}\ and\ \citenamefont
  {Guissani}(1998)}]{guillot1998quantum}%
  \BibitemOpen
  \bibfield  {author} {\bibinfo {author} {\bibfnamefont {B.}~\bibnamefont
  {Guillot}}\ and\ \bibinfo {author} {\bibfnamefont {Y.}~\bibnamefont
  {Guissani}},\ }\bibfield  {title} {\bibinfo {title} {{Quantum effects in
  simulated water by the Feynman--Hibbs approach}},\ }\href@noop {} {\bibfield
  {journal} {\bibinfo  {journal} {J.~Chem.~Phys.}\ }\textbf {\bibinfo {volume}
  {108}},\ \bibinfo {pages} {10162} (\bibinfo {year} {1998})}\BibitemShut
  {NoStop}%
\bibitem [{\citenamefont {Feynman}(1972)}]{feynman1972statistical}%
  \BibitemOpen
  \bibfield  {author} {\bibinfo {author} {\bibfnamefont {R.~P.}\ \bibnamefont
  {Feynman}},\ }\href@noop {} {\emph {\bibinfo {title} {{Statistical Mechanics;
  A set of lectures}}}}\ (\bibinfo  {publisher} {W.A. Benjamin, Reading, MA},\
  \bibinfo {year} {1972})\BibitemShut {NoStop}%
\bibitem [{\citenamefont {{White}}\ and\ \citenamefont
  {{Gugan}}(1992)}]{1992Metro..29...37W}%
  \BibitemOpen
  \bibfield  {author} {\bibinfo {author} {\bibfnamefont {M.~P.}\ \bibnamefont
  {{White}}}\ and\ \bibinfo {author} {\bibfnamefont {D.}~\bibnamefont
  {{Gugan}}},\ }\bibfield  {title} {\bibinfo {title} {{Direct Measurements of
  the Dielectric Virial Coefficients of $^{4}$He between 3 K and 18 K}},\
  }\href {https://doi.org/10.1088/0026-1394/29/1/006} {\bibfield  {journal}
  {\bibinfo  {journal} {Metrologia}\ }\textbf {\bibinfo {volume} {29}},\
  \bibinfo {pages} {37} (\bibinfo {year} {1992})}\BibitemShut {NoStop}%
\bibitem [{\citenamefont {{Wolfram Research{,} Inc.}}(2008)}]{mathematica}%
  \BibitemOpen
  \bibfield  {author} {\bibinfo {author} {\bibnamefont {{Wolfram Research{,}
  Inc.}}},\ }\href {https://www.wolfram.com/mathematica} {\bibinfo {title}
  {Mathematica, {V}ersion 7.0}},\ \bibinfo {howpublished} {{C}hampaign, IL}
  (\bibinfo {year} {2008})\BibitemShut {NoStop}%
\end{thebibliography}%
\end{document}